\documentclass[superscriptaddress,nofootinbib,prd,reprint]{revtex4-2}
 
\usepackage[dvipdfmx]{graphicx}
\usepackage{amsmath}
\usepackage{subfigure}
\usepackage[dvipsnames]{xcolor}
\usepackage{hyperref}
\usepackage{braket}
\usepackage{here}
\usepackage{ulem}
\usepackage{cancel}
\hypersetup{
 bookmarksnumbered=true,
 colorlinks=true,
 linkcolor=[rgb]{0.098,0.098,0.439},
 citecolor=[rgb]{0.098,0.098,0.439},
 urlcolor=[rgb]{0.098,0.098,0.439}
}

\begin{document}

\title{\boldmath Searching for new physics effects in future $W$ mass and $\sin^2\theta_W (Q^2)$ determinations}
\author{Hooman Davoudiasl}
\affiliation{High Energy Theory Group, Physics Department, Brookhaven National Laboratory, Upton, NY 11973, USA}
\author{Kazuki Enomoto}
\affiliation{Department of Physics, KAIST, Daejeon 34141, Korea}
\author{Hye-Sung Lee}
\affiliation{Department of Physics, KAIST, Daejeon 34141, Korea}
\author{Jiheon Lee}
\affiliation{Department of Physics, KAIST, Daejeon 34141, Korea}
\author{William J. Marciano}
\affiliation{High Energy Theory Group, Physics Department, Brookhaven National Laboratory, Upton, NY 11973, USA}
\date{September 2023}

\begin{abstract}
We investigate the phenomenology of the dark $Z$ boson, $Z_d$, which is associated with a new Abelian gauge symmetry and couples to the standard model particles via kinetic mixing $\varepsilon$ and mass mixing $\varepsilon_Z^{}$. We examine two cases: (i) $Z_d$ is lighter than the $Z$ boson, and (ii) $Z_d$ is heavier than that. In the first case, it is known that $Z_d$ causes a deviation in the weak mixing angle at low energies from the standard model prediction. We study the prediction in the model and compare it with the latest experimental data. In the second case, the $Z$-$Z_d$ mixing enhances the $W$ boson mass. We investigate the effect of $Z_d$ on various electroweak observables including the $W$ boson mass using the $S$, $T$, and $U$ parameters. We point out an interesting feature: in the limit $\varepsilon \to 0$, the equation $S = - U$ holds independently of the mass of $Z_d$ and the size of $\varepsilon_Z^{}$, while $|S|\gg |U|$ in many new physics models. We find that the dark $Z$ boson with a mass of $\mathcal{O}(100)~\mathrm{GeV}$ with a relatively large mass mixing can reproduce the CDF result within $2\sigma$ while avoiding all other experimental constraints. Such dark $Z$ bosons are expected to be tested at future high-energy colliders. 
\end{abstract}

\maketitle

\section{Introduction} 

The standard model (SM) is excellent in describing the behavior of the particles and various experimental results~\cite{PDG2022}. However, there exist some mysterious phenomena which are unexplained in the SM such as neutrino oscillations~\cite{Super-Kamiokande:1998kpq, SNO:2001kpb}, dark matter~\cite{Planck:2018vyg}, baryon asymmetry of the Universe~\cite{Planck:2018vyg, Fields:2019pfx}, and some experimental anomalies in the measurements of the $g_\mu-2$~\cite{Muong-2:2021ojo}, the $W$ boson mass~\cite{CDF:2022hxs}, and so on.   
They suggest new physics beyond the SM. 
Thus, it is important to investigate how such new physics models can be tested in current and future experiments. 

In Ref.~\cite{Davoudiasl:2012ag}, a new physics model with a new dark gauge symmetry $U(1)_d$ has been proposed, which is called the dark $Z$ model. 
Although the SM particles do not have a dark charge, they can interact with the dark gauge boson $Z_d$ via kinetic and mass mixing. 
In this model, in addition to the gauge sector, the Higgs sector is also extended by adding the second Higgs doublet $\Phi_2$ and the dark singlet $\Phi_d$ which carry the dark charge.  
$Z_d$ acquires the mass from the vacuum expectation values (VEVs) of $\Phi_2$ and $\Phi_d$, which is a quite different feature from the typical dark photon model of the kinetic mixing~\cite{Holdom:1985ag}, where only the VEV of $\Phi_d$ induces the mass of the dark gauge boson. 
The VEV of $\Phi_2$ causes the mass mixing which is independent of the kinetic mixing and provides a new source of the parity violation in four fermion processes~\cite{Davoudiasl:2012ag}.   
The phenomenology of the dark $Z$ model has been studied in Refs.~\cite{Davoudiasl:2012ag, Davoudiasl:2012qa, Lee:2013fda, Davoudiasl:2013aya,  Davoudiasl:2014kua, Davoudiasl:2014mqa, Davoudiasl:2015bua, Xu:2015wja, San:2022uud, Goyal:2022vkg, Datta:2022zng}.  

In this paper, we examine the phenomenological impact of the model on the measurements of the running weak mixing angle and the $W$ boson mass. 
First, we focus on the dark $Z$ bosons whose mass is smaller than the $Z$ boson mass. 
As studied in the previous works~\cite{Davoudiasl:2012ag, Davoudiasl:2012qa, Davoudiasl:2015bua}, such light dark $Z$ bosons can make a deviation in the weak mixing angle at low energies. 
We update results from previous works with the latest experimental data. 

Second, as a main new part of this paper, we consider the dark $Z$ bosons which are heavier than the $Z$ boson. 
Such dark $Z$ bosons can enhance the prediction of the $W$ boson mass via the deviations in the mass and the gauge couplings of the $Z$ boson induced by the kinetic mixing $\varepsilon$ and the mass mixing $\varepsilon_Z^{}$~\cite{Holdom:1990xp, Burgess:1993vc, Babu:1997st}. 
The deviation in the SM gauge sector is described by the $S$, $T$, and $U$ paramters~\cite{Peskin:1990zt}. 
We derive the $S$, $T$, and $U$ parameters in the model and discuss the effect of $Z_d$ on various electroweak observables including the $W$ boson mass. 
We investigate their behavior in detail and find a remarkable fact that the equation $S = - U$ holds in the limit $\varepsilon \to 0$ independently of the mass of $Z_d$ and the size of the mass mixing $\varepsilon_Z^{}$, while $|S| \gg |U|$ in many new physics models~\cite{Grinstein:1991cd}. 
We show that the effect of heavy $Z_d$ can be large enough to explain the $W$ boson mass anomaly reported by the CDF collaboration~\cite{CDF:2022hxs} while avoiding the constraint from electroweak global fit. 
Such dark $Z$ bosons are expected to have a mass of $\mathcal{O}(100)~\mathrm{GeV}$ and relatively large mass mixing. 
We also discuss the direct searches of $Z_d$ at LHC. 
The model can explain the $W$ boson anomaly, consistent with the current LHC data in some mass regions of $Z_d$. 
Such a dark $Z$ boson is expected to be tested at future high-energy colliders.

The rest of this paper is organized as follows. 
In Sec.~\ref{sec: model}, we overview the dark $Z$ model. 
In Sec.~\ref{sec: running_WMA}, we review the effect of the dark $Z$ boson on the running weak mixing angle with the latest experimental results. 
In Sec.~\ref{sec: Wboson_anomaly}, the $W$ boson mass in the dark $Z$ model is discussed. We will show that the latest CDF-II result, which is significantly different from the SM prediction, can be explained in the model under the current constraints from the electroweak precision measurement and collider searches. 
The summary and conclusion are presented in Sec.~\ref{sec: summary}. 
Some formulas and discussions of the running weak mixing angle are shown in Appendix~\ref{app: WMA}. 
The effect of a heavy $Z_d$ on the $S$, $T$, and $U$ parameters is discussed in Appendix~\ref{app: SMEFT} using higher-dimensional operators.

\section{\boldmath Dark $Z$ model}
\label{sec: model}

In this section, we review the dark $Z$ model proposed in Ref.~\cite{Davoudiasl:2012ag} and investigated in Refs.~\cite{Davoudiasl:2012ag, Davoudiasl:2012qa, Lee:2013fda, Davoudiasl:2013aya,  Davoudiasl:2014kua, Davoudiasl:2014mqa, Davoudiasl:2015bua, Xu:2015wja, San:2022uud, Goyal:2022vkg}. 
This model has a new Abelian gauge symmetry denoted by $U(1)_d$ in addition to the gauge symmetries in the SM, $SU(3)_C \times SU(2)_L \times U(1)_Y$. 
Fermionic fields in the model and their quantum numbers are the same as those in the SM, i.e. the quarks and leptons do not carry the charge of $U(1)_d$, which is denoted by $Q_d$ in the following. 

Although the SM fermions do not carry the $U(1)_d$ charge, they interact with a new gauge boson via mixings including a kinetic mixing~\cite{Holdom:1985ag}.
The kinetic terms for the Abelian gauge bosons are given by 
\begin{equation}
\mathcal{L}_{\text{kin}} = 
- \frac{ 1 }{ 4 } \hat{B}^{\mu\nu} \hat{B}_{\mu\nu}
+ \frac{ \varepsilon }{ 2 \cos \theta_W } \hat{B}^{\mu\nu} \hat{Z}_{d \mu\nu}
- \frac{ 1 }{ 4 } \hat{Z}_d^{\mu\nu} \hat{Z}_{d \mu\nu}, 
\end{equation}
where $\hat{F}_{\mu\nu} = \partial_\mu \hat{F}_\nu - \partial_\nu \hat{F}_\mu$ with $F=B$ or $Z_d$, and $\hat{B}_\mu$ and $\hat{Z}_{d \mu}$ are gauge bosons associated with $U(1)_Y$ and $U(1)_d$ symmetries, respectively. 
The angle $\theta_W$ is the weak mixing angle defined as in the SM;
 \begin{equation}
 \label{eq: WMA_tree}
\tan \theta_W = \frac{ g^\prime }{ g }, 
\end{equation} 
where $g$ and $g^\prime$ are coupling constants of the $SU(2)_L$ and $U(1)_Y$ gauge interactions, respectively. 
The kinetic mixing is described by the dimensionless free parameter $\varepsilon$, whose normalization follows Ref.~\cite{Davoudiasl:2012ag}. 
The kinetic terms can be diagonalized by the following $GL(2, R)$ transformation; 
\renewcommand{\arraystretch}{1.5}
\begin{equation}
\label{eq: mixing_hat_tilde}
\begin{pmatrix}
\tilde{B}_\mu \\
\tilde{Z}_{d \mu} \\
\end{pmatrix}
=
\begin{pmatrix}
1 & - \varepsilon/c_W^{} \\[3pt]
0 & \sqrt{ 1 - \varepsilon^2/ c_W^2}\\
\end{pmatrix}
\begin{pmatrix}
\hat{B}_\mu \\
\hat{Z}_{d \mu} \\
\end{pmatrix}, 
\end{equation}
where $c_W^{} = \cos \theta_W$.

The discussion so far is common with the dark photon~\cite{Holdom:1985ag}. 
However, the nature of the electroweak symmetry breaking in the dark $Z$ model is quite different from that in the dark photon model. 
The dark $Z$ model includes three scalar fields: two isospin doublets with the hypercharge $Y=1/2$ ($\Phi_1$ and $\Phi_2$) and an isospin singlet with $Y=0$ ($\Phi_d$). 
They are color singlets, and their $U(1)_d$ charges are given by $Q_d[\Phi_1] = 0$ and $Q_d[\Phi_2] = Q_d[\Phi_d] = 1$. 
They acquire VEVs as follows;
\renewcommand{\arraystretch}{1.0}
\begin{equation}
\Phi_i = \frac{ 1 }{ \sqrt{2} }
\begin{pmatrix}
0 \\
v_i^{} \\
\end{pmatrix}
, \quad 
\Phi_d = \frac{ 1 }{ \sqrt{2} } v_d^{}, 
\end{equation}
where $i=1,2$. 
The electroweak symmetry is broken by $v_1^{}$ and $v_2^{}$, while the $U(1)_d$ symmetry is broken by $v_2^{}$ and $v_d^{}$. The dark photon model corresponds to the limit $v_2^{} \to 0$. 
For a detailed study of the Higgs sector in the model, see Ref.~\cite{Lee:2013fda}.\footnote{
The notation of the scalar fields in this paper is different from that in Ref.~\cite{Lee:2013fda}, where $Q_d[\Phi_1]=1$ and $Q_d[\Phi_2] = 0$ is assumed. 
The notation in this paper follows Ref.~\cite{Davoudiasl:2012ag}. } 

The VEVs $v_1^{}$, $v_2^{}$, and $v_d^{}$ give masses to the electroweak and dark gauge bosons. 
Their mass terms are given by
\begin{equation}
\mathcal{L}_{\text{mass}} = \tilde{m}_W^2 W^+_\mu W^{- \mu} 
	+ \frac{ 1 }{ 2 } \bigl( \tilde{Z}^\mu, \tilde{Z}_d^\mu \bigr) M_V^2 
	\begin{pmatrix}
	\tilde{Z}_\mu \\[5pt]
	\tilde{Z}_{d\mu} \\
	\end{pmatrix}, 
\end{equation}
where $\tilde{Z}_\mu = - \sin \theta_W \tilde{B}_\mu + \cos \theta_W \hat{W}^3_\mu$, 
$W^\pm_\mu = (\hat{W}^1_\mu \mp i \hat{W}^2_\mu ) / \sqrt{2}$ with $\hat{W}^a_\mu$ ($a = 1$, 2, 3) being the gauge fields of the $SU(2)_L$ symmetry. 
$W_\mu^\pm$ correspond to the $W$ bosons with the mass $\tilde{m}_W^{} = g v / 2 = 80.4~\mathrm{GeV}$, where $v = \sqrt{v_1^2 + v_2^2}\simeq 246~\mathrm{GeV}$.
The squared masses of the neutral gauge bosons are given by a $2 \times 2$ symmetric matrix $M_V^2$, 
whose elements are given by
\begin{align}
& (M_V^2)_{11} = \tilde{m}_Z^2, \\
& (M_V^2)_{12} =  (M_V^2)_{21} = - \tilde{m}_Z^2 \eta ( \varepsilon_Z^{} + \varepsilon t_W^{} ), \\
& (M_V^2)_{22} = \tilde{m}_{Z_d}^2 + \tilde{m}_Z^2 \eta^2 (2 \varepsilon_Z^{} \varepsilon t_W + \varepsilon^2 t_W^2 ),  
\end{align}
with $\tilde{m}_Z^2$ and $\tilde{m}_{Z_d}^2$ defined by
\begin{equation}
\tilde{m}_Z^2 = \frac{ 1 }{ 4 } g_Z^2 v^2, \quad 
	\tilde{m}_{Z_d}^2 = \eta^2 g_d^2 \bigl( v_2^2 + v_d^2 \bigr), 
\end{equation}
where $\eta = (1 - \varepsilon^2 / c_W^2)^{-1/2}$, $t_W^{} = \tan \theta_W^{}$, 
$g_Z = \sqrt{ g^2 + g^{\prime 2}}$, and $g_d$ is the coupling constant of the $U(1)_d$ gauge interaction.
Off-diagonal terms are proportional to a mass mixing parameter $\varepsilon_Z^{}$ given by 
\begin{equation}
\label{eq: EpsilonZ}
\varepsilon_Z^{} = \frac{ 2 g_d^{} }{ g_Z^{} } \sin^2 \beta , 
\end{equation}
where the angle $\beta$ satisfies $\tan \beta = v_2^{} / v_1^{}$.
The mass mixing $\varepsilon_Z$ is independent of the kinetic mixing $\varepsilon$. 
Therefore, $\varepsilon_Z^{}$ provides a new source of the interactions between the SM fermions and the additional gauge boson. 

$M_V^2$ can be diagonalized by an appropriate $SO(2)$ rotation with an angle $\xi$; 
\renewcommand{\arraystretch}{1.5}
\begin{equation}
\begin{pmatrix}
Z_\mu \\
Z_{d\mu} \\
\end{pmatrix}
= 
\begin{pmatrix}
\cos \xi & - \sin \xi \\
\sin \xi & \cos \xi \\
\end{pmatrix}
\begin{pmatrix}
\tilde{Z}_\mu \\
\tilde{Z}_{d\mu} \\
\end{pmatrix}. 
\end{equation}
We identify $Z_\mu$ and $Z_{d \mu}$ as the observed $Z$ boson and the dark $Z$ boson, respectively. 
The mixing angle $\xi$ then satisfies
\begin{equation}
\sin 2 \xi = - \frac{ 2 (M_V^2)_{12} }{ m_Z^2 - m_{Z_d}^2 }, 
\end{equation}
where $m_Z^{} \simeq 91.2~\mathrm{GeV}$ and $m_{Z_d}$ are the masses of the $Z$ boson and the dark $Z$ boson, respectively. 

In the following, we assume $\varepsilon$ and $\varepsilon_Z^{}$ are small and use perturbative expansions. 
Up to the quadratic order, $m_Z^2$, $m_{Z_d}^2$, $\sin \xi$ and $\cos \xi$ are given by
\begin{align}
\label{eq: Z_mass}
& m_Z^2 \simeq \tilde{m}_Z^2 
	\bigg(
		1 + \frac{ (\varepsilon_Z^{} + \varepsilon t_W)^2 }{ 1 - \tilde{r}^2 }
	\bigg), \\[10pt]
\label{eq: Zd_mass}
& m_{Z_d}^2 \simeq \tilde{m}_{Z_d}^2
	\bigg(
		1 - \frac{ \varepsilon_Z^2 }{ \tilde{r}^2 ( 1 - \tilde{r}^2 ) }
		- \frac{ \varepsilon^2 t_W^2 + 2 \varepsilon_Z^{} \varepsilon t_W }{ 1 - \tilde{r}^2 }
	\bigg), \\[10pt]
\label{eq: sin_Xi}
& \sin \xi \simeq \frac{ \varepsilon_Z^{} + \varepsilon t_W }{ 1 - r^2 }, \\[10pt]
\label{eq: cos_Xi}
& \cos \xi \simeq 1 - \frac{ 1 }{ 2 } \biggl( \frac{ \varepsilon_Z^{} + \varepsilon t_W }{ 1 - r^2 } \biggr)^2, 
\end{align}
where $\tilde{r} = \tilde{m}_{Z_d}^{} / \tilde{m}_Z^{}$, and $r = m_{Z_d}^{} / m_Z^{}$. 
These formulas give a good approximation in the case that both $|\varepsilon_Z^{}|$ and $|\varepsilon t_W^{}|$ are sufficiently smaller than one and $|1 - \tilde{r}^2|$.
We extended the results of the previous works, which used only the leading order terms,
to include the higher order terms\footnote{Equation~(\ref{eq: Zd_mass}) coincides with the result in Ref.~\cite{Lee:2013fda}.} as they may be crucial in discussing the $W$ boson mass studied in Sec.~\ref{sec: Wboson_anomaly}.
In the limit $\varepsilon_Z^{} \to 0$, 
Eqs.~(\ref{eq: Z_mass})-(\ref{eq: cos_Xi}) reproduce the results in the simplest dark photon model~\cite{Holdom:1990xp, Cheng:2022aau}. 

One should note that $r$ is different from $\tilde{r}$ at the quadratic order.
While $r$ is the ratio of the physical masses $m_Z^{}$ and $m_{Z_d}^{}$, 
$\tilde{r}$ has a physical meaning only in the no mixing limit, $\varepsilon_Z^{} \to 0$ and $\varepsilon \to 0$. 
The relation between them is given by 
\begin{equation}
\label{eq: r_and_rtilde}
\tilde{r}^2 = r^2 + \frac{ 1 + r^2 }{ 1 - r^2 } \varepsilon_Z^2 + \frac{ 2 r^2 }{ 1 - r^2 } ( \varepsilon^2 t_W^2 + 2 \varepsilon_Z \varepsilon t_W^{} ). 
\end{equation}

The kinetic and mass mixings lead to the interaction between the dark $Z$ boson and the SM fermions as follows;
\begin{equation}
\mathcal{L}_{\text{d}} = 
	- \bigg(
		e \eta \varepsilon c_\xi^{} J_{\text{em}}^\mu 
		+ \frac{ g_Z^{} }{ 2 } (s_\xi - \eta \varepsilon t_W c_\xi ) J_{\text{NC}}^\mu 
	\bigg) Z_{d\mu}, 
\end{equation}
where $s_\xi = \sin \xi$, $c_\xi = \cos \xi$, $e$ is the coupling constant of the electromagnetic interaction, and $J_{\text{em}}^\mu$ and $J_{\text{NC}}^\mu$ are the electromagnetic and the weak neutral currents, respectively.
At the leading order of the mixing parameters, 
the current interactions can be approximated by
\begin{equation}
\label{eq: current_darkZ}
\mathcal{L}_{\text{d}} \simeq
	- \bigg(
		e \varepsilon J_{\text{em}}^\mu 
		+ \frac{ g_Z^{} }{ 2 } 
            \Bigl(\frac{ \varepsilon_Z^{} + r^2 \varepsilon t_W  }{ 1 - r^2 } \Bigr)
            J_{\text{NC}}^\mu 
	\bigg) Z_{d\mu}. 
\end{equation}
The $\varepsilon_Z^{}$ induces an additional parity-violating source compared to the dark photon model ($\varepsilon_Z^{} \to 0$). 
It gives a particularly important effect when $m_{Z_d}^{} \ll m_Z^{}$. 
In that case, $\mathcal{L}_d$ is approximately given by
\begin{equation}
\mathcal{L}_{\text{d}} \simeq
	- \bigg(
		e \varepsilon J_{\text{em}}^\mu 
		+ \frac{ g_Z^{} }{ 2 } 
            \bigl(\varepsilon_Z^{} + r^2 \varepsilon t_W \bigr)
            J_{\text{NC}}^\mu 
	\bigg) Z_{d\mu}. 
\end{equation}
Although the parity-violating effect induced by $\varepsilon$ is much suppressed by $r^2 = m_{Z_d}^2 / m_Z^2 \ll 1$, 
that proportional to $\varepsilon_Z^{}$ does not have such a suppression. 
Since $\varepsilon_Z^{}$ is independent of $\varepsilon$, 
the dark $Z$ model can provide a larger parity violation compared to the dark photon model, with a light gauge boson.
Such an effect can be tested by precisely measuring the weak mixing angle at low-energy experiments as discussed in Refs.~\cite{Davoudiasl:2012ag, Davoudiasl:2012qa, Davoudiasl:2015bua} and the next section. 

Before closing this section, we describe a parameter $\delta$ satisfying 
\begin{equation}
\varepsilon_Z^{} = \frac{ m_{Z_d}^{} }{ m_Z^{} } \delta, 
\end{equation}
which allows smooth $m_{Z_d}^{} \to 0$ behavior for $\varepsilon_Z^{}$-induced amplitudes involving $Z_d^{}$~\cite{Davoudiasl:2012ag}. 
By using $\delta$, the neutral current interaction is given by 
$- (g_Z^{}/2) r \delta^\prime J_\mathrm{NC}^\mu Z_{d\mu}$, where 
\begin{equation}
\delta^\prime = \delta + \frac{ m_{Z_d}^{} }{ m_Z^{} } \varepsilon \tan \theta_W^{},  
\end{equation}
in the case of $m_{Z_d}^{}$ not too small compared to $m_Z^{}$~\cite{Davoudiasl:2015bua}.
The parameter $\delta^\prime$ is particularly convenient when we investigate the phenomenology of the light dark $Z$ boson.

\section{\boldmath Running weak mixing angle and the light dark $Z$ boson} 
\label{sec: running_WMA}

As discussed in previous works~\cite{Davoudiasl:2012ag, Davoudiasl:2012qa, Davoudiasl:2015bua}, 
the light dark $Z$ boson can shift the value of the weak mixing angle $\sin^2 \theta_W^{}$ from the SM prediction at low energies $|q^2| \lesssim m_{Z_d}^{2}$. 
Such a deviation can be tested by precise measurements of the running of the weak mixing angle. 
In this section, we show the latest experimental values of the weak mixing angle and compare them with the SM prediction at $m_Z$ and also using the running weak mixing angle, which provides an update of previous work~\cite{Davoudiasl:2015bua}, using the latest experimental data. 

\subsection{Running weak mixing angle}
\label{sec: WMAdata}

First, we shortly review the current situation of the weak mixing angle measurements. 
There are several ways to define the weak mixing angle~\cite{PDG2022}. 
We employ the one defined in the modified minimal subtraction ($\overline{\mathrm{MS}}$) scheme; 
\begin{equation}
\sin^2 \hat{\theta}_W^{} (\mu) = \frac{ \hat{e}^2 (\mu) }{ \hat{g}^2 (\mu) },
\end{equation}
where $\hat{e}(\mu)$ and $\hat{g}(\mu)$ are the gauge couplings of the electromagnetic and $SU(2)_L$ gauge interactions, respectively, evaluated at the relevant mass scale $\mu$. 
The renormalization scale $\mu$ is usually set to be $m_Z^{}$, 
and the SM prediction is given by~\cite{PDG2022}
\begin{equation}
\label{eq: WMA_SM}
    (\mathrm{SM}) \quad \sin^2 \hat{\theta}_W^{} (m_Z^{}) = 0.23122(04) . 
\end{equation}

The value of $\sin^2 \hat{\theta}_W^{}$ at the $Z$ pole is measured by high-energy colliders, LEP~\cite{ALEPH:2005ab}, SLC~\cite{ALEPH:2005ab}, Tevatron~\cite{CDF:2016cei, D0:2017ekd, CDF:2018cnj} and LHC~\cite{ATLAS:2015ihy, ATLAS:2018gqq, CMS:2018ktx, LHCb:2015jyu}. 
In these experiments, the value of the effective angle for leptons $\sin^2 \theta_\mathrm{eff}^\mathrm{lept}$~\cite{LEP:1991hsu} was measured. 
We note that the value of $\sin^2 \theta_\mathrm{eff}^\mathrm{lept}$ is different from that of $\sin^2 \hat{\theta}_W^{}(m_Z^{})$ although both are defined at the $Z$ pole. 
The relation between them was investigated in Ref.~\cite{Gambino:1993dd} and is numerically given by~\cite{PDG2022}
\begin{equation}
    \sin^2 \theta_\mathrm{eff}^\mathrm{lept} \simeq \sin^2 \hat{\theta}_W^{}(m_Z^{}) + 0.00032. 
\end{equation}
We use this equation to derive the value of $\sin^2 \hat{\theta}_W^{}(m_Z^{})$ corresponding to the observed $\sin^2 \theta_\mathrm{eff}^\mathrm{lept}$ values in each experiment.

At LEP and SLC, various asymmetries at final states were precisely measured, which are sensitive to parity violation. 
The average values of $\sin^2 \hat{\theta}_W^{}$ in leptonic and hadronic (semileptonic) processes are given by
\begin{align}
\label{eq: leptonic_angle}
& \text{(leptonic)} \quad
	\ \sin^2 \hat{\theta}_W^{}(m_Z^{}) = 0.23081(21), \\
\label{eq: hadronic_angle}
& \text{(hadronic)} \quad
	 \sin^2 \hat{\theta}_W^{}(m_Z^{}) = 0.23190(27), 
\end{align}
respectively~\cite{ALEPH:2005ab}.
The former is derived from observations of the lepton forward-backward (FB)  asymmetry of a charged lepton $\ell$ ($A_\mathrm{FB}^{0, \ell}$) and the $\tau$ polarization ($P_\tau^{}$) at LEP and the lepton asymmetry parameter ($\mathcal{A}_\ell$) at SLC. 
The latter is derived from $A_\mathrm{FB}^{0, b}$, $A_\mathrm{FB}^{0, c}$ and the hadronic charge asymmetry $Q_\mathrm{FB}^\mathrm{had}$ at LEP. 
Results from each observable are shown in Fig.~7.6 of Ref.~\cite{ALEPH:2005ab}. 
The intriguing point is that the average values in leptonic and hadronic processes differ by $3.2\sigma$~\cite{ALEPH:2005ab}. 
A new physics scenario to explain this tension was proposed in Ref.~\cite{Choudhury:2001hs}. 
However, the cause of it is still unclear. 

At Tevatron and LHC, the weak mixing angle was measured in the FB asymmetry of $e^+e^-$ and $\mu^+\mu^-$. 
Although their center-of-mass beam energy is much higher than $m_Z^{}$, 
the observed quantity is at the $Z$ pole because the invariant mass of the final state is dominated by values around the $Z$ pole. 
Their results are given by 
\begin{align}
\text{(Tevatron)}\quad \sin^2 \hat{\theta}_W^{}(m_Z^{}) = 0.23116(33), \\ 
\text{(LHC)} \quad  \sin^2  \hat{\theta}_W^{}(m_Z^{}) = 0.23097(33),  
\end{align}
where the former is the combined result of CDF~\cite{CDF:2016cei} and D0~\cite{D0:2017ekd} experiments given in Ref.~\cite{CDF:2018cnj}, and the latter is the average of ATLAS~\cite{ATLAS:2015ihy, ATLAS:2018gqq}, CMS~\cite{CMS:2018ktx} and LHCb~\cite{LHCb:2015jyu} experiments given in Ref.~\cite{PDG2022}.

The weak mixing angle also has been measured in low-energy experiments such as atomic parity violation (APV) in ${}^{133}\mathrm{Cs}$~\cite{Wood:1997zq, Guena:2004sq} and low-energy accelerators: Qweak ($e^- - p$ elastic scattering)~\cite{Qweak:2018tjf}, E158 (M\o ller scattering)~\cite{SLACE158:2005uay}, and PVDIS (deep inelastic scattering of $e^-$ and deuteron)~\cite{PVDIS:2014cmd};  
\begin{align}
\label{eq: WMA_APV}
\text{(APV)} \quad & \sin^2 \hat{\theta}_W^{}(m_Z^{}) = 0.2293(17), \\
\label{eq: WMA_Qweak}
\text{(Qweak)} \quad & \sin^2 \hat{\theta}_W^{}(m_Z^{}) = 0.2308(11), \\
\text{(E158)} \quad & \sin^2 \hat{\theta}_W^{}(m_Z^{}) = 0.2330(15), \\
\label{eq: WMA_PVDIS}
\text{(PVDIS)} \quad & \sin^2 \hat{\theta}_W^{}(m_Z^{}) = 0.2299(43),    
\end{align}
where the relevant energy scales of each experiment are given by $Q\simeq $ $2.4~\mathrm{MeV}$~\cite{Bouchiat:1983uf}, $157~\mathrm{MeV}$~\cite{Qweak:2018tjf}, $161~\mathrm{MeV}$~\cite{SLACE158:2005uay}, and $1.38~\mathrm{GeV}$~\cite{PVDIS:2014cmd}, respectively.\footnote{
The result of APV experiments in Eq.~(\ref{eq: WMA_APV}) is derived by using the value shown in Ref.~\cite{PDG2022}, $\sin^2 \hat{\theta}_W^{}(2.4~\mathrm{MeV}) = 0.2367(18)$, and the renormalization group equation in Ref.~\cite{Erler:2004in}. 
In the same way, we translated the Qweak result in Ref.~\cite{Qweak:2018tjf}, $\sin^2 \hat{\theta}_W^{}(0) = 0.2383(11)$, into the value at the $Z$ pole in Eq.~(\ref{eq: WMA_Qweak}).}
Although there are other low-energy measurements of the weak mixing angle, we do not show their results here because of their relatively large uncertainties \cite{PDG2022}. 

\begin{figure}
\subfigure[]{
\includegraphics[width=0.43\textwidth]{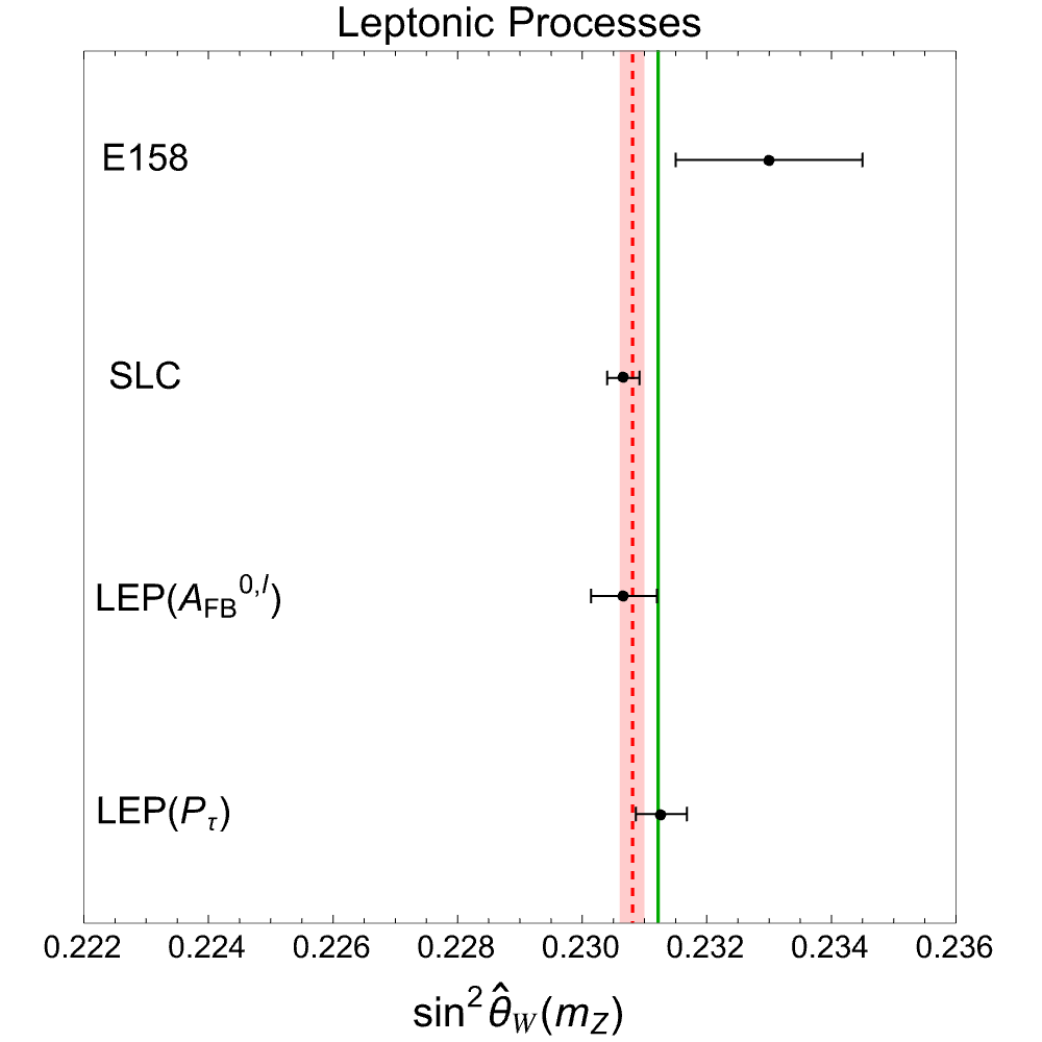}
\label{fig: leptonic_angle}
}
\subfigure[]{
\includegraphics[width=0.43\textwidth]{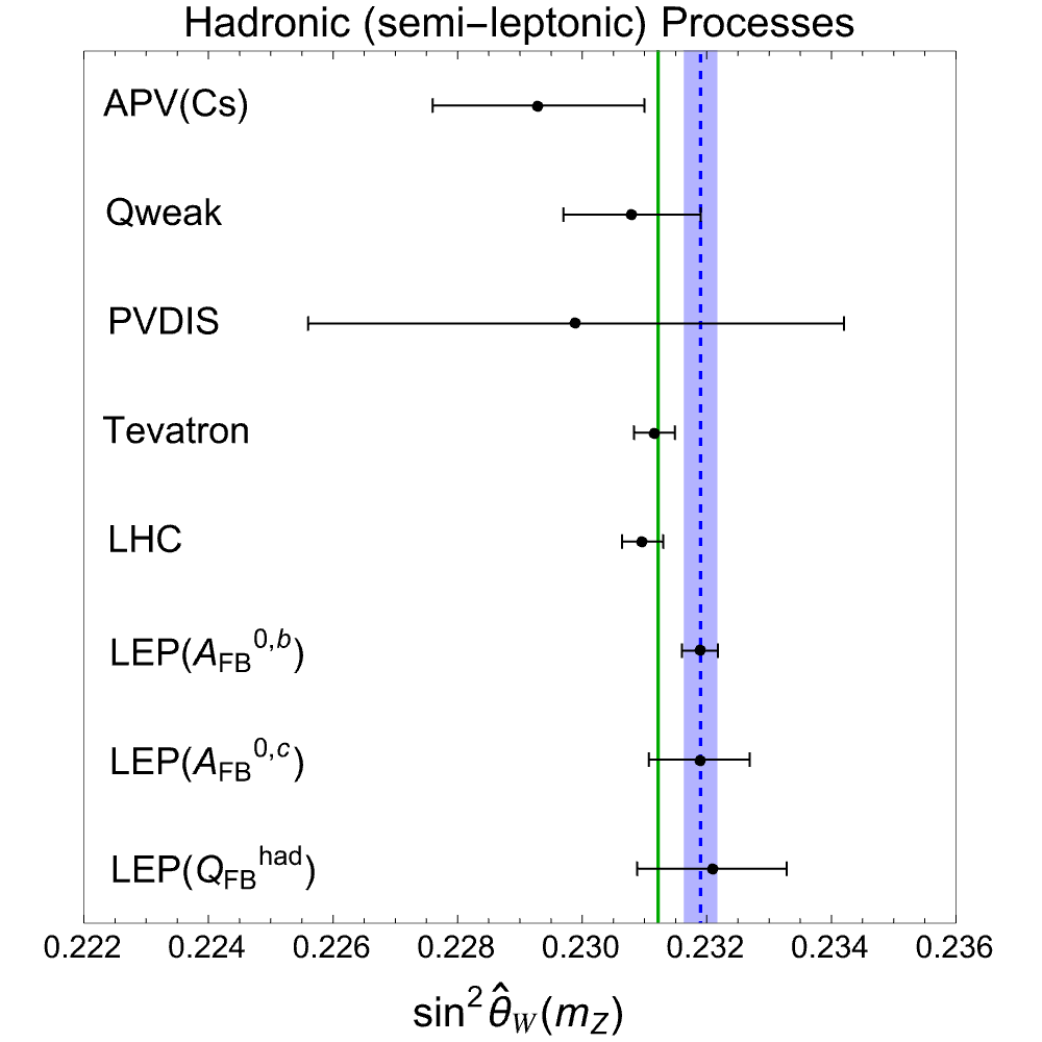}
\label{fig: hadronic_angle}
}
\caption{The weak mixing angle measured in (a) leptonic and (b) hadronic processes. 
The red (blue) band represents the average value in Eq.~\eqref{eq: leptonic_angle} [Eq.~\eqref{eq: hadronic_angle}] with $1\sigma$ uncertainties. 
The SM prediction is also shown with the green line.}
\end{figure}

In Figs.~\ref{fig: leptonic_angle} and~\ref{fig: hadronic_angle}, 
we summarize the experimental values of $\sin^2 \hat{\theta}_W^{}(m_Z^{})$ and the SM prediction. 
In Fig.~\ref{fig: leptonic_angle} [Fig.~\ref{fig: hadronic_angle}], 
$\sin^2 \hat{\theta}_W^{} (m_Z^{})$ in leptonic (hadronic) processes are shown with error bars of $1\sigma$. 
The red (blue) band in Fig.~\ref{fig: leptonic_angle} [Fig.~\ref{fig: hadronic_angle}] shows the LEP and SLC average value in Eq.~(\ref{eq: leptonic_angle}) [Eq.~(\ref{eq: hadronic_angle})] with $1\sigma$ uncertainties. 
The dashed lines in the bands represent the central values. 
The green solid line in both figures shows the SM prediction in Eq.~(\ref{eq: WMA_SM}) with a very small $1\sigma$ uncertainty. 
All experimental results are consistent with the SM prediction. 
However, it is interesting that the leptonic and hadronic average values deviate from the SM prediction in opposite directions. 

Another way to compare the SM prediction and the experimental values is to use the running of the weak mixing angle~\cite{Marciano:1980be, Czarnecki:1998xc, Czarnecki:2000ic, Ferroglia:2003wa, Erler:2004in, Erler:2017knj}; 
\begin{equation}
\label{eq: runningWMA}
    \sin^2\hat{\theta}_W^{}(q^2) = 
        \hat{\kappa}(q^2) \sin^2 \hat{\theta}_W^{}(m_Z^{}), 
\end{equation}
where $\hat{\kappa}(q^2)$ is the form factor including the electroweak radiative corrections evaluated in the $\overline{\mathrm{MS}}$ scheme.

We employ $\hat{\kappa}(q^2)$ discussed in Refs.~\cite{Czarnecki:1998xc, Czarnecki:2000ic}, which is defined in the context of the radiative correction in M\o ller scattering~\cite{Czarnecki:1995fw}. 
Then, the main radiative correction comes from the $\gamma$-$Z$ self-energy and the anapole moment~\cite{Czarnecki:1998xc, Czarnecki:2000ic,Czarnecki:1995fw}. 
The concrete formula for $\hat{\kappa}(q^2)$ is shown in Appendix.~\ref{app: WMA}. 
We also provide a careful comparison of the running weak mixing angle formula we adopted and another formula based on the pinch technique~\cite{Ferroglia:2003wa} in the Appendix~\ref{app: WMA}.

By using Eq.~(\ref{eq: runningWMA}) and the formula of $\hat{\kappa}(q^2)$, we can evaluate the SM prediction of the running as a function of $q^2$, the momentum transfer of neutral current processes.  
We can also evaluate $\sin^2\hat{\theta}_W^{} (q_\mathrm{exp}^2)$ corresponding to the measured $\sin^2 \hat{\theta}_W^{}(m_Z)$ values, where $q^2_\mathrm{exp}$ is typical energy scale of each experiment.

\begin{figure}[b]
\begin{center}
    \includegraphics[width=0.48\textwidth]{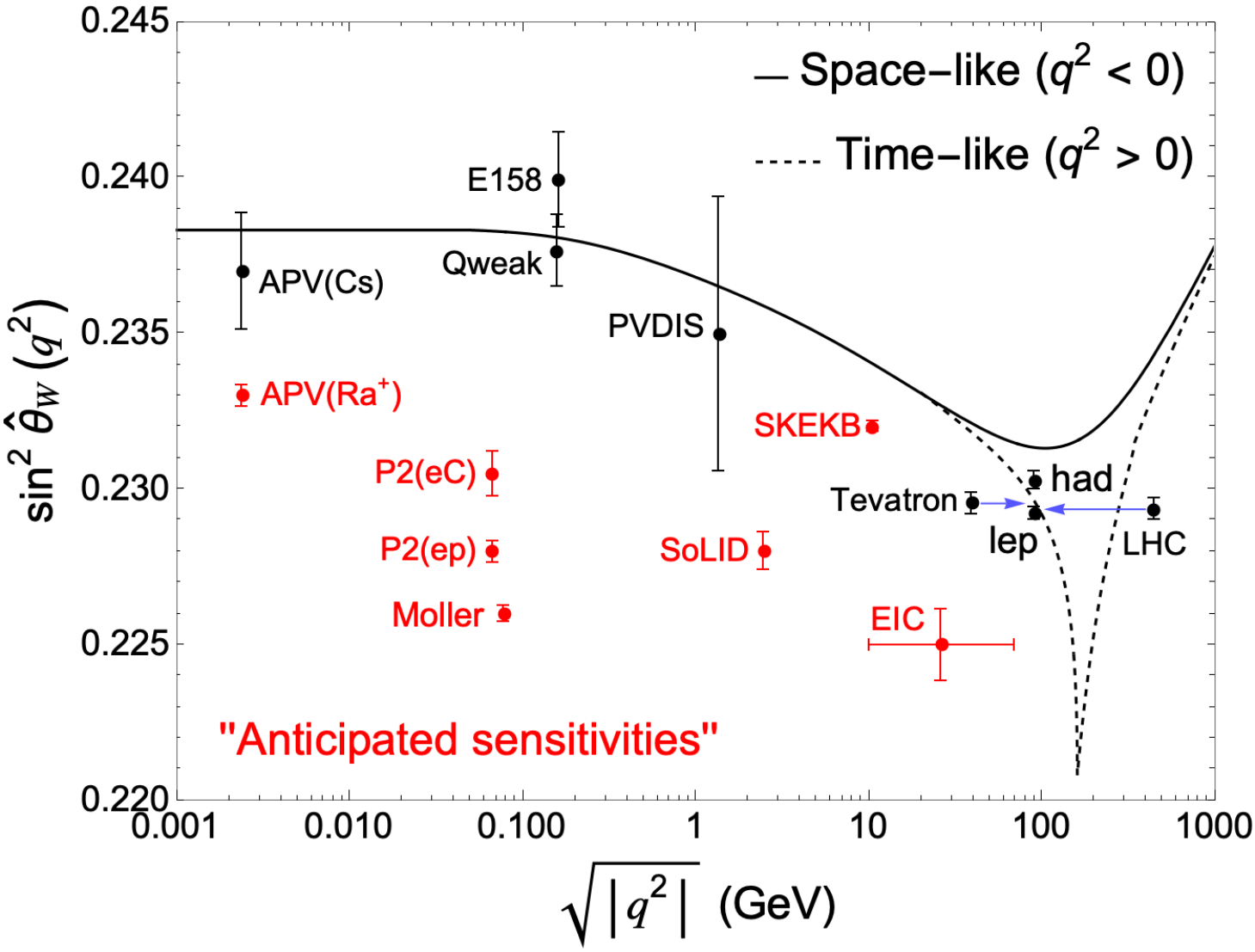}
    \caption{Comparing the experimental values and the SM prediction of the running weak mixing angle in the $\overline{\mathrm{MS}}$ scheme [Eq.~(\ref{eq: runningWMA})]. The black solid (dashed) lines are the SM prediction for spacelike (timelike) momenta. 
    The black points are existing experimental data, and the red points are future anticipated sensitivities. The curves for timelike momenta are shown only in a domain $\sqrt{|q^2|} > 20~\mathrm{GeV}$. (See Appendix~\ref{app: WMA}.).}
    \label{fig: RunningWMA}
\end{center}
\end{figure}

In Fig.~\ref{fig: RunningWMA}, 
we show the SM prediction and the experimental values of the running $\sin^2 \hat{\theta}_W^{}(q^2)$. 
The black solid (dashed) lines are the SM prediction for spacelike (timelike) momenta. 
The black points are experimental values corresponding to $\sin^2\hat{\theta}_W^{}(m_Z^{})$ shown above with $1\sigma$ uncertainties.  
The point ``lep'' (``had'') represents the LEP and SLC average values in Eq.~(\ref{eq: leptonic_angle}) [Eq.~(\ref{eq: hadronic_angle})]. 
The blue arrows mean that the points ``Tevatron'' and ``LHC'' are data at the $Z$ pole. 
The points at the $Z$ pole are obtained from measurements of processes with timelike momentum transfer ($q^2 > 0$). 
On the other hand, the points at low energies are data from processes with spacelike momentum transfer ($q^2 < 0$). 

In Fig.~\ref{fig: RunningWMA}, we also show the anticipated sensitivity in future experiments: APV in $\mathrm{Ra}^+$~\cite{RaAPV}, P2 with the option of the proton target and the carbon target~\cite{Becker:2018ggl}, Moller~\cite{MOLLER:2014iki}, 
SoLID~\cite{Chen:2014psa}, SuperKEKB with a polarized electron beam~\cite{USBelleIIGroup:2022qro}, and Electron Ion Collider (EIC)~\cite{AbdulKhalek:2021gbh, AbdulKhalek:2022hcn} with the red points ``APV($\mathrm{Ra}^+$),'' ``P2(ep),'' ``P2(eC),'' ``Moller,'' ``SOLID,'' ``SKEKB,'' and ``EIC,'' respectively. 
The EIC has several beam options~\cite{AbdulKhalek:2021gbh, AbdulKhalek:2022hcn}, and it is represented by the error bar for $\sqrt{|q^2|}$ at the point ``EIC.''

\subsection{\boldmath Effects of a light dark $Z$ boson}
\label{sec: WMA_and_DarkZ}

In this section, we discuss the effect of a light dark $Z$ boson ($m_{Z_d}^{} = \mathcal{O}(10)~\mathrm{GeV}$) on the running weak mixing angle. 
A light dark $Z$ boson can change the running from the SM prediction at low momentum transfer $|q^2| \lesssim m_{Z_d}^2$~\cite{Davoudiasl:2012ag, Davoudiasl:2015bua}. 
Such a deviation is expected to be tested in future low-energy experiments shown in Fig.~\ref{fig: RunningWMA}. 
In this section, we are interested in low-energy experiments, 
and we consider neutral current processes with spacelike momentum transfers $q^2 = -Q^2 <0$ and a light dark $Z$ boson of $m_{Z_d}^{} = \mathcal{O}(10)~\mathrm{GeV}$.

As explained in Sec.~\ref{sec: model}, the dark $Z$ boson interacts with the SM fermions via kinetic and mass mixing. 
Thus, it modifies the weak neutral current amplitudes and the weak mixing angle. 
The modification in the weak mixing angle is described by a factor $\kappa_d(Q^2)$~\cite{Davoudiasl:2012ag, Davoudiasl:2015bua} given by
\begin{equation}
\label{eq: kappa_factor}
\kappa_d^{}(Q^2) = 1 - \varepsilon \delta^\prime \frac{ m_Z^{} }{ m_{Z_d}^{} } \cot \theta_W^{} \frac{ m_{Z_d}^2 }{ Q^2 + m_{Z_d}^2 }.  
\end{equation}
By using $\kappa_d$, the running weak mixing angle in the dark $Z$ model is given by $\kappa_d^{}(Q^2) \sin^2 \hat{\theta}_W^{}(-Q^2)_\mathrm{SM}^{}$, where $\sin^2 \hat{\theta}_W^{}(-Q^2)_\mathrm{SM}^{}$ is the SM prediction evaluated with Eq.~(\ref{eq: runningWMA}). 
At $Q^2 \ll m_{Z_d}^2$, the deviation from the SM is evaluated by~\cite{Davoudiasl:2015bua} 
\begin{align}
\Delta \sin^2 \hat{\theta}_W^{}(Q^2) = & 
	(\kappa_d^{} (Q^2) - 1)  \sin^2 \hat{\theta}_W^{}(-Q^2)_\mathrm{SM}^{} \nonumber \\
	\simeq & -0.42 \varepsilon \delta^\prime \frac{ m_Z^{} }{ m_{Z_d}^{} }.
\end{align}

The dark $Z$ boson effect on the running weak mixing angle is proportional to $\varepsilon \delta^\prime$. 
Thus, we consider the experimental bound for the combination of the mixing parameters $\varepsilon \delta^\prime$. 
The kinetic mixing is constrained by the electroweak precision tests~\cite{Davoudiasl:2012ag}. 
Following Refs.~\cite{Hook:2010tw, Curtin:2014cca}, 
we employ the bound
\begin{equation}
\label{eq: constraint_EWPT}
    |\varepsilon | \lesssim 0.03, 
\end{equation}
for $m_{Z_d} = \mathcal{O}(10)~\mathrm{GeV}$.

As the constraint on $\delta^\prime$--$m_{Z_d}$ plane, Refs.~\cite{Davoudiasl:2012ag, Davoudiasl:2013aya, Davoudiasl:2015bua} investigated the rare Higgs decay $H \to Z Z_d \to \ell_1^+ \ell^-_1 \ell_2^+ \ell_2^-$ ($\ell_{1,2}^{} = e, \mu$). 
We revisit this constraint with the latest result of the rare Higgs decay searches at the ATLAS~\cite{ATLAS:2021ldb} and CMS~\cite{CMS:2021pcy} experiments.
The branching ratio of $H \to Z Z_d^{}$ is given by 
\begin{align}
\label{eq: HZZd}
& \mathrm{BR}[H \to Z Z_d^{}] = 
\frac{ 1 }{ \Gamma_H^{} }
\frac{ \sqrt{ \lambda (m_H^2, m_Z^2, m_{Z_d}^2) } }{ 16 \pi m_H^3 } \left( \frac{ g m_Z^{} }{ \cos \theta_W^{} } \right)^2 
\nonumber \\
& \times \left( \delta^\prime \frac{ m_{Z_d} }{ m_Z } \frac{s_{\beta- \alpha}}{t_\beta} \right)^2
\left(\frac{ (m_H^2 - m_Z^2 - m_{Z_d}^2)^2 }{4 m_Z^2 m_{Z_d}^2 } + 2 \right),
\end{align}
where $t_\beta = \tan \beta^{}$, $s_{\beta - \alpha}^{} = \sin ( \beta- \alpha)$, $\Gamma_H^{}$ is the decay width of $H$, and $\lambda(x,y,z) = x^2 + y^2 + z^2 -2xy -2xz -2yz$. 

The mixing angle $\alpha$ is defined in the context of two Higgs doublet models and diagonalizes neutral CP-even scalar states~\cite{Branco:2011iw}.\footnote{We neglect the mixing between the Higgs doublets and the singlet scalar for simplicity.}
We here assume that the heavier CP-even scalar boson $H$ is the SM Higgs boson with the mass $125~\mathrm{GeV}$. 
In the formula of $\mathrm{BR}[H \to Z Z_d]$ shown in the previous works~\cite{Davoudiasl:2012ag, Davoudiasl:2013aya, Davoudiasl:2015bua}, 
$s_{\beta - \alpha}^{}/t_\beta = 1$ is implicitly assumed, while it is retained Eq.~(\ref{eq: HZZd}), which is consistent with that in Ref.~\cite{Lee:2013fda}.
In the following, we consider the constraint in the case that $s_{\beta - \alpha}^{}/t_\beta = 1$ according to Refs.~\cite{Davoudiasl:2012ag, Davoudiasl:2013aya, Davoudiasl:2015bua}. 
However, it is interesting to note that the branching ratio for $H \to Z Z_d^{}$ vanishes at tree level in the alignment limit $\cos(\beta- \alpha) = 1$, where the Higgs boson couplings coincide with the SM prediction at tree level.

The branching ratio $\mathrm{Br}(H\to Z Z_d \to 4 \ell)$ is constrained by the current data at the ATLAS~\cite{ATLAS:2021ldb} and CMS~\cite{CMS:2021pcy} experiments. 
The CMS result gives a slightly stronger bound on the $\delta^\prime$-$m_{Z_d}^{}$ plane. 
By using $\mathrm{Br}(Z_d^{} \to e^+ e^- ) + \mathrm{Br}(Z_d^{} \to \mu^+ \mu^- ) \simeq 0.3$~\cite{Batell:2009yf}, the current bound on $\delta^\prime$ with $s_{\beta-\alpha} / t_\beta^{} = 1$ is given by 
\begin{equation}
\label{eq: constraint_HiggsRare}
|\delta^\prime| \lesssim 0.004, 
\end{equation}
for $15~\mathrm{GeV} < m_{Z_d} < 35~\mathrm{GeV}$.
For lighter dark $Z$ bosons, for example, $m_{Z_d} = 10~\mathrm{GeV}$, there is almost no constraint from $H \to Z Z_d^{}$ because of contamination from heavy quarkonia~\cite{ATLAS:2021ldb, CMS:2021pcy}. 
We note that this constraint becomes weaker if we consider a dark decay channel of $Z_d^{}$ or $|s_{\beta - \alpha}^{}/t_\beta^{}| < 1$. 
 
Although we have obtained the constraint for $|\varepsilon|$ [Eq.~(\ref{eq: constraint_EWPT})] and $|\delta^\prime|$ [Eq.~(\ref{eq: constraint_HiggsRare})] for $m_{Z_d}^{} = \mathcal{O}(10)~\mathrm{GeV}$, we will comment on other conceivable experimental constraints in the following. 

Refs.~\cite{Davoudiasl:2012ag, Datta:2022zng} investigated the rare meson decays $K \to \pi Z_d$ and $B \to K Z_d$. These are generated at one-loop level~\cite{Davoudiasl:2012ag} because the tree-level flavor changing neutral current via the additional Higgs bosons is prohibited by $U(1)_d$ symmetry~\cite{Glashow:1976nt}. These processes give an effective constraint if $m_{Z_d}$ is smaller than $|m_K - m_\pi| \simeq \mathcal{O}(100)~\mathrm{MeV}$ and $|m_B - m_K|\simeq \mathcal{O}(1)~\mathrm{GeV}$~\cite{Davoudiasl:2012ag}. However, in the relevant mass region ($m_{Z_d} = \mathcal{O}(10)~\mathrm{GeV}$), the constraint is expected to be inefficient.

For the relevant mass region, a strong bound comes from the dimuon resonance search of $Z_d$ at CMS~\cite{CMS:2019buh} and LHCb~\cite{LHCb:2019vmc} experiments. In the pure dark photon case ($\varepsilon_Z^{}=0$), 
it is given by 
\begin{equation}
\label{eq: constraint_Dimuon}
    \varepsilon^2 \lesssim (0.5\text{--}1)\times 10^{-6}, 
\end{equation}
for $m_{Z_d}^{} = \mathcal{O}(10)~\mathrm{GeV}$ if $Z_d$ decays only into the SM particles. 
This bound is stronger than that from the electroweak precision measurements in Eq.~(\ref{eq: constraint_EWPT}). 
However, this can be diluted if a decay channel into new light dark sector particles $\chi$ opens. Since the decay rate $Z_d \to \chi \bar{\chi}$ is not suppressed by the kinetic and mass mixing, it can be a dominant decay mode, and $\mathrm{BR}[Z_d\to \mu^+\mu^-]$ can be small enough to avoid the constraint.

The light $Z_d$ can also be generated via the $Z$ boson decay resulting in the four-lepton final state: $Z \to Z_d \ell^+ \ell^- \to 4 \ell$~\cite{Park:2015gdo}.
If $Z_d$ decays only into the SM particles, the prediction on the decay obtains a constraint from Eqs.~(\ref{eq: constraint_HiggsRare}) and (\ref{eq: constraint_Dimuon}). For $m_{Z_d} = 10~\mathrm{GeV}$, it is given by $\mathrm{BR}(Z\to Z_d \ell^+\ell^-) \lesssim 1 \times 10^{-9}$. 
We used \texttt{MadGraph5\_aMC@NLO}~\cite{Alwall:2014hca} in the evaluation. 
It yields $\mathrm{BR}(Z\to Z_d \ell^+ \ell^- \to 4 \ell) \lesssim 3 \times 10^{-10}$. 
If $Z_d$ predominantly decays into $\chi \bar{\chi}$, 
the constraints in Eqs.~(\ref{eq: constraint_EWPT}) and (\ref{eq: constraint_HiggsRare}) lead to $\mathrm{BR}(Z\to Z_d \ell^+ \ell^- \to 4 \ell) \lesssim 2 \times 10^{-7} \times \mathrm{BR}(Z_d \to \ell^+ \ell^-)$ with $\mathrm{BR}(Z_d \to \ell^+ \ell^-) \ll 1$. 
On the other hand, the current data for the four-lepton decay of the $Z$ boson is $\mathrm{BR}(Z \to 4 \ell) = (4.55\pm 0.17)\times 10^{-6}$~\cite{PDG2022}.
Thus, in both cases, the signal from the on-shell $Z_d$ is hidden in the uncertainties, and we expect no constraint from this decay channel. 

In the following, we consider constraints available in the case that $Z_d$ predominantly decays into $\chi \bar{\chi}$, where the severe constraint in Eq.~(\ref{eq: constraint_Dimuon}) can be avoided. Then, the exotic Higgs decays $H \to Z(\gamma) Z_d \to Z(\gamma) \cancel{E}$ are induced, where $\chi \bar{\chi}$ are identified as missing energy. 
$H \to Z Z_d$ is generated at tree level as shown in Eq.~(\ref{eq: HZZd}), and we expect that the experimental constraint from this channel is weaker than Eq.~(\ref{eq: constraint_HiggsRare}) because the final state $Z \cancel{E}$ includes the missing energy. 
On the other hand, $H \to \gamma Z_d$ is generated at one-loop level via the current and gauge interactions as $H \to \gamma \gamma$ or $H \to \gamma Z$ in the SM. 
If $Z_d$ is much lighter than $H$, the branching ratio is approximately evaluated by $\mathrm{BR}[H \to \gamma Z_d] \simeq 2\varepsilon^2 \mathrm{BR}[H \to \gamma \gamma]^\mathrm{SM} + \varepsilon_Z^2 \mathrm{BR}[H \to \gamma Z]^\mathrm{SM}|_{m_Z = 0} \simeq (2\varepsilon^2 + \varepsilon_Z^2) \times 10^{-3}$, where $\mathrm{BR}[\cdots]^\mathrm{SM}$ is the branching ratio calculated in the SM. Here, we have neglected the interference terms because they are not expected to drastically change the result. The experimental upper limit of $\mathrm{BR}[H \to \gamma Z_d]$ for $Z_d$ up to $40~\mathrm{GeV}$ is set to be $2.7\text{--}3.1\%$ at $95\%$ CL~\cite{ATLAS:2022xlo}. Thus, this process also does not give a constraint on the relevant parameter region.

The light dark sector particle $\chi$ may cause a deviation in the invisible decay rate of the $Z$ boson, which is strongly constrained~\cite{ALEPH:2005ab}, through the $Z$-$Z_d$ mixing.
However, we found that this constraint is weaker than the above constraints.\footnote{
Using the LEP bound $\Gamma[Z \to \mathrm{inv.}] < 0.6~\mathrm{MeV}$~\cite{ALEPH:2005ab} and the relevant inputs $\delta^\prime = 0.006$ and $m_{Z_d} = 10~\mathrm{GeV}$, 
we have a bound $\Gamma[Z_d \to \chi \bar{\chi}] \simeq (m_Z^{}/m_{Z_d}^{}) (\delta^{\prime})^{-2} \Gamma[Z \to \chi \bar{\chi}] \lesssim 150~\mathrm{GeV}$, which gives no constraint for $m_{Z_d}^{} = \mathcal{O}(10)~\mathrm{GeV}$. 
For smaller values of $|\delta^\prime|$, this constraint is weaker.}
Such a light dark sector particle could be searched for in beam dump experiments such as the proposed BDX~\cite{BDX:2016akw}. 

As a result, we consider the case of $\mathrm{BR}(Z_d \to \chi \bar{\chi}) \simeq 1$ in the following and employ the bounds in Eqs.~(\ref{eq: constraint_EWPT}) and (\ref{eq: constraint_HiggsRare}) for $\delta^\prime$ and $\varepsilon$, respectively.
We thus have the following current experimental bound on $\varepsilon \delta^\prime$;
\begin{equation}
\label{eq: constraint_EpsilonDelta}
|\varepsilon \delta^\prime| \lesssim 0.00012.
\end{equation}
We note that this bound is not effective for $m_{Z_d}^{} \lesssim 10~\mathrm{GeV}$ because of no constraint from $\mathrm{BR}[H \to Z Z_d]$ in this mass region.

We perform the $\chi^2$ fitting of $\varepsilon \delta^\prime$ and 
the measured $\sin^2\hat{\theta}_W^{}(-Q^2)$ at low energy experiments: APV in $^{133}\mathrm{Cs}$, E158, Qweak and PVDIS.\footnote{In a previous work~\cite{Davoudiasl:2015bua}, the fitting used the result by NuTeV collaboration~\cite{NuTeV:2001whx}, which was $3.0\sigma$ higher than the SM prediction. However, we do not use it here because of several concerns about the interpretation of the NuTeV result~\cite{ParticleDataGroup:2016lqr}.} 
Input data are shown in Table~\ref{table: low-energy_WMA}. 
As a benchmark point, we consider two cases $m_{Z_d}^{} = 15~\mathrm{GeV}$ and $m_{Z_d}^{} = 10~\mathrm{GeV}$. 
We found the following best-fit values of $\varepsilon \delta^\prime$
\begin{align}
\label{eq: best-fit_15GeV}
& \varepsilon \delta^\prime = -0.00025(29) \quad \text{for $m_{Z_d}^{} = 15~\mathrm{GeV}$}, \\
\label{eq: best-fit_10GeV}
& \varepsilon \delta^\prime = -0.00016(19) \quad \text{for $m_{Z_d}^{} = 10~\mathrm{GeV}$}, 
\end{align}
with $\chi^2 = 3.4$ in both cases. 
The fitted values are consistent with the SM prediction ($\varepsilon \delta^\prime = 0$) within $1\sigma$ because the experimental values do not deviate significantly from the SM prediction. 
However, with future improved accuracy,  
it may be possible to observe the effects of such light dark $Z$ bosons unless $\varepsilon \delta^\prime$ is too small. 

\begin{table}[t]
\begin{center}
\begin{tabular}{|c| c c c c|}\hline
 	& APV(${}^{133}$Cs) & Qweak & E158 & PVDIS \\\hline
 $Q^2$ ($\mathrm{GeV^2}$) & $5.8\times10^{-6}$ & 0.0248 & 0.026 & 1.901 \\
$ \sin^2 \hat{\theta}_W^{}(-Q^2)$ & 0.2370(18) & 0.2377(11) & 0.2399(15) & 0.2352(44) \\ \hline
\end{tabular}
\caption{Input data for the $\chi^2$ fitting of $\varepsilon \delta^\prime$.}
\label{table: low-energy_WMA}
\end{center}
\end{table} 

\begin{figure}
\subfigure[]{
\includegraphics[width=0.45\textwidth]{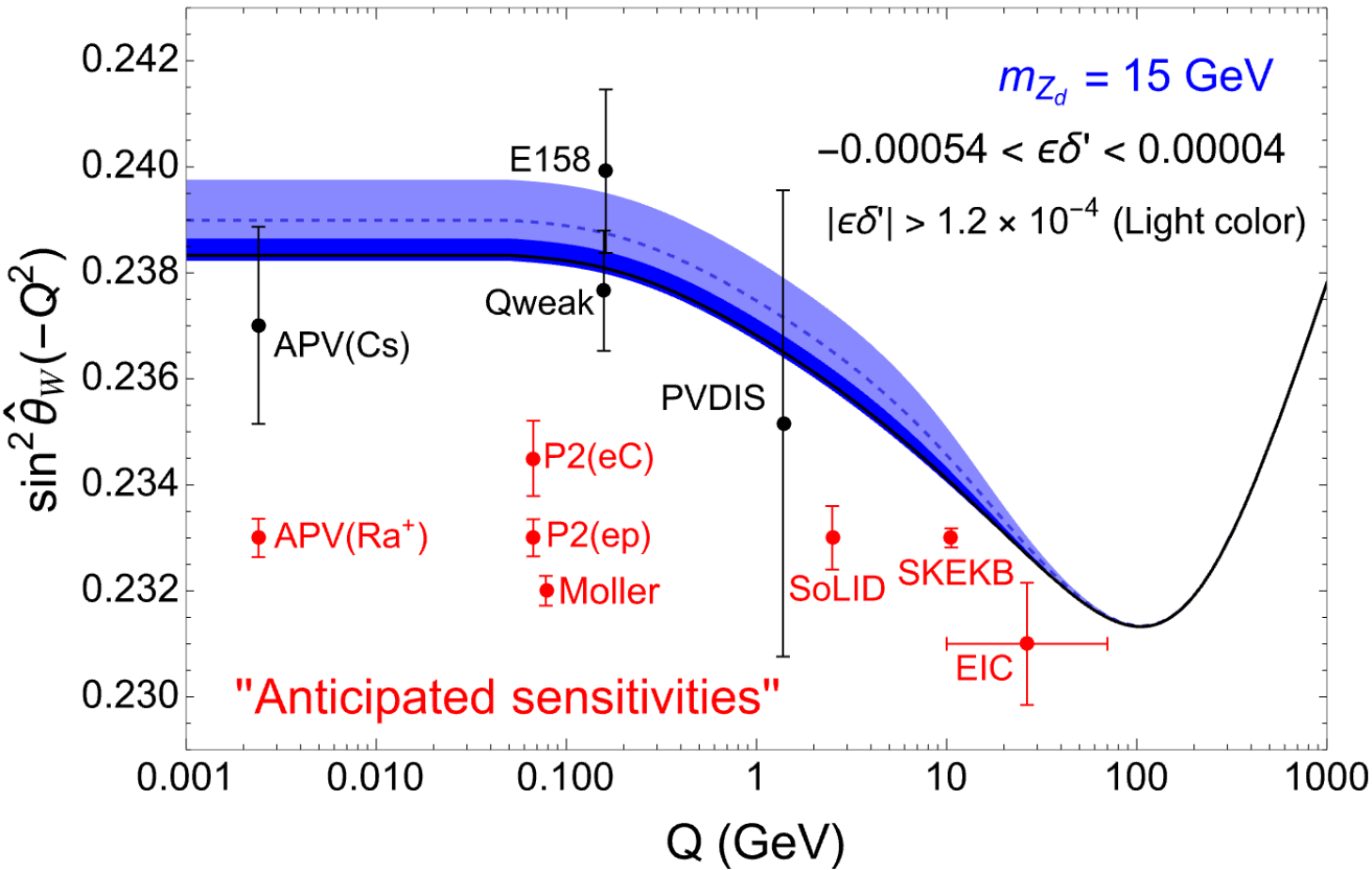}
\label{fig: fitting15GeV}
}
\subfigure[]{
\includegraphics[width=0.45\textwidth]{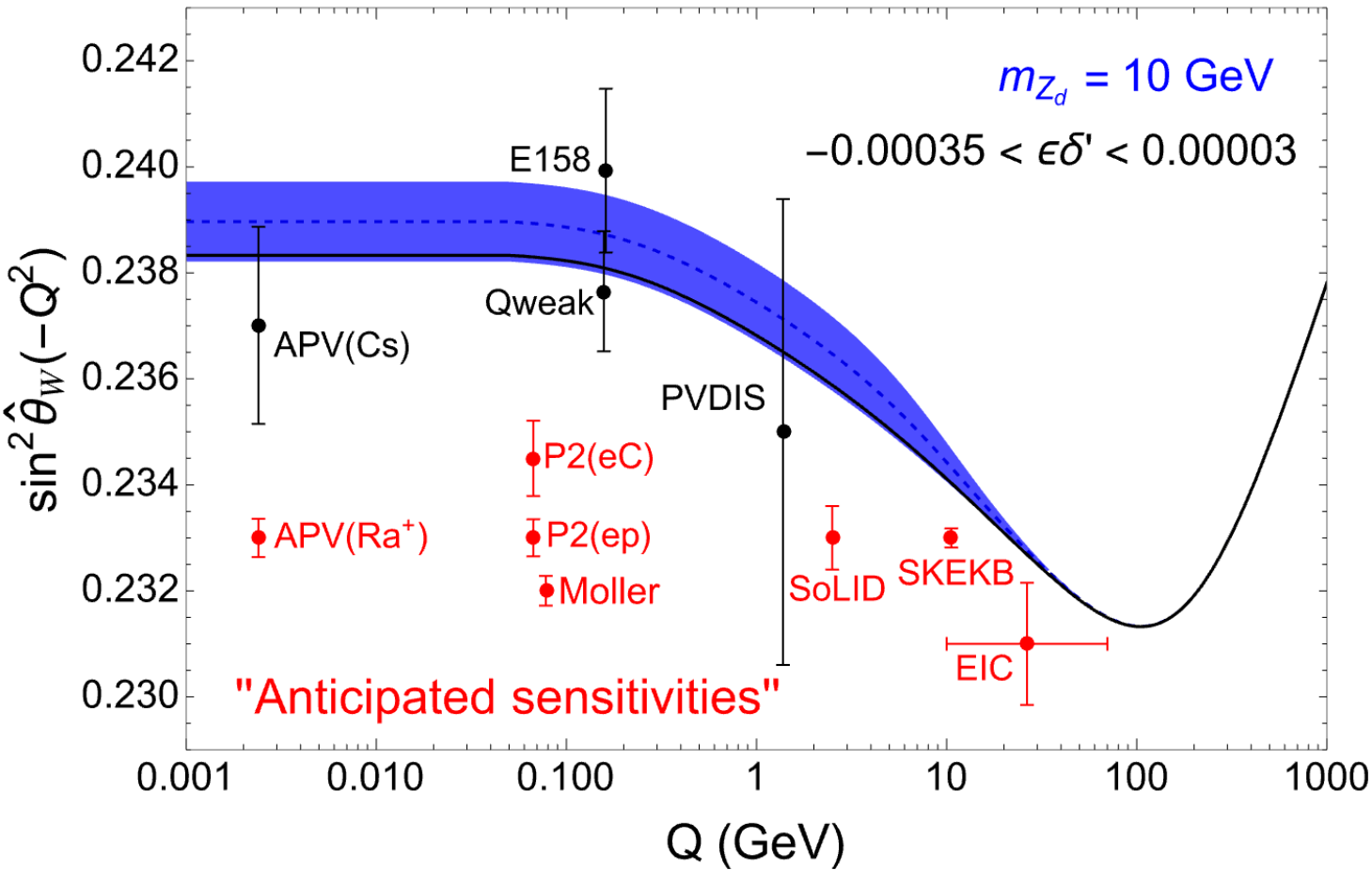}
\label{fig: fitting10GeV}
}
\caption{The effect of the dark $Z$ boson on the running weak mixing angle in the cases that (a) $m_{Z_d}^{} = 15~\mathrm{GeV}$ and (b) $m_{Z_d}^{} = 10~\mathrm{GeV}$. 
The black solid curve is the SM prediction in the $\overline{\mathrm{MS}}$ scheme. The light and dark blue bands represent the result of the $\chi^2$ fitting with $1\sigma$ uncertainties. The dashed curve is the prediction with the best-fit value of $\varepsilon \delta^\prime$. Points on the light blue band are excluded by the experimental bound in Eq.~(\ref{eq: constraint_EpsilonDelta}).}
\label{fig: fittingWMA}
\end{figure}

We show the results of the fitting for $m_{Z_d}^{} = 15~\mathrm{GeV}$ and $10~\mathrm{GeV}$ in Figs.~\ref{fig: fitting15GeV} and~\ref{fig: fitting10GeV}, respectively. 
Here, the black solid curves represent the SM prediction for spacelike momenta.
The effect of the dark $Z$ boson with the fitted $\varepsilon \delta^\prime$ is shown with the blue bands with $1\sigma$ uncertainties. 
The dashed curves in the bands are the result with the best-fit value of $\varepsilon \delta^\prime$. 
The darker blue bands represent points within the $1\sigma$ regions of the fitting which satisfy the experimental bound in Eq.~(\ref{eq: constraint_EpsilonDelta}). 
On the other hand, points in the lighter blue band are within the $1\sigma$ region but excluded by Eq.~(\ref{eq: constraint_EpsilonDelta}). 
We note that there is no light blue (excluded) region in Fig.~\ref{fig: fitting10GeV} ($m_{Z_d}^{} = 10~\mathrm{GeV}$) because of the aforementioned absence of constraint on $\delta^\prime$.

Before closing this section, we briefly comment on the constraints on the Higgs potential in the model. Because we assume no mixing between the doublet scalars and the singlet scalar, the latter does not couple to the SM fermions and gauge bosons at tree level. 
If the singlet scalar is lighter than $m_H/2$, it can induce the Higgs invisible decay, which is constrained by the current LHC data~\cite{PDG2022}. 
This constraint can be avoided by considering a singlet scalar heavier than $m_H/2$, or a sufficiently small Higgs portal coupling. 

The Higgs potential of the model includes that of the two Higgs doublet model. 
Since the Yukawa coupling of $\Phi_2$ is prohibited by the $U(1)_d$ symmetry, the Higgs doublets possess the same couplings as those in the type-I two Higgs doublet model~\cite{Lee:2013fda}. 
The mass of the additional Higgs bosons and the mixing angles $\alpha$ and $\beta$ are constrained by the electroweak precision measurements (including the Higgs measurements)~\cite{Bernon:2015qea, Haller:2018nnx, Chen:2019pkq}, direct searches for the additional Higgs bosons~\cite{Bernon:2015qea, Arbey:2017gmh, Chen:2019pkq, Aiko:2020ksl}, flavor observables~\cite{Enomoto:2015wbn, Arbey:2017gmh, Haller:2018nnx}, and theoretical constraints such as vacuum stability~\cite{Deshpande:1977rw,Klimenko:1984qx,Sher:1988mj,Nie:1998yn,Ferreira:2004yd}, triviality bound~\cite{Flores:1982pr,Kominis:1993zc,Kanemura:1999xf,Ferreira:2009jb}, and perturbative unitarity~\cite{Kanemura:1993hm,Akeroyd:2000wc,Ginzburg:2005dt,Kanemura:2015ska}. 

The above constraints can be avoided by appropriately choosing the parameters in the Higgs potential. 
In particular, in the alignment limit, we can take the decoupling limit of the new particles~\cite{Gunion:2002zf}, where all of their effects are described by higher-dimensional operators suppressed by the new physics scale. 
In such a limit, heavy additional Higgs bosons can naturally avoid all the constraints. 
In addition, the $125~\mathrm{GeV}$ Higgs couplings coincide with the SM ones in this limit. 
Although the deviation from the alignment limit is severely constrained by the current LHC data, there still exist allowed parameter regions near (but not) the alignment limit~\cite{Chen:2019pkq, Bernon:2015qea}. 
Since the above analysis on the running weak mixing angle is almost independent of the parameters in the Higgs potential,\footnote{Although the mass mixing is proportional to $\sin^2 \beta$, the choice of it does not conflict with the constraints as long as $|\varepsilon_Z^{}|$ is small. The Yukawa couplings of the charged and the CP-odd Higgs bosons are proportional to $\tan \beta$. That of the additional CP-even Higgs boson is also approximately proportional to $\tan \beta$ near the alignment limit. 
Thus, smaller $\sin \beta$ makes experimental constraints weaker.
Our definition of $\tan \beta$ is the inverse of the typical definition in the two Higgs doublet models in Ref.~\cite{Lee:2013fda}.} we can always choose the allowed parameters without changing $\sin^2 \hat{\theta}(q^2)$. 
Higgs phenomenology in the model is further discussed in Ref.~\cite{Lee:2013fda}.

\section{\boldmath $W$ boson mass and a heavy dark $Z$ boson} 
\label{sec: Wboson_anomaly}

In this section, we discuss the effect of a dark $Z$ boson on the $W$ boson mass. 
The mixing between $Z$ and $Z_d$ induces deviations in the SM gauge sectors. 
Thus, predictions of the masses and the couplings of the gauge bosons change from the SM ones. 
In particular, as shown below, heavy dark $Z$ bosons enhance the $W$ boson mass, 
and it can explain the anomaly found by the CDF collaboration~\cite{CDF:2022hxs}.

\subsection{\boldmath Anomaly in the $W$ boson mass measurement}

In this section, we briefly review the current situation of the $W$ boson mass measurements, in particular, the anomaly reported by the CDF collaboration~\cite{CDF:2022hxs}. 
We also briefly discuss the new physics interpretation of this anomaly.

The $W$ boson mass has been measured in both $e^+e^-$ and hadron collider experiments. 
Before April 2022, the world average value was given by
\begin{equation}
\label{eq: Before_CDF}
m_W^{\text{World Ave.}} = 80.377(12) ~\mathrm{GeV}. 
\end{equation}
The above is in good agreement with the SM prediction from the electroweak global fit~\cite{PDG2022}; 
\begin{equation}
\label{eq: SM_Wmass}
m_W^{\text{SM}} = 80.356(06)~\mathrm{GeV}. 
\end{equation}
In April 2022, the CDF collaboration reported the result of the $W$ boson mass measurement using the full Run-II data~\cite{CDF:2022hxs} as follows;
\begin{equation}
\label{eq: CDF_Wmass}
m_W^{\text{CDF-II}} = 80.4335(94)~\mathrm{GeV}. 
\end{equation}
This value is significantly different from both the world average value and the SM prediction.  
Because of the reduced uncertainty in the CDF-II result, it is a $7.0\sigma$ deviation from the SM prediction~\cite{CDF:2022hxs}. 
Although the tension is reported by only one group, 
the deviation is considerably large, making it a potentially interesting hint for new physics.

Using the oblique parameters $S$, $T$ and $U$~\cite{Peskin:1990zt}, 
we can generally discuss the new physics effect on the $W$ boson mass. 
The deviation in the squared $W$ boson mass from the SM one is given by~\cite{Peskin:1990zt}
\begin{equation}
\label{eq: mW_and_STU}
    \Delta m_W^2 = m_Z^2 c_W^2
    \left(
        - \frac{ \alpha S }{ 2 ( c_W^2 - s_W^2 ) }
        + \frac{ \alpha T }{ 1 - t_W^2 }
        + \frac{ \alpha U }{ 4 s_W^2 } 
    \right), 
\end{equation}
where $s_W^2 = \sin^2 \theta_W^{}$. 
In order to reproduce the central value of the CDF-II result, 
\begin{equation}
    \Delta m_W^2 \simeq 12.5~\mathrm{GeV}^2 > 0, 
\end{equation}
is required.

Since $S$, $T$, and $U$ also cause deviations in other electroweak observables, 
we have a constraint on them to keep the other observables consistent with experimental data. 
As a result of the electroweak global fit using the CDF-II result~\cite{Lu:2022bgw, Asadi:2022xiy}, the following results are obtained~\cite{Lu:2022bgw}\footnote{When $U=0$ is assumed, the result is given by $S=0.15(08)$ and $T=0.27(06)$ in Table~III of Ref.~\cite{Lu:2022bgw}. Since the $U$ parameter can be as large as the $S$ and $T$ parameters in the dark $Z$ model, we employ the result in Eq.~(\ref{eq: STU_constraints}).};
\begin{equation}
\label{eq: STU_constraints}
    S = 0.06(10), \quad T = 0.11(12), \quad U = 0.14(09). 
\end{equation}

In previous works~\cite{Thomas:2022gib, Cheng:2022aau, Harigaya:2023uhg}, it has been revealed that the simple dark photon model cannot reproduce the CDF-II result within $2\sigma$ because of tight constraints from electroweak precision measurements in Eq.~(\ref{eq: STU_constraints}). 
On the other hand, it has been shown that the effect of mass mixing can be sizable enough to reproduce the CDF-II result~\cite{Lee:2022nqz}, where the authors investigate an extension of the dark $Z$ model with additional vector-like leptons~\cite{Lee:2021gnw}. 
However, they consider only the $\rho$ parameter, or equivalently the $T$ parameter. 
The $S$ and $U$ parameters are ignored, and the constraint on the dark $Z$ boson from the electroweak precision measurements is not investigated in a detailed fashion. 
In addition, only a limited parameter space is studied in Ref.~\cite{Lee:2022nqz}; they assume $\varepsilon = 0$ and show the results for a few benchmark values of $m_{Z_d}$. 

In this paper, we consider all the $S$, $T$, and $U$ parameters and discuss their behavior in detail.
The electroweak-global-fit constraint on these parameters is found by using Eq.~(\ref{eq: STU_constraints}). 
Considering this and collider bounds on $Z_d$, we exhaustively examine parameter space where the $W$ boson mass anomaly can be explained while avoiding all the experimental constraints. 

\subsection{\boldmath Effect of dark $Z$ bosons on the electroweak observables}
\label{sec: STU}
 
In this section, we discuss the dark $Z$ boson effect on various electroweak observables using the $S$, $T$, and $U$ parameters. 
We give formulas for the $S$, $T$, and $U$ paramters and discuss their behavior in detail.

In the dark $Z$ model, 
kinetic mixing and mass mixing induce shifts in the mass and the current interaction of the $Z$ boson from those in the SM at tree level. 
The effect of such shifts on various observables can be described in terms of the $S$, $T$ and $U$ parameters~\cite{Holdom:1990xp}. 
In the following, we do not consider other new physics effects in four-fermion processes for simplicity, for example, oblique corrections induced by additional Higgs bosons and the dark $Z$ mediation induced by the current interaction in Eq.~(\ref{eq: current_darkZ}). 
Since the former is a 1-loop effect, we can consider a parameter region such that their contribution is subleading and small enough.\footnote{Large mass differences among the additional Higgs bosons can enhance the $S$ and $T$ parameters large enough to explain the W boson mass anomaly~\cite{Lu:2022bgw, Song:2022xts, Heo:2022dey, Ahn:2022xax, Lee:2022gyf}. Here, we do not consider such parameter regions to investigate the effect of the kinetic and mass mixing. We choose the parameters in the Higgs potential such that all the constraints on them can be avoided as explained at the end of Sec.~\ref{sec: WMA_and_DarkZ}.}
Although the latter is a tree-level effect, 
it is expected to have little impact on observables at the $Z$ pole unless $m_{Z_d}^{}$ is very close to $m_Z^{}$~\cite{Harigaya:2023uhg}. 

In the dark $Z$ model, the mass terms and the current interactions of the SM electroweak gauge bosons are given by
\begin{align}
\label{eq: deviation_DarkZ}
    \mathcal{L}_{\text{EW}}^{} = 
    & \, m_W^2 W^{+\mu} W^-_\mu 
    + \frac{ 1 }{ 2 } \tilde{m}_Z^2 \bigl( 1 + \Delta_1^{} \bigr) Z^\mu Z_\mu
\nonumber \\[5pt]
    & - \frac{ g }{ \sqrt{2} } \bigl( J_{\text{CC}}^\mu W_\mu^+ + \mathrm{h.c.} \bigr)
    - e J_{\text{em}}^\mu A_\mu 
\nonumber \\[5pt]
    & - \frac{ g }{ 2 c_W^{} } \bigl( 1 + \Delta_2^{} \bigr)
        J_{\mathrm{NC}}^\mu Z_\mu
    - e \Delta_3^{} J_{\mathrm{em}}^\mu Z_\mu. 
\end{align}
Deviations from the SM are described by $\Delta_1^{}$, $\Delta_2^{}$ and $\Delta_3^{}$. 
Up to the quadratic order of $\bar{\varepsilon}$ ($\bar{\varepsilon} = \varepsilon$ or $\varepsilon_Z^{}$), 
they are evaluated as 
\begin{align}
\label{eq: Delta1}
\Delta_1^{} = & \frac{ m_Z^2 }{ \tilde{m}_Z^2 } - 1 
        = \frac{ (1 - r^2) \sin^2 \xi }{ 1 + \sin^2 \xi (r^2 - 1 )}
        %
        %
        \simeq \frac{ (\varepsilon_Z^{} + \varepsilon t_W )^2 }{ 1 - r^2 }, \\[5pt]
\label{eq: Delta2}
\Delta_2^{} = & (\cos \xi - 1) +  \eta \varepsilon t_W \sin \xi
\nonumber \\[5pt]
        \simeq & - \frac{ 1 }{ 2 } 
        \left( \frac{ \varepsilon_Z^{} + \varepsilon t_W }{ 1 - r^2 } \right)
        \left( \frac{ \varepsilon_Z^{} + (2 r^2 - 1 ) \varepsilon t_W }{ 1 - r^2 }\right), \\[5pt]
\label{eq: Delta3}
\Delta_3^{} = & - \eta \varepsilon \sin \xi 
        \simeq - \varepsilon \left( \frac{ \varepsilon_Z^{} + \varepsilon t_W }{ 1 - r^2 } \right). 
\end{align}
These expansions are valid as long as $|\varepsilon t_W^{}|$ and $|\varepsilon_Z^{}|$ are sufficiently smaller than one and $|1-r^2|$. 
In Eq.~(\ref{eq: Delta1}), we have replaced $\tilde{r}^2$ in $m_Z^2$ given by Eq.~(\ref{eq: Z_mass}) with $r^2$ because the difference is higher-order of $\bar{\varepsilon}$ as one can see in Eq.~\eqref{eq: r_and_rtilde}.

The $S$, $T$ and $U$ parameters can be found by using $\Delta_1$, $\Delta_2$, and $\Delta_3$ as follows~\cite{Holdom:1990xp, Burgess:1993vc};
\begin{align}
\label{eq: S}
    \alpha S = &\, 8 s_W^2 c_W^2 \Delta_2^{} - 4 s_W c_W (c_W^2 - s_W^2) \Delta_3^{} \nonumber \\
        \simeq & - 4 s_W^{} c_W^{}
            \left( \frac{ \varepsilon_Z^{} + \varepsilon t_W }{ 1 - r^2 } \right)
            \left\{
                s_W^{} c_W^{} \frac{ \varepsilon_Z^{} + \varepsilon t_W^{} }{ 1 - r^2 } - \varepsilon 
            \right\}, \\[3pt]
\label{eq: T}
    \alpha T = & - \Delta_1^{} + 2 \Delta_2^{} 
\nonumber \\
        \simeq & - \left( \frac{ \varepsilon_Z^{} + \varepsilon t_W }{ 1 - r^2 } \right)
            \left( \frac{ (2-r^2) \varepsilon_Z^{} + r^2 \varepsilon t_W^{} }{ 1 - r^2 } \right), \\[5pt]
\label{eq: U}
    \alpha U = & -8 s_W^2 c_W ( c_W \Delta_2^{} + s_W^{} \Delta_3 ) 
\nonumber \\
        \simeq & \, 4 s_W^2 c_W^2 \left( \frac{ \varepsilon_Z^{} + \varepsilon t_W }{ 1 - r^2 } \right)^2. 
\end{align}
Equation~(\ref{eq: T}) is consistent with a formula for the $\rho$ parameter induced by the mass and kinetic mixings in Refs.~\cite{Lee:2021gnw, Bian:2017xzg}. 
We also note that the $U$ parameter in Eq.~(\ref{eq: U}) is always positive which is preferred to explain $W$ boson mass anomaly~\cite{Lu:2022bgw, Asadi:2022xiy} [see Eq.~(\ref{eq: STU_constraints})].

In order to investigate the behavior of the $S$, $T$, and $U$ parameters, 
we consider two significant cases: the dark photon (DP) limit ($\varepsilon_Z^{} \to 0$) and the pure dark $Z$ (DZ) limit ($\varepsilon \to 0$). 
First, we discuss the DP limit. 
Then, $S$, $T$, and $U$ are given by
\begin{align}
\label{eq: S_DPlimit}
& \alpha S_\mathrm{DP} = 4 s_W^2 \bigl( c_W^2 - r^2 \bigr) \left( \frac{ \varepsilon }{ 1 - r^2 } \right)^2, \\[5pt]
\label{eq: T_DPlimit}
& \alpha T_\mathrm{DP} = - t_W^2 r^2 \left( \frac{ \varepsilon }{ 1 - r^2 } \right)^2, \\[5pt]
\label{eq: U_DPlimit}
& \alpha U_\mathrm{DP} = 4 s_W^4 \left( \frac{ \varepsilon }{ 1 - r^2 } \right)^2, 
\end{align}
where a subscript DP means that the quantities are evaluated in the DP limit. 
These formulas are consistent with those in the dark photon model in Refs.~\cite{Holdom:1990xp, Cheng:2022aau}.\footnote{Although Ref.~\cite{Harigaya:2023uhg} also studies the $S$, $T$, and $U$ parameters in the dark photon model, their formulas do not coincide with Eqs. (\ref{eq: S_DPlimit})-(\ref{eq: U_DPlimit}) even accounting for differences in notation. This is because Ref.~\cite{Harigaya:2023uhg} takes the formulas from Ref.~\cite{Babu:1997st}, which uses perturbative expansions by $\sin \xi \simeq \xi$, not $\bar{\varepsilon}$, up to the linear order including a part of quadratic terms. 
By using this method, the $S$, $T$, and $U$ parameters in the dark $Z$ model are given by $\alpha S \simeq 4s_W^{} c_W^{} \varepsilon (\varepsilon_Z^{} + \varepsilon t_W^{})/(1-r^2)$, $\alpha T \simeq (\varepsilon^2 t_W^2 - \varepsilon_Z^2)/(1-r^2)$, and $\alpha U \simeq 0$. However, these formulas are valid only in the case of $r^2 \gg 1$ ($m_{Z_d} \gg m_Z$) because some terms proportional to $\xi^2$ are dropped while the term $(r^2 - 1)\xi^2$ remains in the calculation.}
We see that $S_\mathrm{DP}$, $T_\mathrm{DP}$, and $U_\mathrm{DP}$ are all proportional to $(\varepsilon/(1-r^2))^2$. 

\begin{figure}[b]
\subfigure[]{\includegraphics[width=0.48\textwidth]{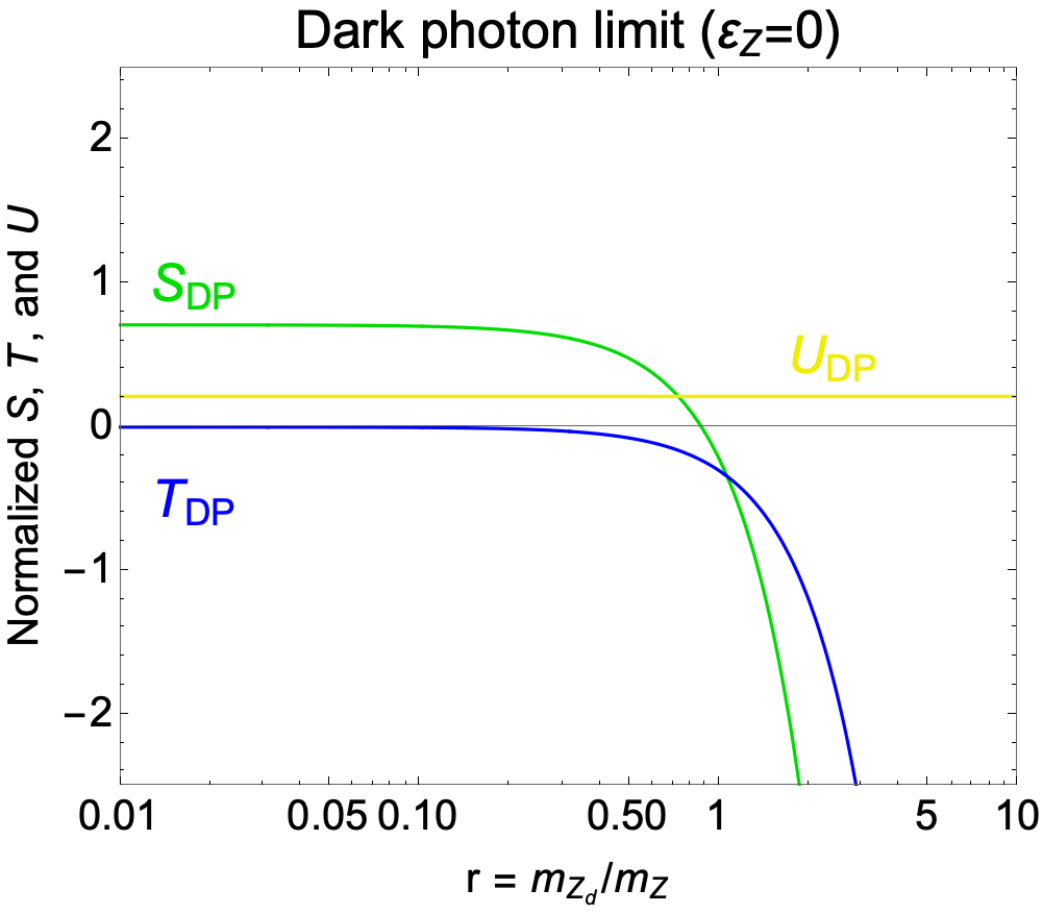}
\label{fig: STU_DPlimit}}
\subfigure[]{\includegraphics[width=0.48\textwidth]{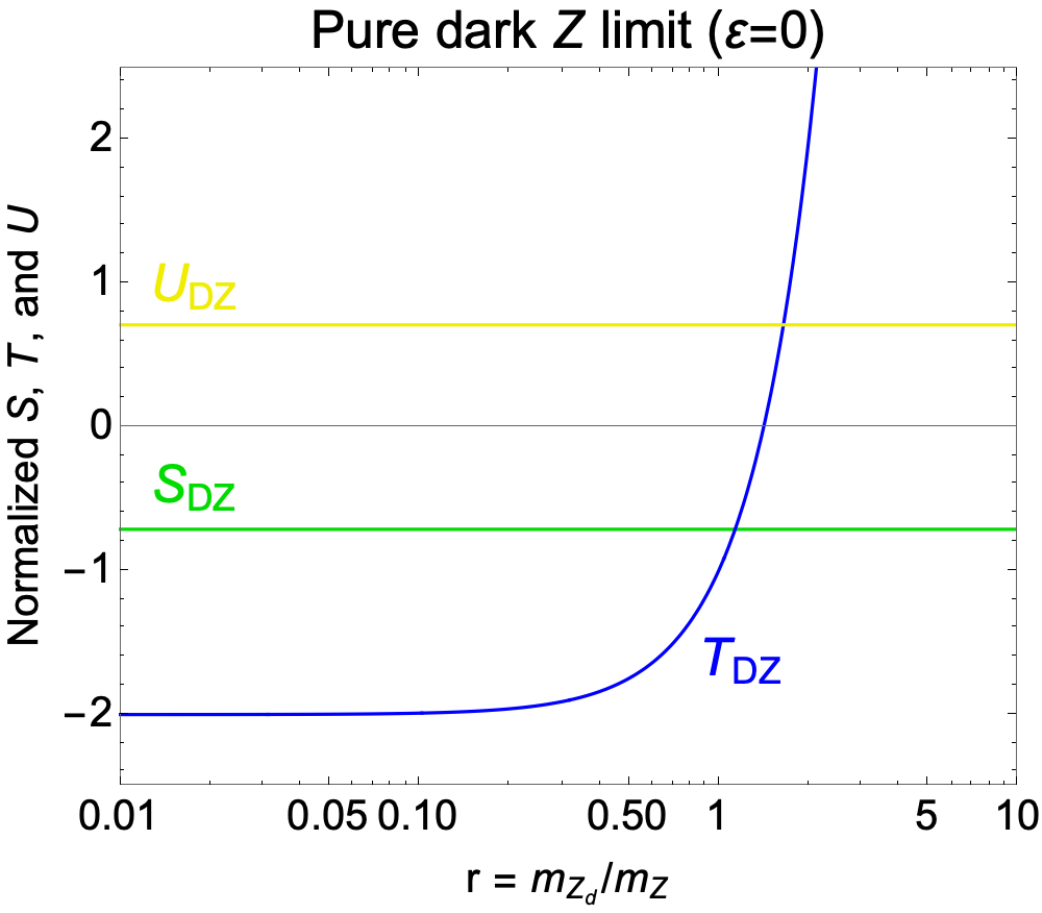}
\label{fig: SUT_DZlimit}}
    \caption{The behavior of the $S$, $T$, and $U$ parameters in (a) the DP limit and (b) the pure DZ limit with small mixing, $|\bar{\varepsilon}/(1-r^2)| \ll 1$. }
\label{fig: STUparameters}
\end{figure}

In Fig.~\ref{fig: STU_DPlimit}, we show the behavior of $S_\mathrm{DP}$, $ T_\mathrm{DP}$, and $U_\mathrm{DP}$ normalized by the positive common factor $\alpha^{-1}(\varepsilon/(1-r^2))^2$. 
$T_\mathrm{DP}$ is always negative, while $U_\mathrm{DP}$ is positive. 
We note that the negative $T$ parameter is not favored to explain the $W$ boson mass anomaly [See Eq.~(\ref{eq: STU_constraints})]. 
The sign of $S_\mathrm{DP}$ depends on $r$. 
When $r$ is larger (smaller) than $c_W \simeq 0.88$, $S_\mathrm{DP}$ is negative (positive). 
We note that the above behavior is correct only when $|\varepsilon/(1-r^2)| \ll 1$ is satisfied because we use the perturbative expansion. 

For heavy $Z_d$, $|S_\mathrm{DP}|$ and $|T_\mathrm{DP}|$ are much larger than $|U_\mathrm{DP}|$. It is because for $r^2 \gg 1$,
they behave as  
\begin{equation}
\label{eq: STU_decouplingDP}
\begin{array}{l}
\alpha S_\mathrm{DP} (r^2 \gg  1) \simeq - 4 s_W^2 \varepsilon^2 r^{-2} \propto 1/m_{Z_d}^2, \\
\alpha T_\mathrm{DP} (r^2 \gg  1) \simeq - t_W^2 \varepsilon^2 r^{-2} \propto 1/m_{Z_d}^2, \\
\alpha U_\mathrm{DP} (r^2 \gg  1) \simeq 4 s_W^4 \varepsilon^2 r^{-4} \propto 1/m_{Z_d}^4. 
\end{array}
\end{equation}
Therefore, $|U_\mathrm{DP}|$ is suppressed compared to $|S_\mathrm{DP}|$ and $|T_\mathrm{DP}|$ in the case of heavy $Z_d$.
In addition, for $r^2 \gg 1$, we have a simple relation between $S_\mathrm{DP}$ and $T_\mathrm{DP}$; $S_\mathrm{DP}/T_\mathrm{DP} \simeq 4 c_W^2$~\cite{Harigaya:2023uhg}. 

The suppression of $U_\mathrm{DP}$ is understood in terms of the standard model effective field theory (SMEFT). In the SMEFT, the $S$ and $T$ parameters are generated by dimension-six operators, while the $U$ parameter is generated by a dimension-eight operator~\cite{Grinstein:1991cd}. Thus, $U$ is generally expected to have a relative suppression factor $\Lambda^{-2}$ compared to $S$ and $T$, where $\Lambda$ is the cut-off scale. 
This is the case for the decoupling limit of many new physics models including the dark photon model. 
In Appendix~\ref{app: SMEFT}, we present a detailed discussion of higher-dimensional operators and the $S$, $T$, and $U$ parameters in the DP limit of the model.

Next, we discuss the pure DZ limit. 
In this limit, the $S$, $T$, and $U$ parameters are given by
\begin{align}
\label{eq: S_pDZlimit}
& \alpha S_\mathrm{DZ} = - 4 s_W^2 c_W^2 \left( \frac{ \varepsilon_Z^{} }{ 1 - r^2 } \right)^2 \\[5pt]
\label{eq: T_pDZlimit}
& \alpha T_\mathrm{DZ} = (r^2 - 2) \left( \frac{ \varepsilon_Z^{} }{ 1 - r^2 } \right)^2, \\[5pt]
\label{eq: U_pDZlimit}
& \alpha U_\mathrm{DZ} = 4 s_W^2 c_W^2 \left( \frac{ \varepsilon_Z^{} }{ 1 - r^2 } \right)^2, 
\end{align}
where a subscript DZ means that the quantities are evaluated in the pure DZ limit. $S_\mathrm{DZ}$, $T_\mathrm{DZ}$, and $U_\mathrm{DZ}$ are all proportional to $(\varepsilon_Z^{}/(1-r^2))^2$. 

In Fig.~\ref{fig: SUT_DZlimit}, we show the behavior of $S_\mathrm{DZ}$, $T_\mathrm{DZ}$, and $U_\mathrm{DZ}$ normalized by the positive common factor $\alpha^{-1} (\varepsilon_Z^{}/(1-r^2))^2$. 
We can see that it is quite different compared to that in the DP limit.
$S_\mathrm{DZ}$ is always negative, while $U_\mathrm{DZ}$ is positive. 
The sign of $T_\mathrm{DZ}$ depends on $r$. 
When $r$ is larger (smaller) than $\sqrt{2} \simeq 1.4$, $T_\mathrm{DZ}$ is positive (negative). Thus, heavy $Z_d$ is favored to satisfy the constraint in Eq.~(\ref{eq: STU_constraints}). 
We note that the above behavior is correct only when $|\varepsilon_Z^{} /(1-r^2)| \ll 1$ is satisfied like in the case of the DP limit. 

In addition, we find an interesting relation, 
\begin{equation}
S_\mathrm{DZ} = - U_\mathrm{DZ}, 
\end{equation}
in the pure DZ limit.
This is not a result of perturbative expansions. 
For $\varepsilon = 0$, $\hat{B}_\mu = \tilde{B}_\mu$ and $\hat{Z}_{d\mu} = \tilde{Z}_{d\mu}$ hold. 
Then, $\tilde{Z}_\mu$ couples to only the neutral current as in the SM, and 
$\tilde{Z}_{d\mu}$ does not couples to the SM fermions. 
Since $\varepsilon_Z^{}$ induces the mixing between $\tilde{Z}_\mu$ and $\tilde{Z}_{d \mu}$, the term $\Delta_3 J_\mathrm{em}^\mu Z_\mu$ is not generated by $\varepsilon_Z^{}$. 
Thus, $\Delta_3$ is zero at all orders of $\varepsilon_Z^{}$ in the pure DZ limit. 
This yields $\alpha S_\mathrm{DZ} = - \alpha U_\mathrm{DZ} = 8 s_W^2 c_W^2 \Delta_2$.  

The relation between $S_\mathrm{DZ}$ and $U_\mathrm{DZ}$ leads to a unique behavior for them in the decoupling limit of $Z_d$ ($r^2 \gg 1$). 
In this limit, we obtain 
\begin{equation}
\label{eq: STU_decouplingDZ}
\begin{array}{l}
\alpha S_\mathrm{DZ} (r^2 \gg  1) \simeq - 4 s_W^2 c_W^2 \varepsilon_Z^2  r^{-4} \propto 1/m_{Z_d}^4, \\
\alpha T_\mathrm{DZ} (r^2 \gg  1) \simeq \varepsilon^2_Z r^{-2} \propto 1/m_{Z_d}^2, \\
\alpha U_\mathrm{DZ} (r^2 \gg  1) \simeq 4 s_W^2 c_W^2 \varepsilon_Z^2 r^{-4} \propto 1/m_{Z_d}^4. 
\end{array}
\end{equation}
$T_\mathrm{DZ}$ and $U_\mathrm{DZ}$ are proportional to $r^{-2}$ and $r^{-4}$, respectively, which is the same as in the DP limit. 
On the other hand, $S_\mathrm{DZ}$ is proportional to $r^{-4}$ not $r^{-2}$. 
It is suppressed by $r^{-2}$ compared to $S_\mathrm{DP}$. 
In terms of the SMEFT, it indicates that $S_\mathrm{DZ}$ is generated by a dimension-eight operator not a dimension-six one.  
This is an intriguing feature not seen in many other new physics models.
The absence of a dimension-six operator which is a source of the $S$ parameter is due to the fact that $\tilde{Z}_{d\mu}$ has no couplings with the SM fermions at tree level, as discussed in detail in Appendix~\ref{app: SMEFT}.

\subsection{ \boldmath The $W$ boson anomaly in the dark $Z$ model}
\label{sec: Wboson_in_darkZ}

In this section, we discuss the $W$ boson mass in the dark $Z$ model using the $S$, $T$, and $U$ parameters derived in the previous section. 
We will show the dark $Z$ model can explain the $W$ boson mass anomaly under the constraint from the electroweak global fit in Eq.~(\ref{eq: STU_constraints}) and constraints from the direct search for $Z_d$. 

By using $S$, $T$ and $U$ parameters in Eqs~(\ref{eq: S})--(\ref{eq: U}) and Eq.~(\ref{eq: mW_and_STU}), $\Delta m_W^2$ is given by 
\begin{equation}
\label{eq: mW}
\Delta m_W^2 = -m_Z^2 \left( \frac{ c_W^4 }{ c_W^2 - s_W^2 } \right) \left( \frac{ 1 }{ 1 - r^2 } \right)
    \Bigl( \varepsilon_Z + \varepsilon t_W^{} \Bigr)^2. 
\end{equation}
We note that the sign of $\Delta m_W^2$ is determined by whether $r^2 = (m_{Z_d}^{} / m_Z^{})^2$ is larger than one or not. 
The deviation $\Delta m_W^2$ is positive with heavy dark $Z$ bosons ($r^2 > 1$) and negative with light dark $Z$ bosons ($r^2 < 1$). 
As mentioned earlier, in order to explain the $W$ boson mass anomaly, 
$\Delta m_W^2 \simeq 12.5~\mathrm{GeV}^2 > 0$ is required. 
Therefore, light dark $Z$ bosons cannot explain the $W$ boson mass anomaly, while heavy dark $Z$ bosons can. 
This is already known in the context of the dark photon model~\cite{Thomas:2022gib, Cheng:2022aau, Harigaya:2023uhg}. 

Using $m_Z^{} = 91.1876~\mathrm{GeV}$ and $s_W^2 = 0.23122$~\cite{PDG2022}, we can estimate the values of $m_{Z_d}^{}$, $\varepsilon$, and $\varepsilon_Z^{}$ required to reproduce the CDF-II result ($\Delta m_W^2 = 12.5~\mathrm{GeV}^2$) by the following equation\footnote{Here, we employ the value of $s_W^2$ in the $\overline{\mathrm{MS}}$ scheme. The difference using another definition is a higher-order effect in the perturbative expansion.}; 
\begin{equation}
\Bigl( \varepsilon_Z^{} + 0.55 \varepsilon \Bigr)^2 \simeq (1.6 \times 10^{-3}) \times \Bigg( \biggl( \frac{ m_{Z_d} }{ 100~\mathrm{GeV} } \biggr)^2 - 0.83 \Bigg).
\end{equation}
For example, in the case of $m_{Z_d} = 200~\mathrm{GeV}$, 
the CDF-II result can be explained by satisfying $|\varepsilon_Z^{} + 0.55 \varepsilon | \simeq 0.07$. 
For a heavier dark $Z$ boson, a larger value of $|\varepsilon_Z^{} + \varepsilon t_W^{}|$ is required.  
Thus, relatively large mixing, $|\varepsilon_Z^{}|$ and/or $|\varepsilon| \simeq \mathcal{O}(0.1)$, are necessary to explain the $W$ boson mass anomaly.

\begin{figure}[t]
\begin{center}
    \includegraphics[width=0.48\textwidth]{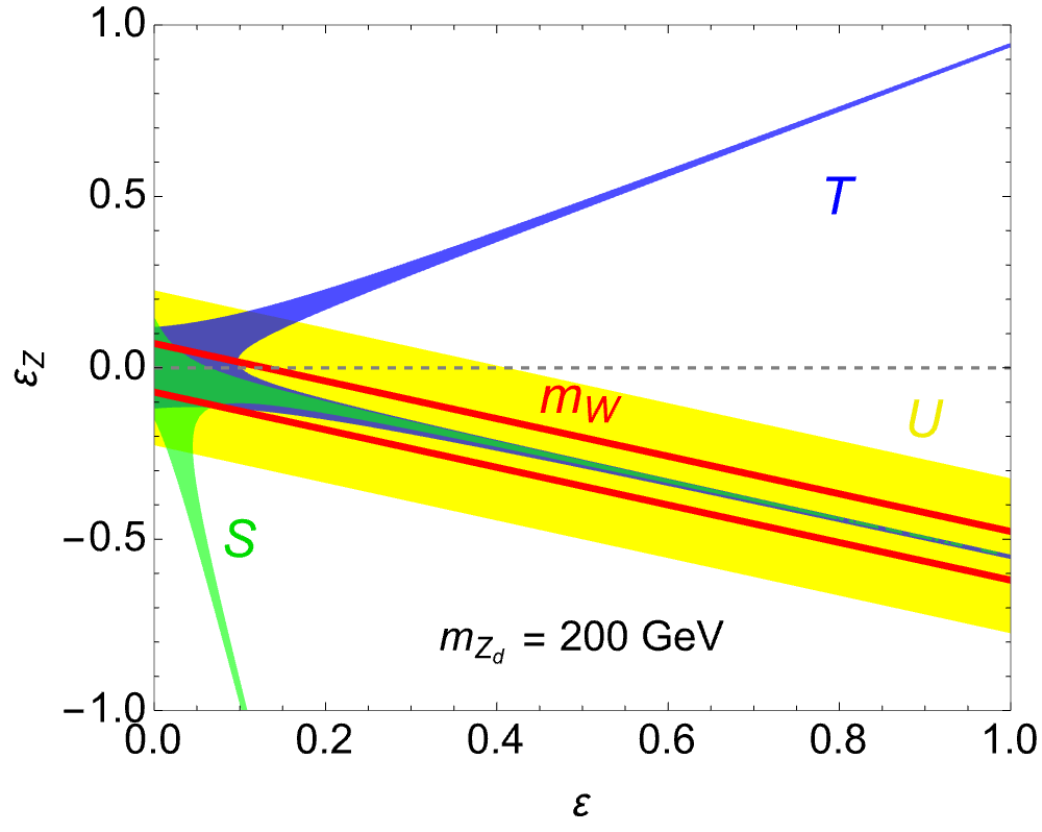}
    \caption{Allowed regions for the $S$ (green), $T$ (blue) and $U$ (yellow) parameters within $2\sigma$ in the case of $m_{Z_d} = 200~\mathrm{GeV}$. On the two red bands (not including the region between the bands), the CDF-II result can be reproduced within $2\sigma$. The dashed line represents $\varepsilon_Z = 0$.}
\label{fig: $W$ boson mass anomaly1}
\end{center}
\end{figure}

In Fig.~\ref{fig: $W$ boson mass anomaly1}, we show regions in $\varepsilon$-$\varepsilon_Z^{}$ plane where the constraint on $S$, $T$ and $U$ parameters are satisfied within $2\sigma$ uncertainties with the colors green, blue and yellow, respectively in the case of $m_{Z_d}^{} = 200~\mathrm{GeV}$ as a benchmark point. 
In the red region, the CDF-II results can be reproduced within $2\sigma$. 

There are two red bands because $\Delta m_W^2$ is proportional to $(\varepsilon_Z + \varepsilon t_W)^2$. 
All colored regions are invariant by the simultaneous change of the sign $\varepsilon_Z^{} \to - \varepsilon_Z$ and $\varepsilon \to - \varepsilon$ 
because the $S$, $T$, and $U$ parameters and $\Delta m_W^2$ are given by the quadratics of the mixing parameters. 
We can see that there are overlapping regions of all the colored regions for small $|\varepsilon|$, where the CDF-II result can be reproduced under the constraint from the electroweak global fit.  

\begin{figure}[t]
\subfigure[]{\includegraphics[width=0.48\textwidth]{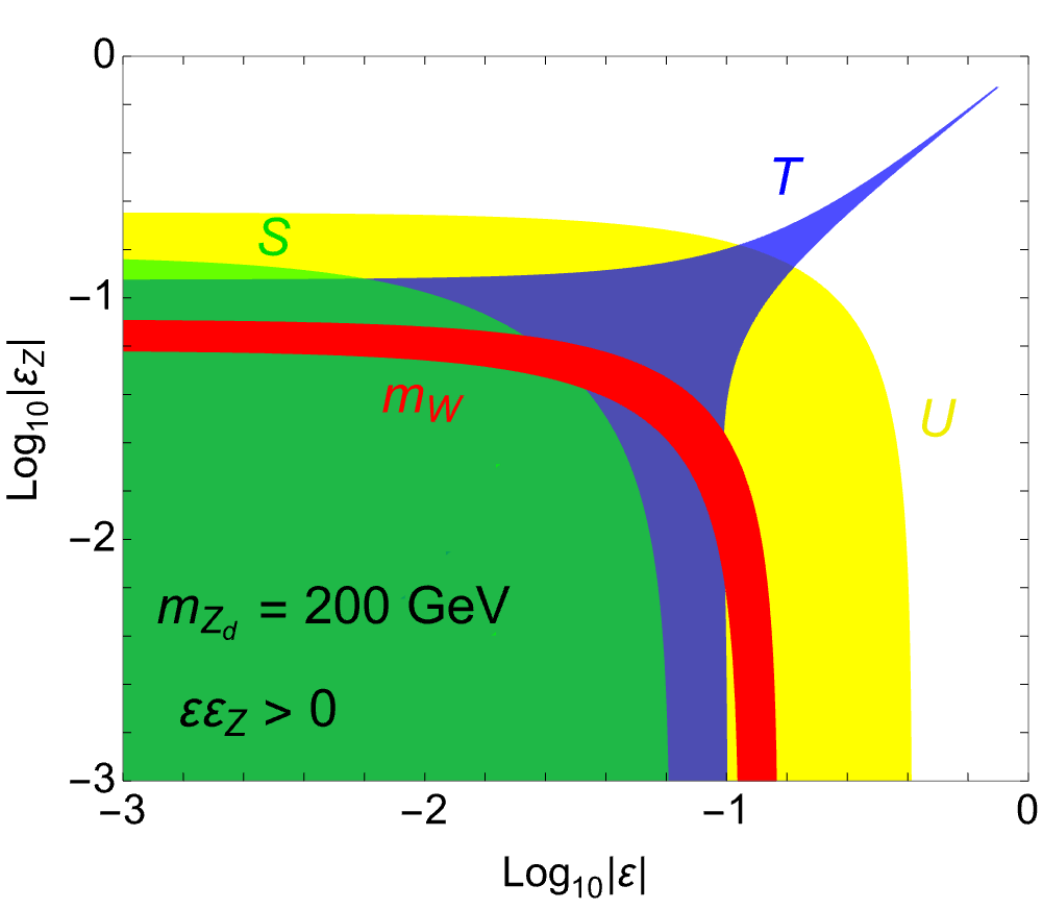}
\label{fig: $W$ boson mass anomaly2_a}}
\subfigure[]{\includegraphics[width=0.48\textwidth]{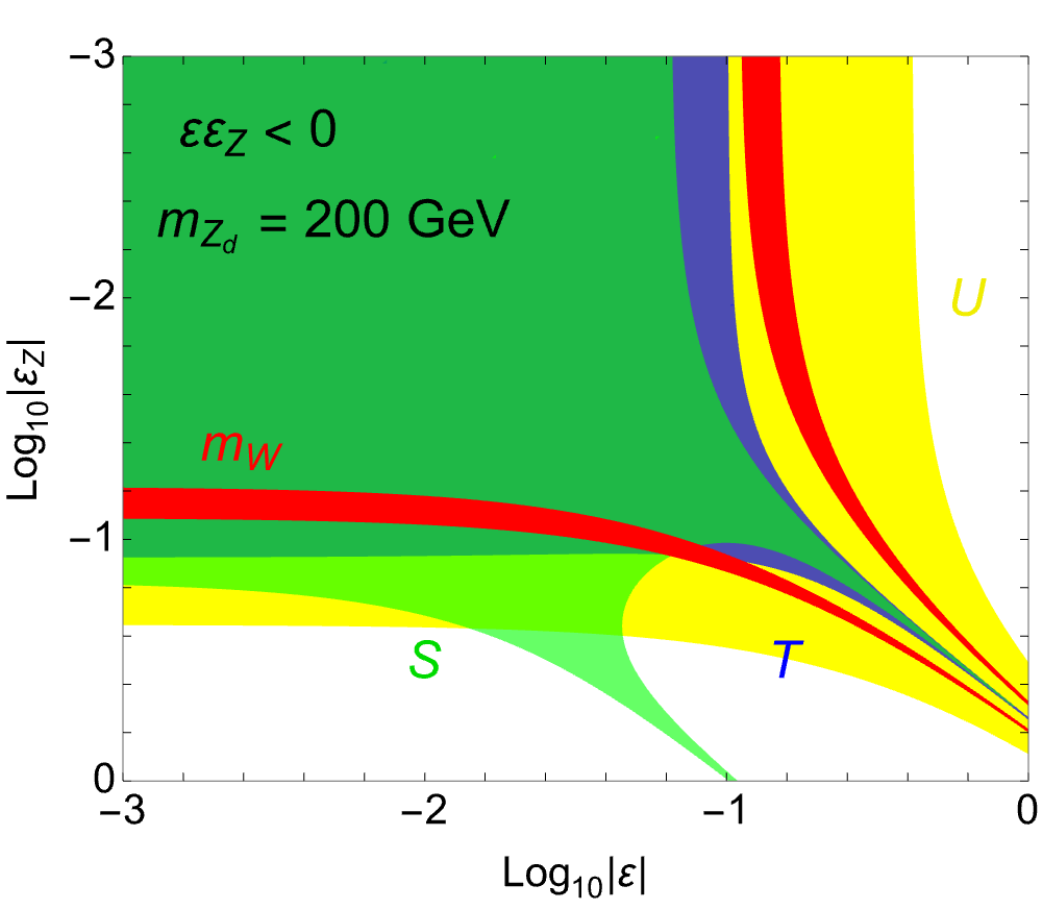}
\label{fig: $W$ boson mass anomaly2_b}}
    \caption{The same as Fig.~\ref{fig: $W$ boson mass anomaly1} but in the log scale in the cases of (a) $\varepsilon \varepsilon_Z^{}>0$ and (b) $\varepsilon \varepsilon_Z^{} < 0$. }
\label{fig: $W$ boson mass anomaly2}
\end{figure}

In Fig.~\ref{fig: $W$ boson mass anomaly2}, we show the same plots as Fig.~\ref{fig: $W$ boson mass anomaly1} but in the log scale. 
Figure~\ref{fig: $W$ boson mass anomaly2_a} [Fig.~\ref{fig: $W$ boson mass anomaly2_b}] illustrates the case where $\varepsilon$ and $\varepsilon_Z^{}$ have the same (opposite) signs, i.e. the upper (lower) half of Fig.~\ref{fig: $W$ boson mass anomaly1}.
We can see that if the mass mixing parameter $|\varepsilon_Z^{}|$ is very small ($|\varepsilon_Z^{}|\lesssim 0.01$), the behavior of the $S$, $T$ and $U$ parameters and the $W$ boson mass is almost the same with those in the dark photon model ($\varepsilon_Z^{} \to 0$). 
In such regions, the constraints from the electroweak precision measurements cannot be satisfied on the red band within $2\sigma$.\footnote{One may think of using the fitting result assuming $U=0$, which is $S=0.15(08)$ and $T=0.27(06)$~\cite{Lu:2022bgw}, instead of Eq.~(\ref{eq: STU_constraints}) because $|U|$ is much smaller than $|S|$ and $|T|$ in the DP limit $r^2 \gg 1$. However, it cannot help the situation because $S$ and $T$ have negative large value in such a case [See Fig.~\ref{fig: STU_DPlimit}].} 
It is the same situation in the dark photon model~\cite{Thomas:2022gib, Cheng:2022aau, Harigaya:2023uhg}. 
On the other hand, 
if the mass mixing parameter is $\mathcal{O}(0.1)$, 
there are regions where the CDF-II result can be reproduced while satisfying the constraints. 
Therefore, the $W$ boson mass anomaly can be explained in the dark $Z$ model with relatively large mass mixing and heavy dark $Z$ bosons. 

\begin{figure}[t]
\begin{center}
    \includegraphics[width=0.48\textwidth]{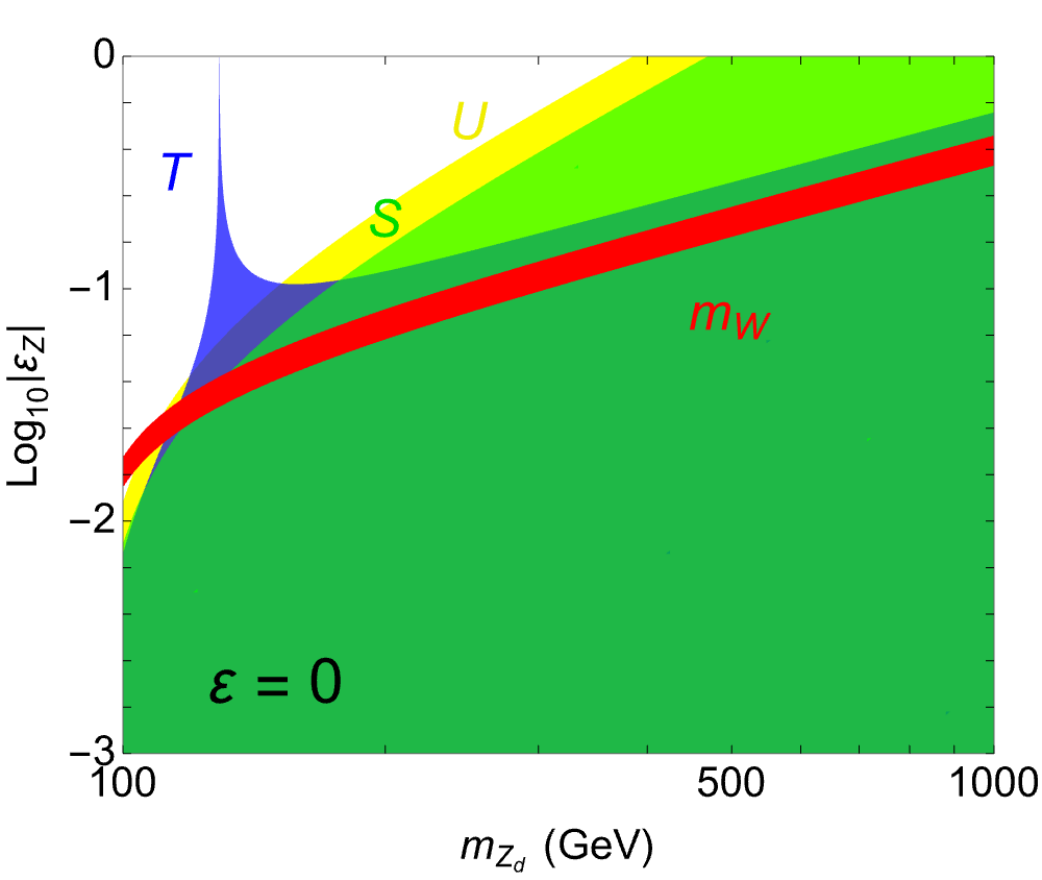}
    \caption{The allowed regions of the mass $m_{Z_d}$ and the mass mixing $\varepsilon_Z$ for $\varepsilon = 0$.
    The color code is the same as Fig.~\ref{fig: $W$ boson mass anomaly1}.
    The large $|\varepsilon_Z|$ region of $\mathcal{O}(1)$ may not be valid as the perturbation in $\varepsilon_Z$ is used.}
\label{fig: $W$ boson mass anomaly3}
\end{center}
\end{figure}

As shown in Figs.~\ref{fig: $W$ boson mass anomaly1} and \ref{fig: $W$ boson mass anomaly2}, 
the kinetic mixing $\varepsilon$ is not important to explain the $W$ boson mass anomaly while maintaining  agreement with precision electroweak data. 
In Fig.~\ref{fig: $W$ boson mass anomaly3}, we thus show the allowed regions of $S$, $T$ and $U$ parameters and the CDF-II result within $2\sigma$ in the pure DZ limit ($\varepsilon \to 0$) by using the same colors as those in Figs.~\ref{fig: $W$ boson mass anomaly1} and \ref{fig: $W$ boson mass anomaly2}. 
In the overlapping regions, the CDF-II result can be reproduced under the constraint from the electroweak global fit. 
We can see that the minimum value of $m_{Z_d}^{}$ to explain the $W$ boson mass anomaly is given by about $130~\mathrm{GeV}$ with $|\varepsilon_Z^{}| \simeq 0.03$.
Larger mass mixing is required in the case of heavier dark $Z$ bosons, 
for example, $|\varepsilon_Z^{}| \gtrsim 0.4$ for $m_{Z_d}^{} \gtrsim 1~\mathrm{TeV}$. 
However, the perturbative expansion in $\varepsilon_Z^{}$ becomes worse for such a large $|\varepsilon_Z^{}|$, and the result would be less reliable. 
Therefore, in the following, we consider dark $Z$ bosons with the mass $m_{Z_d}^{} = \mathcal{O}(100)~\mathrm{GeV}$. 

\begin{figure}[t]
\subfigure[]{\includegraphics[width=0.48\textwidth]{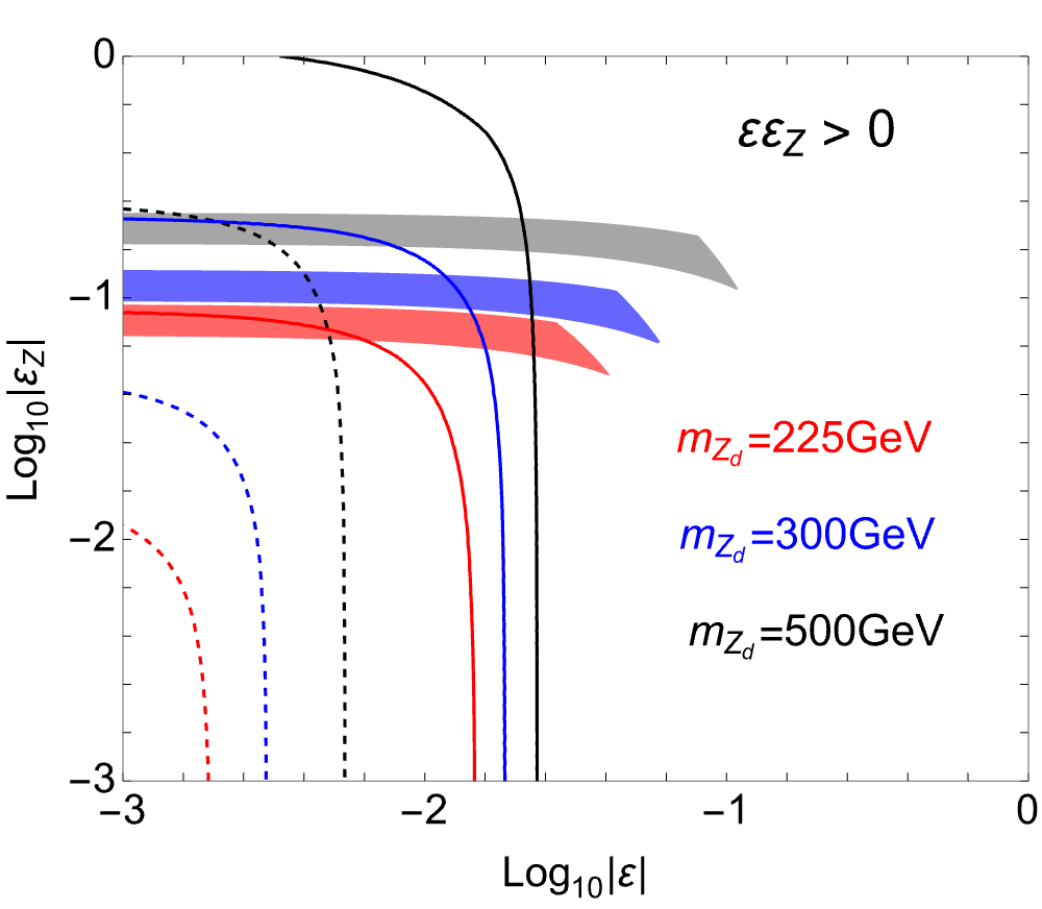}
\label{fig: LHCbound1}}
\subfigure[]{\includegraphics[width=0.48\textwidth]{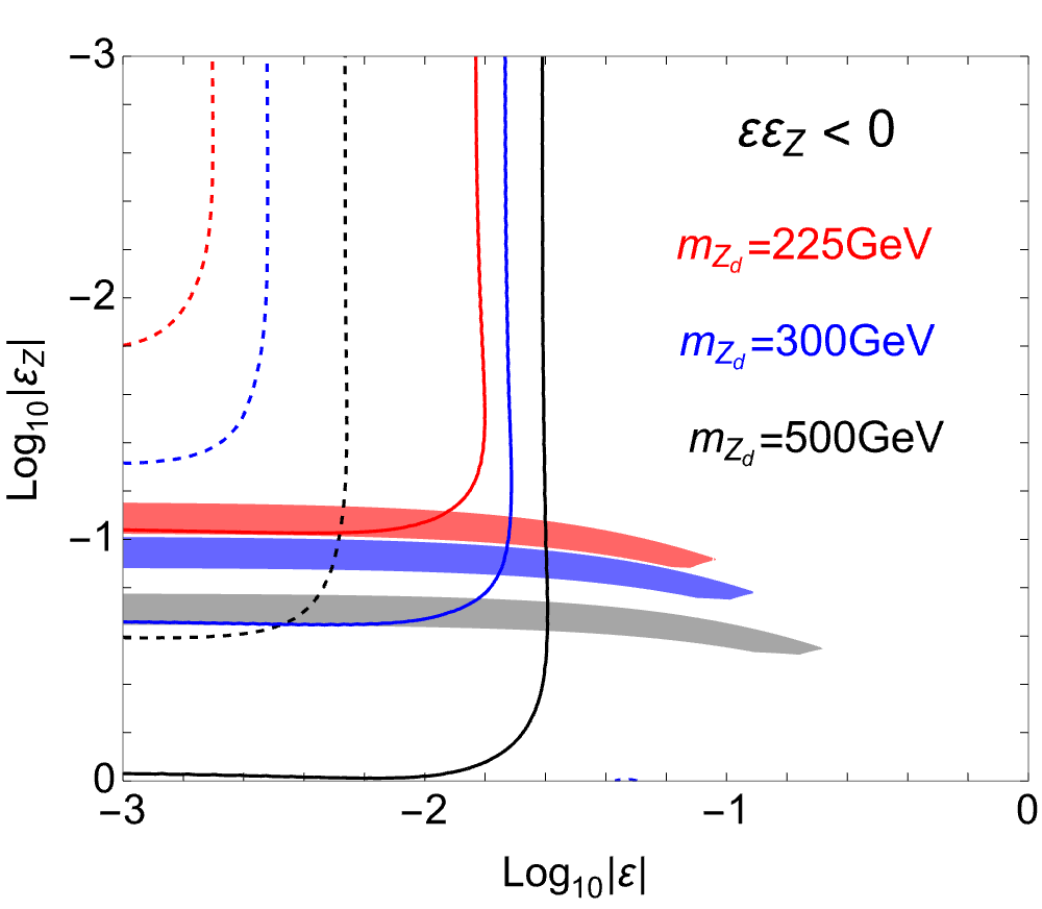}
\label{fig: LHCbound2}}
    \caption{The bound from the dilepton resonance search at LHC in the cases of (a) $\varepsilon \varepsilon_Z^{} > 0$ and (b) $ \varepsilon \varepsilon_Z < 0$.
    In the colored regions, the $W$ boson mass anomaly can be explained under the constraint from the electroweak global fit.
    The solid (dashed) curves are the upper bounds when we do (do not) assume a dark decay channel $Z_d \to \chi\bar\chi$ of $g_d=0.1$ and $m_\chi=50$ GeV. 
    Strong constraints from the dilepton resonance searches are diluted when there is a dominant invisible decay mode and sizable parameter regions survive.}
\end{figure}

Such a dark $Z$ boson is constrained by the dilepton resonance search $Z_d^{} \to \ell^+ \ell^-$ ($\ell = e$, $\mu$) at LHC~\cite{ATLAS:2019erb}. 
The latest result in Ref.~\cite{ATLAS:2019erb} gives a constraint on $\sigma \times \mathrm{Br}[Z_d^{} \to \ell^+ \ell^-]$ in the mass range
\begin{equation}
225~\mathrm{GeV} \leq m_{Z_d}^{} \leq 6000~\mathrm{GeV}, 
\end{equation}
where the lower limit is to avoid the $Z$ boson peak region.
Although the dijet resonance search~\cite{CMS:2019gwf} is also conceivable, 
it gives a weaker constraint in the relevant mass region~\cite{Aboubrahim:2022qln}.

In Fig.~\ref{fig: LHCbound1} [Fig.~\ref{fig: LHCbound2}], 
we show the upper bounds at $95\%$ CL on the mixing parameters from the dilepton resonance search in the case of $\varepsilon \varepsilon_Z^{} > 0$ ($\varepsilon \varepsilon_Z^{} < 0$). 
As benchmark points, we consider the cases of $m_{Z_d}^{} = 225$, $300$, and $500~\mathrm{GeV}$, 
and contours of the upper bound are shown by the red, blue and black dashed curves, respectively. 
In evaluating the bound, we used \texttt{FeynRules}~\cite{Alloul:2013bka} and \texttt{MadGraph5\_aMC@NLO}~\cite{Alwall:2014hca}. 
In the corresponding colored regions, the $W$ boson mass anomaly can be explained while avoiding the constraint from electroweak global fit within $2\sigma$ in each case.
We can see that all the red and blue regions are excluded by the current dilepton resonance search. 
On the other hand, there is an allowed region with small kinetic mixing $\varepsilon\simeq 10^{-3}$ in the case of $m_{Z_d}^{} = 500~\mathrm{GeV}$.

We note that this collider bound can be relaxed by considering a new dark decay mode of the dark $Z$ boson. 
For example, we consider the dark fermion $\chi$ carrying a unit dark charge $Q_d = 1$. 
Then, the dark decay $Z_d \to \chi \bar{\chi}$ can be a dominant decay mode if it is kinematically allowed, and the constraint is relaxed as discussed in Sec.~\ref{sec: WMA_and_DarkZ}.
Here, we consider the case that $m_\chi = 50~\mathrm{GeV}$ and $g_d^{} = 0.1$. 
We note that such a dark fermion does not change the decay mode of the $Z$ boson. 
The upper bounds on $|\varepsilon|$ and $|\varepsilon_Z^{}|$ are shown in Figs.~\ref{fig: LHCbound1} and \ref{fig: LHCbound2} by the solid lines of the same colors as the dashed lines. 
We can see that allowed regions appear in the case of $m_{Z_d} = 225~\mathrm{GeV}$ (the red region) and $300~\mathrm{GeV}$ (the blue region). 
In the case of $m_{Z_d}^{} = 500~\mathrm{GeV}$, the allowed region is extended. 
These allowed regions are expected to be tested by future dilepton resonance searches. 

\section{Summary and conclusion} 
\label{sec: summary}

In this paper, we have investigated the phenomenological impacts of the dark $Z$ boson on the running weak mixing angle and the $W$ boson mass measurements. 
In the first part, we have briefly reviewed the effect of a light dark $Z$ boson on the running weak mixing angle and updated some results from previous works with the latest experimental data.  

In the latter part of our work, we have investigated whether the dark $Z$ model can explain the $W$ boson mass anomaly reported by the CDF experiment along with the constraints from other electroweak observables and collider dilepton resonance searches. We have investigated the effect of $Z_d$ on various electroweak observables including the $W$ boson mass by using the $S$, $T$, and $U$ parameters. We have found that in the pure DZ limit, the equation $S = -U$ holds independently of the mass of $Z_d$ and the size of $\varepsilon_Z^{}$. This is an intriguing feature of the dark $Z$ model not common to many other new physics models including the dark photon model.
It has revealed that heavy dark $Z$ bosons with mass $m_{Z_d}^{} > m_Z^{}$ are required to resolve the $W$ boson mass anomaly. 
The result of the electroweak global fit and the CDF-II result can be compatible in parameter regions with $m_{Z_d}^{}$ larger than $130~\mathrm{GeV}$ and relatively large mass mixing $\varepsilon_Z^{} > \mathcal{O}(0.01)$.
Although the current dilepton resonance searches strongly constrain such parameter regions, we have viable regions for heavy dark $Z$, $m_{Z_d} \gtrsim 500~\mathrm{GeV}$. 
By allowing dark decay channels of $Z_d$, this constraint is relaxed, 
and the allowed regions appear even for lighter $Z_d$ and is extended for heavy $Z_d$. 
Future resonance searches at high-energy colliders would be effective to search for such a dark $Z$ boson.

\section*{Acknowledgement}
This work was supported in part by the U.S. Department of Energy (Grant No. DE-SC0012704) and the National Research Foundation of Korea (Grant No. NRF-2021R1A2C2009718). 
Digital data related to the results presented  in this paper can be found at \cite{digfile}.

\appendix
\section{Formulas for running weak mixing angle}
\label{app: WMA}

In this appendix, we show the formulas required to evaluate the running weak mixing angles in Figs.~\ref{fig: RunningWMA} and \ref{fig: fittingWMA}. 
In addition, we give a comparison between the typical form of the weak mixing angle defined in Refs.~\cite{Czarnecki:2000ic, Czarnecki:1998xc}, which we employ in this paper, and another form using the pinch technique (PT) in Ref.~\cite{Ferroglia:2003wa}.  

Let us first discuss the typical form. 
As shown in Eq.~(\ref{eq: runningWMA}), we need $\hat{\kappa}(q^2)$ to evaluate the running weak mixing angle; 
\begin{equation}
\label{eq: WMA_app}
\sin^2\hat{\theta}(q^2) = \hat{\kappa}(q^2) \sin^2 \hat{\theta}(m_Z^{}). 
\end{equation}
In the following, we use the abbreviations $\hat{s}^2 = \sin^2\hat{\theta}(m_Z^{})$ and $\hat{c}^2 = 1 - \hat{s}^2$.  
The method to evaluate $\hat{\kappa}$ is discussed in Refs.~\cite{Czarnecki:2000ic, Czarnecki:1998xc} in the context of M\o ller scattering. 
The factor $\hat{\kappa}$ is separated into the fermionic contribution $\hat{\kappa}_f^{} (q^2)$ and the bosonic contribution $\hat{\kappa}_b^{} (q^2)$ as follows; 
\begin{equation}
\hat{\kappa}(q^2) = 1 + \hat{\kappa}_f^{} (q^2) + \hat{\kappa}_b^{} (q^2). 
\end{equation}
As discussed in Refs.~\cite{Czarnecki:2000ic, Czarnecki:1998xc}, 
the dominant contribution to $\hat{\kappa}$ comes from the $\gamma$-$Z$ self-energy and the anapole moment. 
The fermionic loop diagrams of $\gamma$-$Z$ self-energy are included in $\hat{\kappa}_f^{}$, while 
the $W$ loop diagram of $\gamma$-$Z$ self-energy and the anapole moment are included in $\hat{\kappa}_b^{}$. 

The formula for spacelike momenta ($q^2 <0$) is given in Refs.~\cite{Czarnecki:2000ic, Czarnecki:1998xc}. 
The formula for timelike momenta ($q^2>0$) can be obtained by appropriately replacing functions in the spacelike formula. 
As a result, the fermionic contribution for both spacelike and timelike momenta is given by the following single formula; 
\begin{align}
\label{eq: kappa_fermionic}
\hat{\kappa}_f^{}(q^2) & =  - \frac{ \alpha }{ 2 \pi \hat{s}^2 }
	\biggl\{
		 \frac{ 1 }{ 3 } \sum_f \bigl(T_{3f}^{} Q_f - 2 \hat{s}^2 Q_f^2 \bigr)
		\nonumber \\
		& \times \biggl[	
			\log \frac{ m_f^2 }{ m_Z^2 }
			- \frac{ 5 }{ 3 }  - \frac{ 4 m_f^2 }{ q^2 }
			+ 2 \Bigl( 1 + \frac{2 m_f^2 }{ q^2 } \Bigr) \Lambda(D_f^{})
		\biggr]
	\biggr\},  
\end{align}
where $m_f^{}$, $Q_f^{}$ and $T_{3f}^{}$ are the mass, the electric charge, and the third component of the isospin of the SM fermion $f$, respectively, 
$D_f^{} = 4m_f^2 / q^2 - 1$, and the function $\Lambda (x)$ is defined as~\cite{Sirlin:1989qz}
\begin{equation}
\Lambda (x) = \left\{
\begin{array}{ll}
\displaystyle{ \sqrt{x} \, \mathrm{arctan}\Bigl( \frac{ 1 }{ \sqrt{x}} \Bigr) } & (x>0), \\[15pt]
\displaystyle{\frac{ \sqrt{-x} }{ 2 } \log \biggl| \frac{ \sqrt{-x} + 1 }{ \sqrt{-x} - 1 } \biggr|} & (x<0). 
\end{array}
\right.
\end{equation}
The summation $\sum_f^{}$ includes quarks' color degree of freedom. 
In order to partly include the non-perturbative effect of the QCD, 
we employ the effective quark masses~\cite{Jegerlehner1991, Marciano:1993jd}; 
$m_u^{} \simeq 62~\mathrm{MeV}$, $m_d^{} \simeq 83~\mathrm{MeV}$, $m_s^{} \simeq 215~\mathrm{MeV}$, $m_c^{} \simeq 1.5~\mathrm{GeV}$, $m_b^{} \simeq 4.5~\mathrm{GeV}$ and multiply the QCD correction factor $(1 + \alpha_s^{} / \pi) \simeq 1.042$ in contributions from quarks. We use the on-shell mass for the top quark $m_t \simeq 172.5~\mathrm{GeV}$~\cite{PDG2022}.

The bosonic contribution is given by 
\begin{align}
\label{eq: kappa_bosonic}
\hat{\kappa}_b^{}(q^2) =  & - \frac{ \alpha }{ 2 \pi \hat{s}^2 }
	\biggl\{
		- \frac{ 42 \hat{c}^2 + 1 }{ 12 } \log \frac{ m_W^2 }{ m_Z^2 }
		+ \frac{ 1 }{ 18 }
		\nonumber \\[5pt]
		& - \bigl( \Lambda (D_W^{}) - 1 \bigr)
		\Bigl[ (7 + 4z ) \hat{c}^2 
			+ \frac{ 1 }{ 6 } ( 1 - 4z ) \Bigr]
		\nonumber \\[5pt]
		& + z \biggl[ \frac{ 3 }{ 4 } + z 
			- 2 \Bigl( z + \frac{ 3 }{ 2 } \Bigr) \Lambda (D_W^{} )
		\nonumber \\
			& \hspace{80pt} -z ( 2 + z )
				\Lambda_2 (D_W^{})
		\biggr] \biggr\}, 
\end{align}
where $z = m_W^2/q^2$, $D_W^{} = 4 m_W^2 / q^2 - 1$, and the function $\Lambda_2(x)$ is given by
\begin{equation}
\Lambda_2 (x) = \left\{
	\begin{array}{ll}
	\displaystyle{ - 4 \mathrm{arctan}^2 \Bigl( \frac{ 1 }{ \sqrt{ x } } \Bigr)} &  (x >0), \\[15pt]
	\displaystyle{\mathrm{Re}\biggl[
		\log^2 \biggl( \frac{ \sqrt{-x} + 1 }{ \sqrt{-x} - 1 } \biggr) \biggr]} &  (x < 0). 
	\end{array}
\right.
\end{equation} 

Now, we discuss the PT form.
In Ref.~\cite{Ferroglia:2003wa}, the running weak mixing angle is discussed with a gauge-invariant and process-independent form factor $\hat{\kappa}^\mathrm{PT}$, which is defined using the PT~\cite{Cornwall:1981zr, Cornwall:1989gv, Papavassiliou:1989zd, Degrassi:1992ue} as follows; 
\begin{equation}
\label{eq: kappa_PT}
    \hat{\kappa}^\mathrm{PT}(q^2, \mu) = 
    1 - \frac{ \hat{c} }{ \hat{s} } \frac{ a_{\gamma Z}^{} (q^2, \mu) }{ q^2 }, 
\end{equation}
where $\mu$ is the renormalization scale, and $a_{\gamma Z}^{}$ is a PT $\gamma$-$Z$ self-energy given by~\cite{Degrassi:1992ue}
\begin{equation}
\label{eq: PT_self_energy}
    a_{\gamma Z}^{} (q^2, \mu) = A_{\gamma Z} (q^2, \mu)
    - \frac {2 e^2}{\hat{c}\hat{s}}
        (2 q^2 \hat{c}^2 - m_W^2 ) I_{WW}^{} (q^2, \mu), 
\end{equation}
where $A_{\gamma Z}$ is the conventional $\gamma$-$Z$ self-energy evaluated in the 't Hooft-Feynman gauge, and $I_{WW}^{}$ is the pinch term from vertex and box diagrams.
In Eq.~(\ref{eq: PT_self_energy}), divergent terms in the self-energy and the pinch term are subtracted using the $\overline{\mathrm{MS}}$ prescription.

Reference~\cite{Ferroglia:2003wa} defines the running weak mixing angle as follows; 
\begin{equation}
    \sin^2 \hat{\theta}_W^\mathrm{PT} (q^2) = 
    \mathrm{Re}[\hat{\kappa}^\mathrm{PT}(q^2, m_Z^{})] \sin^2 \hat{\theta}_W^{}(m_Z^{}). 
\end{equation}
An analytic formula for the real part of the pinch term in the $\overline{\mathrm{MS}}$ scheme is given by~\cite{Degrassi:1992ue} 
\begin{equation}
\label{eq: pinch_term}
    \mathrm{Re}[I_{WW}^{}] =
    \frac{ 1 }{ 8 \pi^2 } 
    \biggl[
        \frac{ 1 }{ 2 } \log \biggl( \frac{ m_W^2 }{ m_Z^2 } \biggr) +  \Lambda(D_W) - 1
    \biggr], 
\end{equation}
at $\mu = m_Z^{}$. In the $\gamma$-$Z$ self-energy, 
the real parts of the fermionic contribution $A_{\gamma Z}^f$~\cite{Marciano:1980pb} and the bosonic contribution $A_{\gamma Z}^b$~\cite{Sirlin:1989qz} are given by
\begin{align}
\label{eq: fermionic_PT}
& \mathrm{Re}[A_{\gamma Z}^f] = 
    \frac{ \alpha q^2 }{ 6 \pi \hat{c} \hat{s} }
    \sum_f \bigl(T_{3f}^{} Q_f - 2 \hat{s}^2 Q_f^2 \bigr)
\nonumber \\
   & \times  \biggl[
        \log \frac{ m_f^2 }{ m_Z^2 } - \frac{ 5 }{  3 } - \frac{ 4 m_f^2 }{ q^2 }
	+ 2 \Bigl( 1 + \frac{ 2 m_f^2 }{ q^2 } \Bigr)\Lambda(D_f^{})
    \biggr], 
\end{align}
and 
\begin{align}
\label{eq: bosonic_PT}
\mathrm{Re}[A_{\gamma Z}^b]  & = 
    - \frac{ \alpha q^2 }{ 4 \pi \hat{s} \hat{c} }
    \biggl\{  - \frac{ 1 }{ 9 }
    +  \frac{ 1 }{ 2 } \log\biggl( \frac{ m_W^2 }{ m_Z^2 } \biggr) 
	\Bigl( 4z + 6 \hat{c}^2 + \frac{ 1 }{ 3 } \Bigr)
\nonumber \\[7pt]
        & 
        + \Bigl( \Lambda(D_W) - 1 \Bigr)\biggl[ 8 z \Bigl( \hat{c}^2 + \frac{ 1 }{ 3 } \Bigr) 
    + 6 \hat{c}^2 + \frac{ 1 }{ 3 } 
        \biggr]
    \biggr\},  
\end{align}
respectively. 
In Eqs.~(\ref{eq: fermionic_PT}) and (\ref{eq: bosonic_PT}), the divergent terms are subtracted using the $\overline{\mathrm{MS}}$ prescription.

The difference between the two definitions $\sin^2\hat{\theta}_W^{}(q^2)$ and $\sin^2\hat{\theta}_W^\mathrm{PT}(q^2)$ is caused by the bosonic contribution: vertex and box contributions mediated by the $W$ boson~\cite{Ferroglia:2003wa}. 
The fermionic contribution is the same because it comes from the same $\gamma$-$Z$ self-energy diagrams. 
Using all the formulas above, the difference $\Delta \hat{s}^2 (q^2)$ is evaluated by 
\begin{align}
\Delta \hat{s}^2 (q^2) & \equiv 
    \sin^2 \hat{\theta}_W^{}(q^2) - \sin^2 \hat{\theta}_W^\mathrm{PT}(q^2)
    \nonumber \\[5pt]
    & = \frac{ \alpha z }{ 2 \pi }
    \Bigl[
        z + \frac{ 9 }{ 4 } 
        + (2z+3)\bigl( \Lambda(D_W) - 1 \bigr)
        \nonumber \\[5pt]
        & \hspace{20pt} + z (2 + z) \Lambda_2(D_W)
    \Bigr]. 
\end{align}
At low energies ($q^2 \to 0$ and $|z| \to \infty$), 
\begin{align}
& \Lambda(D_W) \simeq 1 - \frac{ 1 }{ 12 z } - \frac{ 1 }{ 120 z^2 } \cdots, \\[5pt]
& \Lambda_2(D_W) \simeq -\frac{ 1 }{ z }  - \frac{ 1 }{ 12 z^2 } - \frac{ 1 }{ 90 z^3 } \cdots , 
\end{align}
and we have
\begin{equation}
\label{eq: difference_Low_energies}
\Delta \hat{s}^2 (0) = -\frac{ 2 \alpha }{ 9 \pi } \simeq - 5.5 \times 10^{-4}. 
\end{equation}
At high energies ($|q^2| \to \infty$ and $z \to 0$), 
the two weak mixing angles coincide; 
\begin{equation}
\label{eq: difference_High_energies}
\Delta \hat{s}^2 (q^2) \to 0 \quad (|q^2| \to \infty), 
\end{equation}
where we use 
\begin{equation}
\lim_{|q|^2 \to \infty} z \Lambda(D_W) = 0, \quad 
\lim_{|q|^2 \to \infty} z \Lambda_2(D_W) = 0. 
\end{equation}

\begin{figure}[h]
\begin{center}
    \includegraphics[width=0.45\textwidth]{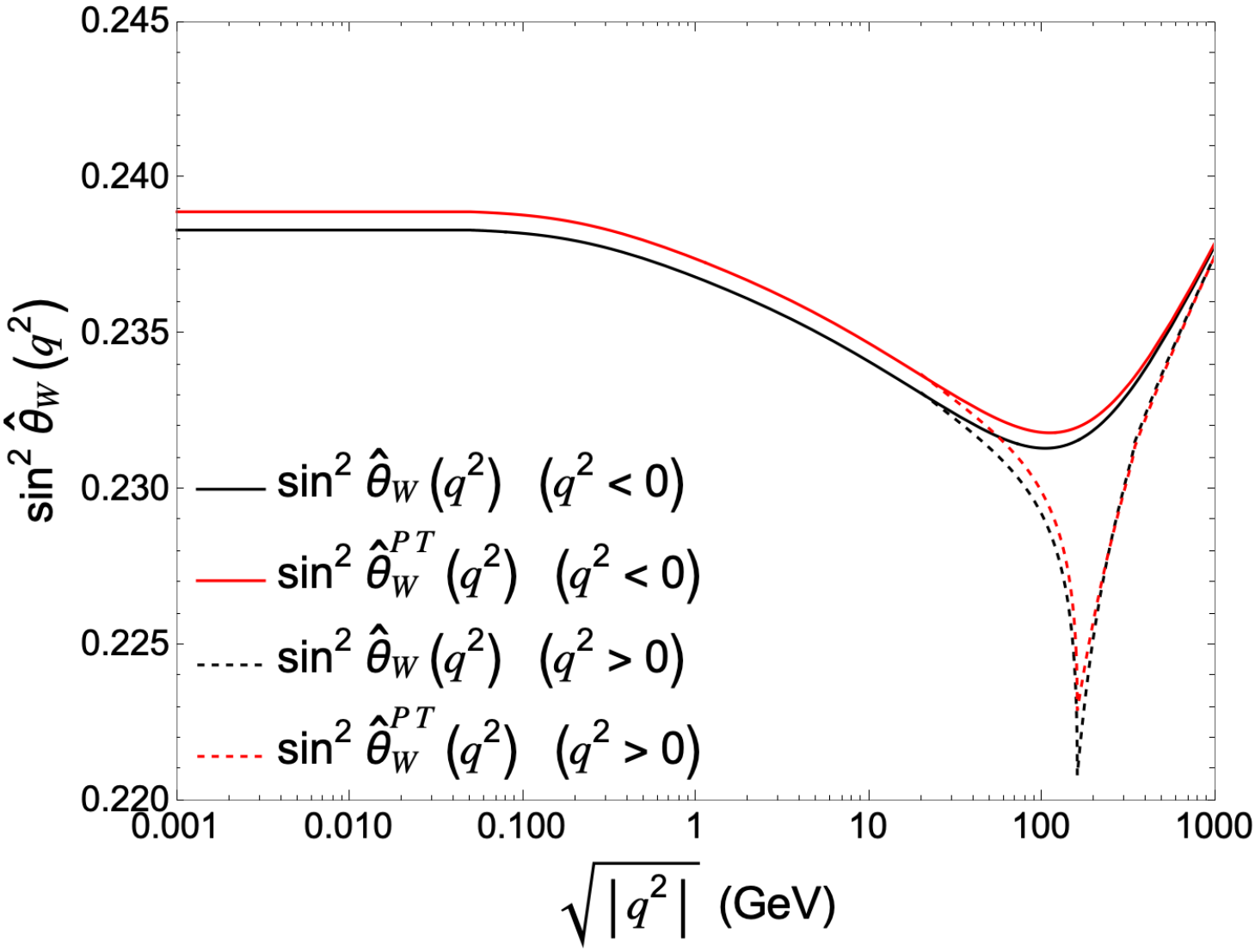}
    \caption{A comparison of typical form $\sin^2 \hat{\theta}_W^{}(q^2)$ in black and PT form $\sin^2 \hat{\theta}^\mathrm{PT}_W(q^2)$ in red. Solid (dashed) curves represent spacelike (timelike) momenta. The curves for timelike momenta are shown only in a domain $\sqrt{|q^2|} > 20~\mathrm{GeV}$.}
    \label{eq: comparison_WMA}
\end{center}
\end{figure}

In Fig.~\ref{eq: comparison_WMA}, 
the two definitions of the weak mixing angle are compared. 
The black-solid (red-solid) and black-dashed (red-dashed) curves show $\sin^2\hat{\theta}_W^{} (q^2)$ [$\sin^2 \hat{\theta}_W^\mathrm{PT}(q^2)$] in the spacelike and timelike domains, respectively. 
The curves for timelike momenta (dashed curves) are shown only in a domain $\sqrt{|q^2|} > 20~\mathrm{GeV}$ to avoid spikes from the (unphysical) effective quark masses ($m_q^{}$) around the region $q^2 \simeq 4 m_q^2$. 
Except for these spikes, the behavior in the time-like domain is in good agreement with that in the spacelike domain at low energies.
In Ref.~\cite{Ferroglia:2003wa}, the running of the electromagnetic coupling $\alpha$~\cite{Marciano:1980be} and the complex mass of the $W$ boson~\cite{Sirlin:1991fd}, which soften the behavior around the $W$-$W$ threshold, were employed in the evaluation of $\sin^2\hat{\theta}_W^{\mathrm{PT}}$.
However, we do not employ them here in order to compare it to $\sin^2 \hat{\theta}_W^{}(q^2)$ which does not use these prescriptions. 
We can see that the difference is larger at low energies and it seems to converge to a certain value, while it vanishes at high energies. 
These behaviors are what is expected from Eqs.~(\ref{eq: difference_Low_energies}) and (\ref{eq: difference_High_energies}). 

\section{\boldmath Decoupling behavior of the $S$, $T$, and $U$ parameters}
\label{app: SMEFT}

In this appendix, we discuss the $S$, $T$, and $U$ parameters in the decoupling limit of the dark $Z$ model ($m_{Z_d} \gg m_Z$). 
They are described by higher-dimensional operators. 
As shown in Sec.~\ref{sec: STU}, the source of $S$ is a dimension-six operator in the DP limit, while it is a dimension-eight operator in the pure DZ limit. 
The purpose of this appendix is to identify the cause of this difference. 

Before considering higher-dimensional operators in the model, 
we review the general discussion in Ref.~\cite{Burgess:1993vc} of the effect of deviations in the gauge sector on electroweak observables. 
The deviation in the gauge sector is parametrized as follows; 
\begin{align}
\Delta \mathcal{L} =&  - \frac{ A }{ 4 } F_{\mu\nu} F^{\mu\nu} - \frac{ B }{ 2 } W_{\mu\nu}^\dagger W^{\mu\nu} - \frac{ C }{ 4 } Z_{\mu\nu} Z^{\mu\nu} \nonumber \\
& + \frac{ G }{ 2 } F_{\mu\nu} Z^{\mu\nu} + w \tilde{m}_W^2 W_\mu^\dagger W^\mu + \frac{ z }{ 2 } \tilde{m}_Z^2 Z_\mu Z^\mu, 
\end{align}
where $F_{\mu\nu}$, $W_{\mu\nu}$, and $Z_{\mu\nu}$ are the field strengths for the photon $A_\mu$, the $W$ boson $W_\mu$, and the $Z$ boson $Z_\mu$, respectively, and 
$\tilde{m}_W$ and $\tilde{m}_Z$ are the masses of the $W$ and $Z$ bosons in the SM. 

Canonical kinetic terms are given by redefining the gauge fields  
\begin{align}
& A_\mu \to \left( 1 - \frac{ A }{ 2 } \right) A_\mu + G Z_\mu, \\
& W_\mu \to \left( 1 - \frac{ B }{ 2 } \right) W_\mu, \\
& Z_\mu \to \left( 1 - \frac{ C }{ 2 } \right) Z_\mu,  
\end{align}
up to linear order corrections in small parameters $A$, $B$, $C$, and $G$. 
Then, the gauge sector is given by 
\begin{align}
\mathcal{L}_\mathrm{gauge} = & - \frac{ 1 }{ 4 } F_{\mu\nu} F^{\mu\nu} - \frac{ 1 }{ 2 } W_{\mu\nu}^\dagger W^{\mu\nu} - \frac{ 1 }{ 4 } Z_{\mu\nu} Z^{\mu\nu}
\nonumber \\[5pt]
& + (1 + w - B ) \tilde{m}_W^2 W_{\mu}^\dagger W^\mu  \nonumber \\[5pt]
& + \frac{ 1 }{ 2 } (1 + z - C) \tilde{m}_Z^2 Z_\mu Z^\mu, 
\end{align}
and the current interactions are given by 
\begin{align}
& \mathcal{L}_\mathrm{em} = - \tilde{e} \left( 1 - \frac{ A }{ 2 } \right) J_\mathrm{em}^\mu A_\mu, \\
& \mathcal{L}_\mathrm{CC} = - \frac{ \tilde{e} }{ \sqrt{2} \tilde{s}_W } \left( 1 - \frac{ B }{ 2 } \right) \bigl( J_{CC}^\mu W_\mu^\dagger + \mathrm{H.c.} \bigr), \\
& \mathcal{L}_\mathrm{NC} = - \frac{ \tilde{e} }{ \tilde{s}_W \tilde{c}_W } \left( 1 - \frac{ C }{ 2 } \right) \left( \frac{ 1 }{ 2 } J_\mathrm{NC}^\mu  + \tilde{s}_W \tilde{c}_W G J_\mathrm{em}^\mu \right)Z_\mu, 
\end{align}
where tildes on $e$, $s_W$, and $c_W$ mean that they are defined by using tree-level gauge coupling as in the SM. 
They are not physical quantities because of the deviations. 
The physical quantities $e$, $s_W$, and $c_W$ satisfy 
\begin{align}
& \tilde{e} = e \left( 1 + \frac{ A }{ 2 } \right), \\
& \tilde{s}_W^2 = s_W^2 \left\{ 1 + \frac{ c_W^2 }{ c_W^2 - s_W^2 }(A - C - w + z) \right\}, \\
& \tilde{c}_W^2 = c_W^2 \left\{ 1 - \frac{ s_W^2 }{ c_W^2 - s_W^2 }(A - C - w + z) \right\}, 
\end{align}
where $s_W^2$ and $c_W^2$ are defined using $\alpha$, $G_F$, and $m_Z$ as 
\begin{equation}
    \frac{ G_F }{ \sqrt{2} } = \frac{ e^2 }{ 8 s_W^2 c_W^2 m_Z^2 }.
\end{equation}

By using these deviations, 
the $S$, $T$, and $U$ parameters are generally given by 
\begin{align}
    & \alpha S = 4 s_W^2 c_W^2 \left( A - C - \frac{ c_W^2 - s_W^2 }{ c_W s_W } G \right), \\
    & \alpha T = w - z, \\
    & \alpha U = 4 s_W^4 \left( A - \frac{ 1 }{ s_W^2 } B + \frac{ c_W^2 }{ s_W^2 } C - 2 \frac{ c_W }{ s_W } G \right). 
\end{align}
In the dark $Z$ model, $z - C = \Delta_1$, $C = - 2 \Delta_2$, $G = \Delta_3$, and the other deviations are zero [See Eq.~(\ref{eq: deviation_DarkZ})]. We note that in this case, the difference between $S$ and $U$ is caused by $G$, and $T$ is generated by only $z$. 

Next, we discuss higher-dimensional operators in the dark $Z$ model. 
We can take the decoupling limit of $Z_d$ by considering $v_d \gg v$. 
$\tilde{Z}_{d\mu}$ obtains a large mass proportional to $v_d$, while other gauge bosons remain massless because $\hat{Z}_{d\mu} = \eta \tilde{Z}_{d\mu}$, which does not include the electroweak gauge boson $\tilde{B}_\mu$. 
In such a case, $\tilde{Z}_{d\mu}$ represents $Z_d$, a heavy mode that should be integrated out at low energies, while $\tilde{Z}_\mu$ represents the $Z$ boson. 
Then, at the electroweak scale, the effect of $Z_d$ is described by higher-dimensional operators including the fields in the SM and $\Phi_2$. In the following, we employ the unitary gauge for $Z_d$ to avoid the nonphysical mode associated with the $U(1)_d$ breaking.

First, we consider higher-dimensional operators in the DP limit ($\varepsilon_Z^{} \rightarrow 0$), which is realized by taking $v_2 = 0$. 
Then, $\Phi_2$ is irrelevant in the present discussion, and we neglect it here. 
$\tilde{Z}_{d}$ has the following interactions;
\begin{equation}
\label{eq: Zd_interaction_DPlimit}
    \mathcal{L}_d = - \frac{ i e }{ 2 c_W^2 }  \eta \varepsilon\bigl( \Phi_1^\dagger \overleftrightarrow{D}_\mu \Phi_1 \bigr) \tilde{Z}_{d}^\mu 
         - \frac{ e }{ c_W^2 }  \eta \varepsilon \sum_f \overline{f} \gamma_\mu Y f \tilde{Z}_{d}^\mu + \cdots, 
\end{equation}
where $\Phi_1^\dagger \overleftrightarrow{D}_\mu \Phi_1$ is defined as $\Phi_1^\dagger D_\mu \Phi_1 - (D_\mu \Phi_1)^\dagger \Phi_1$, $D_\mu$ is the covariant derivative in the SM, $Y$ is the hypercharge, and $f$ is the SM fermions with a definite chirality.\footnote{In a strict sense, $c_W^{}$ and $e$ in Eq.~(\ref{eq: Zd_interaction_DPlimit}) should be $\tilde{c}_W$ and $\tilde{e}$. However, this difference is a higher-order effect in higher-dimensional operators and the $S$, $T$, and $U$ parameters. We also replace $\tilde{m}_Z^{}$ and $\tilde{m}_{Z_d}^{}$ with $m_Z^{}$ and $m_{Z_d}^{}$, respectively when the differences are higher-order effects.}
Therefore, at the leading order in $1/m_{Z_d}^2$, the following effective interactions are generated via $Z_d$-mediated tree-level diagrams
\begin{equation}
\mathcal{L}^{(6)}_\mathrm{DP} = \frac{ 1 }{ 8 m_{Z_d}^2 } \left( \frac{e \eta  \varepsilon}{ c_W^2 } \right)^2 O_{\varphi}^{(6)}
	- \frac{ i }{ 2 m_{Z_d}^2 } \left( \frac{e  \eta \varepsilon}{ c_W^2 } \right)^2  O_{\varphi \psi}^{(6)}, 
\end{equation}
where $O_{\varphi}^{(6)}$ and $O_{\varphi \psi}^{(6)}$ are dimension-six operators defined as 
\begin{equation}
O_{\varphi}^{(6)} = O_1^\mu O_{1\mu}, \quad 
O_{\varphi \psi}^{(6)} = O_{1\mu} \sum_f \overline{f} \gamma^\mu Y f, 
\end{equation}
with $O_1^\mu = \Phi_1^\dagger \overleftrightarrow{D}^\mu \Phi_1$. 
We note that the coefficient of $O_{\varphi}^{(6)}$ includes a symmetry factor $1/2$. 
Although four-fermion interactions are also generated at $\mathcal{O}(m_{Z_d}^2)$, we neglect them because they do not affect the gauge sector. 

After $\Phi_1$ acquires a VEV ($\left< \Phi_1 \right> = v/\sqrt{2}$), 
$O_{\varphi}^{(6)}$ and $O_{\varphi \psi}^{(6)}$ induce deviations in the mass term and the current interactions of the $Z$ boson, respectively. 
They are given by
\begin{align}
\Delta \mathcal{L}_\mathrm{DP}^{(6)} = & 
	\frac{ \tilde{m}_Z^2 }{ 2 } (z_\mathrm{DP}^{(6)} - C_\mathrm{DP}^{(6)}) Z_\mu Z^\mu \nonumber \\
	& + \frac{ C_\mathrm{DP}^{(6)} }{ 4 } \frac{ \tilde{e} }{ \tilde{s}_W \tilde{c}_W } J_\mathrm{NC}^\mu Z_\mu
	- \tilde{e} G_\mathrm{DP}^{(6)} J_\mathrm{em}^\mu Z_\mu, 
\end{align}
where 
\begin{equation}
\begin{array}{c}
C_\mathrm{DP}^{(6)} = 2 ( \eta\varepsilon t_W^{})^2 r^{-2}, \quad  G_\mathrm{DP}^{(6)} = \eta^2\varepsilon^2  t_W^{} r^{-2}, \\
 z_\mathrm{DP}^{(6)} = (\eta\varepsilon  t_W^{})^2 r^{-2}, 
\end{array}
\end{equation}
where $r = m_{Z_d}^{} / m_{Z}^{}$. 
The $S$, $T$, and $U$ parameters at $\mathcal{O}(m_{Z_d}^{-2})$ are given by 
\begin{equation}
\begin{array}{l}
    \alpha S_\mathrm{DP}^{(6)} = - 4 s_W^2\eta^2  \varepsilon^2 r^{-2}, \\ 
    \alpha T_\mathrm{DP}^{(6)} = - t_W^2\eta^2  \varepsilon^2 r^{-2 }, \\
    \alpha U_\mathrm{DP}^{(6)} = 0. 
\end{array}
\end{equation}
The effect of $C_\mathrm{DP}^{(6)}$ and $G_\mathrm{DP}^{(6)}$ in $U_\mathrm{DP}^{(6)}$ are canceled, and the $U$ paramter is not generated at $\mathcal{O}(m_{Z_d}^{-2})$. This cancelation is due to the fact that $C_\mathrm{DP}^{(6)}$ and $G_\mathrm{DP}^{(6)}$ are caused by the deviation in the interaction between the $Z$ boson and the hypercharge current. Since the hypercharge current is given by 
\begin{equation}
\sum_f \bar{f} \gamma^\mu Y f = - \frac{ 1 }{ 2 } J_\mathrm{NC}^\mu + c_W^2 J_\mathrm{EM}, 
\end{equation}
the $C$ and $G$ generated by the deviation in the hypercharge current interaction, which are denoted by $C_Y$ and $G_Y$, satisfy the relation $C_Y = 2 t_W^{} G_Y$. 
Then, $U_Y$, which is the $U$ parameter induced by $C_Y$ and $G_Y$, is given by
\begin{equation}
\label{eq: vanishing_UY}
\alpha U_Y = \ 4 s_W^4 
\left( \frac{ c_W^2 }{ s_W^2 } C_Y - 2 \frac{ c_W^{} }{ s_W^{} } G_Y \right) = 0.
\end{equation}
Here, we note that dimension-six operators do not contribute to the $U$ parameter.  However, a deviation in the hypercharge current interaction does not induce the $U$ parameter, regardless of the dimension of the operator generating the deviation.

At the leading order, the $U$ parameter is generated by a dimension-eight operator whose Feynman diagram is the same as that for $O_{\varphi}^{(6)}$. 
The propagator of $Z_d$ in the unitary gauge is approximately given by
\begin{align}
\bigl< \tilde{Z}_{d}^\mu \tilde{Z}_d^\nu \bigr>_0 = &
\frac{ -i }{ q^2 - m_{Z_d}^2 }\left( g^{\mu\nu} - \frac{ q^\mu q^\nu }{ m_{Z_d}^2 } \right) \nonumber \\[7pt]
\simeq & \frac{ i g^{\mu\nu} }{ m_{Z_d}^2 }
    + \frac{ i }{ m_{Z_d}^4 } \bigl( q^2 g^{\mu\nu} - q^\mu q^\nu \bigr)
    + \mathcal{O}\left( \frac{ 1 }{ m_{Z_d}^6 } \right), 
\end{align}
where $q^\mu$ is the momentum carried by $Z_d$. The first term yields $O_{\varphi}^{(6)}$, while the second term generates the following effective interaction
\begin{equation}
\mathcal{L}_\mathrm{DP}^{(8)} = \frac{ 1 }{ 8 m_{Z_d}^4 }
    \left( \frac{ e \eta \varepsilon }{ c_W^2 } \right)^2 O_\varphi^{(8)}, 
\end{equation}
where $O_\varphi^{(8)}$ is a dimension-eight operator given by
\begin{align}
    O_{\varphi}^{(8)} & = O_{1\mu} \bigl( - \partial^2 g^{\mu\nu} + \partial^\mu \partial^\nu \bigr) O_{1\nu}
\nonumber \\[5pt]
& = \frac{ 1 }{ 2 } O_1^{\mu \nu} O_{1\mu\nu} + \cdots, 
\end{align}
with $O_1^{\mu\nu} = \partial^\mu O_1^\nu - \partial^\nu O_1^\mu$, and $\cdots$ represents terms of total divergence. 
This induces a deviation in the kinetic term of the $Z$ boson after the electroweak symmetry breaking; 
\begin{equation}
\Delta \mathcal{L}_\mathrm{DP}^{(8)} 
	= - \frac{ C_\mathrm{DP}^{(8)} }{ 4 }  Z_{\mu\nu}Z^{\mu\nu}, 
\end{equation}
where $C_\mathrm{DP}^{(8)} =  \bigl( \eta \varepsilon t_W^{} \bigr)^2 r^{-4}$. 
This gives the leading term of the $U$ parameter 
\begin{equation}
\alpha U_\mathrm{DP}^{(8)} = 4 s_W^4 \eta^2 \varepsilon^2 r^{-4} . 
\end{equation}
The deviation in the hypercharge current interaction $Z^\mu \sum_f \bar{f} \gamma^\mu Y f$ is also generated by dimension-eight operators and induces $C$ and $G$ at $\mathcal{O}(m_{Z_d}^{-4})$. 
However, it does not contribute to $U$ as shown in Eq.~(\ref{eq: vanishing_UY}). 

Consequently, in the dark photon limit, the leading terms of the $S$, $T$, and $U$ parameters in the decoupling limit are induced by $O_{\varphi \psi}^{(6)}$, $O_{\varphi}^{(6)}$, and $O_\varphi^{(8)}$, respectively, as follows; 
\begin{equation}
\begin{array}{l}
\alpha S_\mathrm{DP} (r^2 \gg 1) \simeq  - 4 s_W^2 \eta^2\varepsilon^2  r^{-2}, \\
\alpha T_\mathrm{DP} (r^2 \gg 1) \simeq - t_W^2 \eta^2\varepsilon^2  r^{-2}, \\
\alpha U_\mathrm{DP} (r^2 \gg 1) \simeq 4 s_W^4 \eta^2\varepsilon^2  r^{-4}, 
\end{array}
\end{equation}
which coincides with Eqs.~(\ref{eq: STU_decouplingDP}) when $\eta \simeq 1$ is taken.

Next, we consider higher-dimensional operators in the pure DZ limit. 
In this limit, the kinetic mixing $\varepsilon$ vanishes, and $\hat{Z}_d^\mu = \tilde{Z}_d^\mu$ and $\hat{B}_\mu = \tilde{B}_\mu$ hold [See Eq.~(\ref{eq: mixing_hat_tilde})]. 
Thus, $\tilde{Z}_d$ couples to only $\Phi_2$ at tree level, which carries the dark charge $Q_d[\Phi_2] = 1$; 
\begin{equation}
    \mathcal{L}_d = | (D_\mu + i g_d \tilde{Z}_d) \Phi_2 |^2
    = - i g_d \bigl(\Phi_2^\dagger \overleftrightarrow{D}^\mu \Phi_2 \bigr)
        \tilde{Z}_{d_\mu} + \cdots, 
\end{equation}
where $\cdots$ are irrelevant terms in the present discussion. 
As shown in the case of the dark photon limit, 
this term generates the following effective interactions;
\begin{equation}
\mathcal{L}_\mathrm{DZ} = \frac{ g_d^2  }{ 2 m_{Z_d}^2 }O_\varphi^{\prime (6)}
	+ \frac{ g_d^2 }{ 2 m_{Z_d}^4 } O_\varphi^{\prime (8)}, 
\end{equation}
where the dimension-six operator $O_\varphi^{\prime (6)}$ and dimension-eight operator $O_\varphi^{\prime (8)}$ are given by 
\begin{align}
    & O_\varphi^{\prime (6)} =   O_{2 \mu}O_2^{\mu}
    & O_\varphi^{\prime (8)} =   \frac{ 1 }{ 2 } O_{2\mu\nu}O_2^{ \mu\nu}, 
\end{align}
and $O_2^{\mu} = \Phi_2^\dagger \overleftrightarrow{D}^\mu \Phi_2$, and $O_2^{\mu \nu} = \partial^\mu O_2^{\nu} - \partial^\nu O_2^{\mu}$.

After electroweak symmetry breaking ($\left< \Phi_2 \right> = v_2 /\sqrt{2}$), 
these operators induce the following deviation in the gauge sector 
\begin{equation}
    \Delta \mathcal{L}_\mathrm{DZ} = 
    - \frac{ 1 }{ 4 } \varepsilon_Z^2 r^{-4} Z_{\mu\nu} Z^{\mu\nu}
    - \frac{ \tilde{m}_Z^2 }{ 2 } \varepsilon_Z^2 r^{-2} Z_\mu Z^\mu. 
\end{equation}
Thus, we have $C_\mathrm{DZ}^{(8)} = \varepsilon_Z^2 r^{-4}$ and $z_\mathrm{DZ}^{(6)} = - \varepsilon_Z^2 r^{-2}$. 
The deviation in the current interactions is not generated because $\tilde{Z}_{d\mu}$ has no coupling with the SM fermions. Thus, $G$ is zero, and it leads to $S_\mathrm{DZ} = - U_\mathrm{DZ}$ as discussed in Sec.~\ref{sec: STU}. 

The deviation $C_\mathrm{DZ}^{(8)}$ induces $S_\mathrm{DZ}$ and $U_\mathrm{DZ}$, while $z_\mathrm{DZ}^{(6)}$ gives $T_\mathrm{DZ}$. 
Consequently, in the pure DZ limit, the leading terms of the $S$, $T$, and $U$ parameters are given by
\begin{equation}
\begin{array}{l}
\alpha S_\mathrm{DZ} (r^2 \gg 1) \simeq -4 s_W^2 c_W^2 \varepsilon_Z^2 r^{-4}, \\
\alpha T_\mathrm{DZ} (r^2 \gg 1) \simeq \varepsilon_Z^2 r^{-2}, \\
\alpha U_\mathrm{DZ} (r^2 \gg 1) \simeq 4 s_W^2 c_W^2 \varepsilon_Z^2 r^{-4}, 
\end{array}
\end{equation}
which coincide with Eq.~(\ref{eq: STU_decouplingDZ}).


\begin{thebibliography}{99}


\bibitem{PDG2022}
R.~L.~Workman \textit{et al.} [Particle Data Group],
``Review of Particle Physics,''
PTEP \textbf{2022}, 083C01 (2022)

\bibitem{Super-Kamiokande:1998kpq}
Y.~Fukuda \textit{et al.} [Super-Kamiokande],
``Evidence for oscillation of atmospheric neutrinos,''
Phys. Rev. Lett. \textbf{81}, 1562-1567 (1998)
[arXiv:hep-ex/9807003 [hep-ex]].

\bibitem{SNO:2001kpb}
Q.~R.~Ahmad \textit{et al.} [SNO],
``Measurement of the rate of $\nu_e+d \to p+p+e^-$ interactions produced by $^8$B solar neutrinos at the Sudbury Neutrino Observatory,''
Phys. Rev. Lett. \textbf{87}, 071301 (2001)
[arXiv:nucl-ex/0106015 [nucl-ex]]; 
``Direct evidence for neutrino flavor transformation from neutral current interactions in the Sudbury Neutrino Observatory,''
Phys. Rev. Lett. \textbf{89}, 011301 (2002)
[arXiv:nucl-ex/0204008 [nucl-ex]].

\bibitem{Planck:2018vyg}
N.~Aghanim \textit{et al.} [Planck],
``Planck 2018 results. VI. Cosmological parameters,''
Astron. Astrophys. \textbf{641}, A6 (2020)
[erratum: Astron. Astrophys. \textbf{652}, C4 (2021)]
[arXiv:1807.06209 [astro-ph.CO]].

\bibitem{Fields:2019pfx}
B.~D.~Fields, K.~A.~Olive, T.~H.~Yeh and C.~Young,
``Big-Bang Nucleosynthesis after Planck,''
JCAP \textbf{03}, 010 (2020)
[erratum: JCAP \textbf{11}, E02 (2020)]
[arXiv:1912.01132 [astro-ph.CO]].

\bibitem{Muong-2:2021ojo}
B.~Abi \textit{et al.} [Muon g-2],
``Measurement of the Positive Muon Anomalous Magnetic Moment to 0.46 ppm,''
Phys. Rev. Lett. \textbf{126}, no.14, 141801 (2021)
[arXiv:2104.03281 [hep-ex]].

\bibitem{CDF:2022hxs}
T.~Aaltonen \textit{et al.} [CDF],
``High-precision measurement of the $W$ boson mass with the CDF II detector,''
Science \textbf{376}, no.6589, 170-176 (2022)

\bibitem{Davoudiasl:2012ag}
H.~Davoudiasl, H.~S.~Lee and W.~J.~Marciano,
```Dark' Z implications for Parity Violation, Rare Meson Decays, and Higgs Physics,''
Phys. Rev. D \textbf{85}, 115019 (2012)
[arXiv:1203.2947 [hep-ph]].

\bibitem{Holdom:1985ag}
B.~Holdom,
``Two U(1)'s and Epsilon Charge Shifts,''
Phys. Lett. B \textbf{166}, 196-198 (1986)

\bibitem{Davoudiasl:2012qa}
H.~Davoudiasl, H.~S.~Lee and W.~J.~Marciano,
``Muon Anomaly and Dark Parity Violation,''
Phys. Rev. Lett. \textbf{109}, 031802 (2012)
[arXiv:1205.2709 [hep-ph]].

\bibitem{Lee:2013fda}
H.~S.~Lee and M.~Sher,
``Dark Two Higgs Doublet Model,''
Phys. Rev. D \textbf{87}, no.11, 115009 (2013)
[arXiv:1303.6653 [hep-ph]].

\bibitem{Davoudiasl:2013aya}
H.~Davoudiasl, H.~S.~Lee, I.~Lewis and W.~J.~Marciano,
``Higgs Decays as a Window into the Dark Sector,''
Phys. Rev. D \textbf{88}, no.1, 015022 (2013)
[arXiv:1304.4935 [hep-ph]].

\bibitem{Davoudiasl:2014mqa}
H.~Davoudiasl, W.~J.~Marciano, R.~Ramos and M.~Sher,
``Charged Higgs Discovery in the W plus ``Dark'' Vector Boson Decay Mode,''
Phys. Rev. D \textbf{89}, no.11, 115008 (2014)
[arXiv:1401.2164 [hep-ph]].

\bibitem{Davoudiasl:2014kua}
H.~Davoudiasl, H.~S.~Lee and W.~J.~Marciano,
``Muon $g-2$, rare kaon decays, and parity violation from dark bosons,''
Phys. Rev. D \textbf{89}, no.9, 095006 (2014)
[arXiv:1402.3620 [hep-ph]].


\bibitem{Xu:2015wja}
F.~Xu,
``Dark $Z$ Implication for Flavor Physics,''
JHEP \textbf{06}, 170 (2015)
[arXiv:1504.07415 [hep-ph]].

\bibitem{Davoudiasl:2015bua}
H.~Davoudiasl, H.~S.~Lee and W.~J.~Marciano,
``Low $Q^2$ weak mixing angle measurements and rare Higgs decays,''
Phys. Rev. D \textbf{92}, no.5, 055005 (2015)
[arXiv:1507.00352 [hep-ph]].

\bibitem{San:2022uud}
Y.~C.~San, M.~Perelstein and P.~Tanedo,
``Dark Z at the International Linear Collider,''
Phys. Rev. D \textbf{106}, no.1, 015027 (2022)
[arXiv:2205.10304 [hep-ph]].

\bibitem{Goyal:2022vkg}
A.~Goyal, M.~Kumar, S.~Kumar and R.~Rahaman,
``Exploring dark $Z_d$-boson in future large hadron-electron collider,''
Eur. Phys. J. C \textbf{83}, no.2, 132 (2023)
[arXiv:2209.03240 [hep-ph]].

\bibitem{Datta:2022zng}
A.~Datta, A.~Hammad, D.~Marfatia, L.~Mukherjee and A.~Rashed,
``Dark photon and dark Z mediated B meson decays,''
JHEP \textbf{03}, 108 (2023)
[arXiv:2210.15662 [hep-ph]].

\bibitem{Holdom:1990xp}
B.~Holdom,
``Oblique electroweak corrections and an extra gauge boson,''
Phys. Lett. B \textbf{259}, 329-334 (1991)

\bibitem{Burgess:1993vc}
C.~P.~Burgess, S.~Godfrey, H.~Konig, D.~London and I.~Maksymyk,
``Model independent global constraints on new physics,''
Phys. Rev. D \textbf{49}, 6115-6147 (1994)
[arXiv:hep-ph/9312291 [hep-ph]].

\bibitem{Babu:1997st}
K.~S.~Babu, C.~F.~Kolda and J.~March-Russell,
``Implications of generalized Z - Z-prime mixing,''
Phys. Rev. D \textbf{57}, 6788-6792 (1998)
[arXiv:hep-ph/9710441 [hep-ph]].

\bibitem{Peskin:1990zt}
M.~E.~Peskin and T.~Takeuchi,
``A New constraint on a strongly interacting Higgs sector,''
Phys. Rev. Lett. \textbf{65}, 964-967 (1990);
``Estimation of oblique electroweak corrections,''
Phys. Rev. D \textbf{46}, 381-409 (1992).

\bibitem{Grinstein:1991cd}
B.~Grinstein and M.~B.~Wise,
``Operator analysis for precision electroweak physics,''
Phys. Lett. B \textbf{265}, 326-334 (1991)

\bibitem{Cheng:2022aau}
Y.~Cheng, X.~G.~He, F.~Huang, J.~Sun and Z.~P.~Xing,
``Dark photon kinetic mixing effects for the CDF W-mass measurement,''
Phys. Rev. D \textbf{106}, no.5, 055011 (2022)
[arXiv:2204.10156 [hep-ph]].



\bibitem{ALEPH:2005ab}
S.~Schael \textit{et al.} [ALEPH, DELPHI, L3, OPAL, SLD, LEP Electroweak Working Group, SLD Electroweak Group and SLD Heavy Flavour Group],
``Precision electroweak measurements on the $Z$ resonance,''
Phys. Rept. \textbf{427}, 257-454 (2006)
[arXiv:hep-ex/0509008 [hep-ex]].

\bibitem{CDF:2016cei}
T.~A.~Aaltonen \textit{et al.} [CDF],
``Measurement of $\sin^2\theta^{\rm lept}_{\rm eff}$ using $e^+e^-$ pairs from $\gamma^*/Z$ bosons produced in $p\bar{p}$ collisions at a center-of-momentum energy of 1.96 TeV,''
Phys. Rev. D \textbf{93}, no.11, 112016 (2016)
[arXiv:1605.02719 [hep-ex]].

\bibitem{D0:2017ekd}
V.~M.~Abazov \textit{et al.} [D0],
``Measurement of the Effective Weak Mixing Angle in $p\bar{p}\rightarrow Z/\gamma^* \rightarrow \ell^+\ell^-$ Events,''
Phys. Rev. Lett. \textbf{120}, no.24, 241802 (2018)
[arXiv:1710.03951 [hep-ex]].

\bibitem{CDF:2018cnj}
T.~A.~Aaltonen \textit{et al.} [CDF and D0],
``Tevatron Run II combination of the effective leptonic electroweak mixing angle,''
Phys. Rev. D \textbf{97}, no.11, 112007 (2018)
[arXiv:1801.06283 [hep-ex]].

\bibitem{ATLAS:2015ihy}
G.~Aad \textit{et al.} [ATLAS],
``Measurement of the forward-backward asymmetry of electron and muon pair-production in $pp$ collisions at $\sqrt{s}$ = 7 TeV with the ATLAS detector,''
JHEP \textbf{09}, 049 (2015)
[arXiv:1503.03709 [hep-ex]].

\bibitem{LHCb:2015jyu}
R.~Aaij \textit{et al.} [LHCb],
``Measurement of the forward-backward asymmetry in $Z/\gamma^{\ast} \rightarrow \mu^{+}\mu^{-}$ decays and determination of the effective weak mixing angle,''
JHEP \textbf{11}, 190 (2015)
[arXiv:1509.07645 [hep-ex]].

\bibitem{CMS:2018ktx}
A.~M.~Sirunyan \textit{et al.} [CMS],
``Measurement of the weak mixing angle using the forward-backward asymmetry of Drell-Yan events in pp collisions at 8 TeV,''
Eur. Phys. J. C \textbf{78}, no.9, 701 (2018)
[arXiv:1806.00863 [hep-ex]].

\bibitem{ATLAS:2018gqq}
 ATLAS collaboration,
``Measurement of the effective leptonic weak mixing angle using electron and muon pairs from $Z$-boson decay in the ATLAS experiment at $\sqrt s = 8$ TeV,''
ATLAS-CONF-2018-037.

\bibitem{LEP:1991hsu}
G.~Alexander \textit{et al.} [LEP, ALEPH, DELPHI, L3 and OPAL],
``Electroweak parameters of the $Z^0$ resonance and the Standard Model: the LEP Collaborations,''
Phys. Lett. B \textbf{276}, 247-253 (1992)

\bibitem{Gambino:1993dd}
P.~Gambino and A.~Sirlin,
``Relation between $\sin^2\hat{\theta}_W^{}(m_Z^{})$ and $\sin^2 \theta_\mathrm{eff}^\mathrm{lept}$,''
Phys. Rev. D \textbf{49}, 1160-1162 (1994)
[arXiv:hep-ph/9309326 [hep-ph]].

\bibitem{Choudhury:2001hs}
D.~Choudhury, T.~M.~P.~Tait and C.~E.~M.~Wagner,
``Beautiful mirrors and precision electroweak data,''
Phys. Rev. D \textbf{65}, 053002 (2002)
[arXiv:hep-ph/0109097 [hep-ph]].

\bibitem{Wood:1997zq}
C.~S.~Wood, S.~C.~Bennett, D.~Cho, B.~P.~Masterson, J.~L.~Roberts, C.~E.~Tanner and C.~E.~Wieman,
``Measurement of parity nonconservation and an anapole moment in cesium,''
Science \textbf{275}, 1759-1763 (1997)

\bibitem{Guena:2004sq}
J.~Guena, M.~Lintz and M.~A.~Bouchiat,
``Measurement of the parity violating 6S-7S transition amplitude in cesium achieved within $2 \times 10^{-13}$ atomic-unit accuracy by stimulated-emission detection,''
Phys. Rev. A \textbf{71}, 042108 (2005)
[arXiv:physics/0412017 [physics.atom-ph]].

\bibitem{Qweak:2018tjf}
D.~Androi\'c \textit{et al.} [Qweak],
``Precision measurement of the weak charge of the proton,''
Nature \textbf{557}, no.7704, 207-211 (2018)
[arXiv:1905.08283 [nucl-ex]].

\bibitem{SLACE158:2005uay}
P.~L.~Anthony \textit{et al.} [SLAC E158],
``Precision measurement of the weak mixing angle in Moller scattering,''
Phys. Rev. Lett. \textbf{95}, 081601 (2005)
[arXiv:hep-ex/0504049 [hep-ex]].

\bibitem{PVDIS:2014cmd}
D.~Wang \textit{et al.} [PVDIS],
``Measurement of parity violation in electron\textendash{}quark scattering,''
Nature \textbf{506}, no.7486, 67-70 (2014)

\bibitem{Bouchiat:1983uf}
C.~Bouchiat and C.~A.~Piketty,
``Parity Violation in Atomic Cesium and Alternatives to the Standard Model of Electroweak Interactions,''
Phys. Lett. B \textbf{128} (1983), 73

\bibitem{Erler:2004in}
J.~Erler and M.~J.~Ramsey-Musolf,
``The Weak mixing angle at low energies,''
Phys. Rev. D \textbf{72}, 073003 (2005)
[arXiv:hep-ph/0409169 [hep-ph]].

\bibitem{Marciano:1980be}
W.~J.~Marciano and A.~Sirlin,
``Precise SU(5) Predictions for $\sin^2\theta_W^{\exp}$, $m_W$ and $m_Z$,''
Phys. Rev. Lett. \textbf{46}, 163 (1981)

\bibitem{Czarnecki:1998xc}
A.~Czarnecki and W.~J.~Marciano,
``Parity violating asymmetries at future lepton colliders,''
Int. J. Mod. Phys. A \textbf{13}, 2235-2244 (1998)
[arXiv:hep-ph/9801394 [hep-ph]].

\bibitem{Czarnecki:2000ic}
A.~Czarnecki and W.~J.~Marciano,
``Polarized Moller scattering asymmetries,''
Int. J. Mod. Phys. A \textbf{15}, 2365-2376 (2000)
[arXiv:hep-ph/0003049 [hep-ph]].

\bibitem{Ferroglia:2003wa}
A.~Ferroglia, G.~Ossola and A.~Sirlin,
``The Electroweak form-factor $\hat{\kappa}(q^2)$ and the running of $\sin^2 \hat{\theta}_W$,''
Eur. Phys. J. C \textbf{34}, 165-171 (2004)
[arXiv:hep-ph/0307200 [hep-ph]].

\bibitem{Erler:2017knj}
J.~Erler and R.~Ferro-Hern\'andez,
``Weak Mixing Angle in the Thomson Limit,''
JHEP \textbf{03}, 196 (2018)
[arXiv:1712.09146 [hep-ph]].

\bibitem{Czarnecki:1995fw}
A.~Czarnecki and W.~J.~Marciano,
``Electroweak radiative corrections to polarized M\o ller scattering asymmetries,''
Phys. Rev. D \textbf{53}, 1066-1072 (1996)
[arXiv:hep-ph/9507420 [hep-ph]].

\bibitem{RaAPV}
M. Nunez Portela \textit{et al.}, 
``Ra$+$ ion trapping: toward an atomic parity violation measurement and an optical clock,''
Appl. Phys. B \textbf{114}, 173-182 (2014).

\bibitem{Becker:2018ggl}
D.~Becker, R.~Bucoveanu, C.~Grzesik, K.~Imai, R.~Kempf, K.~Imai, M.~Molitor, A.~Tyukin, M.~Zimmermann and D.~Armstrong, \textit{et al.}
``The P2 experiment,''
Eur. Phys. J. A \textbf{54}, no.11, 208 (2018)
[arXiv:1802.04759 [nucl-ex]].

\bibitem{MOLLER:2014iki}
J.~Benesch \textit{et al.} [MOLLER],
``The MOLLER Experiment: An Ultra-Precise Measurement of the Weak Mixing Angle Using M\o ller Scattering,''
[arXiv:1411.4088 [nucl-ex]].

\bibitem{Chen:2014psa}
J.~P.~Chen \textit{et al.} [SoLID],
``A White Paper on SoLID (Solenoidal Large Intensity Device),''
[arXiv:1409.7741 [nucl-ex]].

\bibitem{USBelleIIGroup:2022qro}
D.~M.~Asner \textit{et al.} [US Belle II Group and Belle II/SuperKEKB e- Polarization Upgrade Working Group],
``Snowmass 2021 White Paper on Upgrading SuperKEKB with a Polarized Electron Beam: Discovery Potential and Proposed Implementation,''
[arXiv:2205.12847 [physics.acc-ph]].

\bibitem{AbdulKhalek:2021gbh}
R.~Abdul Khalek, A.~Accardi, J.~Adam, D.~Adamiak, W.~Akers, M.~Albaladejo, A.~Al-bataineh, M.~G.~Alexeev, F.~Ameli and P.~Antonioli, \textit{et al.}
``Science Requirements and Detector Concepts for the Electron-Ion Collider: EIC Yellow Report,''
Nucl. Phys. A \textbf{1026}, 122447 (2022)
[arXiv:2103.05419 [physics.ins-det]].

\bibitem{AbdulKhalek:2022hcn}
R.~Abdul Khalek, U.~D'Alesio, M.~Arratia, A.~Bacchetta, M.~Battaglieri, M.~Begel, M.~Boglione, R.~Boughezal, R.~Boussarie and G.~Bozzi, \textit{et al.}
``Snowmass 2021 White Paper: Electron Ion Collider for High Energy Physics,''
[arXiv:2203.13199 [hep-ph]].

\bibitem{Hook:2010tw}
A.~Hook, E.~Izaguirre and J.~G.~Wacker,
``Model Independent Bounds on Kinetic Mixing,''
Adv. High Energy Phys. \textbf{2011}, 859762 (2011)
[arXiv:1006.0973 [hep-ph]].

\bibitem{Curtin:2014cca}
D.~Curtin, R.~Essig, S.~Gori and J.~Shelton,
``Illuminating Dark Photons with High-Energy Colliders,''
JHEP \textbf{02}, 157 (2015)
[arXiv:1412.0018 [hep-ph]].

\bibitem{ATLAS:2021ldb}
G.~Aad \textit{et al.} [ATLAS],
``Search for Higgs bosons decaying into new spin-0 or spin-1 particles in four-lepton final states with the ATLAS detector with 139 fb$^{-1}$ of $pp$ collision data at $\sqrt{s}=13$ TeV,''
JHEP \textbf{03}, 041 (2022)
[arXiv:2110.13673 [hep-ex]].

\bibitem{CMS:2021pcy}
A.~Tumasyan \textit{et al.} [CMS],
``Search for low-mass dilepton resonances in Higgs boson decays to four-lepton final states in proton\textendash{}proton collisions at $\sqrt{s}=13\,\text {TeV} $,''
Eur. Phys. J. C \textbf{82}, no.4, 290 (2022)
[arXiv:2111.01299 [hep-ex]].

\bibitem{Branco:2011iw}
G.~C.~Branco, P.~M.~Ferreira, L.~Lavoura, M.~N.~Rebelo, M.~Sher and J.~P.~Silva,
``Theory and phenomenology of two-Higgs-doublet models,''
Phys. Rept. \textbf{516}, 1-102 (2012)
[arXiv:1106.0034 [hep-ph]].

\bibitem{Batell:2009yf}
B.~Batell, M.~Pospelov and A.~Ritz,
``Probing a Secluded U(1) at B-factories,''
Phys. Rev. D \textbf{79}, 115008 (2009)
[arXiv:0903.0363 [hep-ph]].

\bibitem{Glashow:1976nt}
S.~L.~Glashow and S.~Weinberg,
``Natural Conservation Laws for Neutral Currents,''
Phys. Rev. D \textbf{15}, 1958 (1977)

\bibitem{CMS:2019buh}
A.~M.~Sirunyan \textit{et al.} [CMS],
``Search for a Narrow Resonance Lighter than 200 GeV Decaying to a Pair of Muons in Proton-Proton Collisions at $\sqrt{s} =$  TeV,''
Phys. Rev. Lett. \textbf{124}, no.13, 131802 (2020)
[arXiv:1912.04776 [hep-ex]].

\bibitem{LHCb:2019vmc}
R.~Aaij \textit{et al.} [LHCb],
``Search for $A'\to\mu^+\mu^-$ Decays,''
Phys. Rev. Lett. \textbf{124}, no.4, 041801 (2020)
[arXiv:1910.06926 [hep-ex]].


\bibitem{Park:2015gdo}
J.~C.~Park, J.~Kim and S.~C.~Park,
``Galactic center GeV gamma-ray excess from dark matter with gauged lepton numbers,''
Phys. Lett. B \textbf{752}, 59-65 (2016)
[arXiv:1505.04620 [hep-ph]].

\bibitem{Alwall:2014hca}
J.~Alwall, R.~Frederix, S.~Frixione, V.~Hirschi, F.~Maltoni, O.~Mattelaer, H.~S.~Shao, T.~Stelzer, P.~Torrielli and M.~Zaro,
``The automated computation of tree-level and next-to-leading order differential cross sections, and their matching to parton shower simulations,''
JHEP \textbf{07}, 079 (2014)
[arXiv:1405.0301 [hep-ph]].

\bibitem{ATLAS:2022xlo}
G.~Aad \textit{et al.} [ATLAS],
``Search for dark photons from Higgs boson decays via $ZH$ production with a photon plus missing transverse momentum signature from $pp$ collisions at $\sqrt{s}$ = 13 TeV with the ATLAS detector,''
JHEP \textbf{07}, 133 (2023)
[arXiv:2212.09649 [hep-ex]].

\bibitem{BDX:2016akw}
M.~Battaglieri \textit{et al.} [BDX],
``Dark Matter Search in a Beam-Dump eXperiment (BDX) at Jefferson Lab,''
[arXiv:1607.01390 [hep-ex]].



\bibitem{NuTeV:2001whx}
G.~P.~Zeller \textit{et al.} [NuTeV],
``A Precise Determination of Electroweak Parameters in Neutrino Nucleon Scattering,''
Phys. Rev. Lett. \textbf{88}, 091802 (2002)
[erratum: Phys. Rev. Lett. \textbf{90}, 239902 (2003)]
[arXiv:hep-ex/0110059 [hep-ex]].

\bibitem{ParticleDataGroup:2016lqr}
``Electroweak model and constraints on new physics'' 
Review in C.~Patrignani \textit{et al.} [Particle Data Group],
``Review of Particle Physics,''
Chin. Phys. C \textbf{40}, no.10, 100001 (2016)

\bibitem{Bernon:2015qea}
J.~Bernon, J.~F.~Gunion, H.~E.~Haber, Y.~Jiang and S.~Kraml,
``Scrutinizing the alignment limit in two-Higgs-doublet models: m$_h$=125  GeV,''
Phys. Rev. D \textbf{92}, no.7, 075004 (2015)
[arXiv:1507.00933 [hep-ph]];
``Scrutinizing the alignment limit in two-Higgs-doublet models. II. m$_H$=125  GeV,''
Phys. Rev. D \textbf{93}, no.3, 035027 (2016)
[arXiv:1511.03682 [hep-ph]].

\bibitem{Haller:2018nnx}
J.~Haller, A.~Hoecker, R.~Kogler, K.~M\"onig, T.~Peiffer and J.~Stelzer,
``Update of the global electroweak fit and constraints on two-Higgs-doublet models,''
Eur. Phys. J. C \textbf{78}, no.8, 675 (2018)
[arXiv:1803.01853 [hep-ph]].

\bibitem{Chen:2019pkq}
N.~Chen, T.~Han, S.~Li, S.~Su, W.~Su and Y.~Wu,
``Type-I 2HDM under the Higgs and Electroweak Precision Measurements,''
JHEP \textbf{08}, 131 (2020)
[arXiv:1912.01431 [hep-ph]].

\bibitem{Arbey:2017gmh}
A.~Arbey, F.~Mahmoudi, O.~Stal and T.~Stefaniak,
``Status of the Charged Higgs Boson in Two Higgs Doublet Models,''
Eur. Phys. J. C \textbf{78}, no.3, 182 (2018)
[arXiv:1706.07414 [hep-ph]].

\bibitem{Aiko:2020ksl}
M.~Aiko, S.~Kanemura, M.~Kikuchi, K.~Mawatari, K.~Sakurai and K.~Yagyu,
``Probing extended Higgs sectors by the synergy between direct searches at the LHC and precision tests at future lepton colliders,''
Nucl. Phys. B \textbf{966}, 115375 (2021)
[arXiv:2010.15057 [hep-ph]].

\bibitem{Enomoto:2015wbn}
T.~Enomoto and R.~Watanabe,
``Flavor constraints on the Two Higgs Doublet Models of Z$_{2}$ symmetric and aligned types,''
JHEP \textbf{05}, 002 (2016)
[arXiv:1511.05066 [hep-ph]].

\bibitem{Deshpande:1977rw}
N.~G.~Deshpande and E.~Ma,
``Pattern of Symmetry Breaking with Two Higgs Doublets,''
Phys. Rev. D \textbf{18}, 2574 (1978)

\bibitem{Klimenko:1984qx}
K.~G.~Klimenko,
``On Necessary and Sufficient Conditions for Some Higgs Potentials to Be Bounded From Below,''
Theor. Math. Phys. \textbf{62}, 58-65 (1985)

\bibitem{Sher:1988mj}
M.~Sher,
``Electroweak Higgs Potentials and Vacuum Stability,''
Phys. Rept. \textbf{179}, 273-418 (1989)

\bibitem{Nie:1998yn}
S.~Nie and M.~Sher,
``Vacuum stability bounds in the two Higgs doublet model,''
Phys. Lett. B \textbf{449}, 89-92 (1999)
[arXiv:hep-ph/9811234 [hep-ph]].

\bibitem{Ferreira:2004yd}
P.~M.~Ferreira, R.~Santos and A.~Barroso,
``Stability of the tree-level vacuum in two Higgs doublet models against charge or CP spontaneous violation,''
Phys. Lett. B \textbf{603}, 219-229 (2004)
[erratum: Phys. Lett. B \textbf{629}, 114-114 (2005)]
[arXiv:hep-ph/0406231 [hep-ph]].

\bibitem{Flores:1982pr}
R.~A.~Flores and M.~Sher,
``Higgs Masses in the Standard, Multi-Higgs and Supersymmetric Models,''
Annals Phys. \textbf{148}, 95 (1983)

\bibitem{Kominis:1993zc}
D.~Kominis and R.~S.~Chivukula,
``Triviality bounds in two doublet models,''
Phys. Lett. B \textbf{304}, 152-158 (1993)
[arXiv:hep-ph/9301222 [hep-ph]].

\bibitem{Kanemura:1999xf}
S.~Kanemura, T.~Kasai and Y.~Okada,
``Mass bounds of the lightest CP even Higgs boson in the two Higgs doublet model,''
Phys. Lett. B \textbf{471}, 182-190 (1999)
[arXiv:hep-ph/9903289 [hep-ph]].

\bibitem{Ferreira:2009jb}
P.~M.~Ferreira and D.~R.~T.~Jones,
``Bounds on scalar masses in two Higgs doublet models,''
JHEP \textbf{08}, 069 (2009)
[arXiv:0903.2856 [hep-ph]].

\bibitem{Kanemura:1993hm}
S.~Kanemura, T.~Kubota and E.~Takasugi,
``Lee-Quigg-Thacker bounds for Higgs boson masses in a two doublet model,''
Phys. Lett. B \textbf{313}, 155-160 (1993)
[arXiv:hep-ph/9303263 [hep-ph]].

\bibitem{Akeroyd:2000wc}
A.~G.~Akeroyd, A.~Arhrib and E.~M.~Naimi,
``Note on tree level unitarity in the general two Higgs doublet model,''
Phys. Lett. B \textbf{490}, 119-124 (2000)
[arXiv:hep-ph/0006035 [hep-ph]].

\bibitem{Ginzburg:2005dt}
I.~F.~Ginzburg and I.~P.~Ivanov,
``Tree-level unitarity constraints in the most general 2HDM,''
Phys. Rev. D \textbf{72}, 115010 (2005)
[arXiv:hep-ph/0508020 [hep-ph]].

\bibitem{Kanemura:2015ska}
S.~Kanemura and K.~Yagyu,
``Unitarity bound in the most general two Higgs doublet model,''
Phys. Lett. B \textbf{751}, 289-296 (2015)
[arXiv:1509.06060 [hep-ph]].

\bibitem{Gunion:2002zf}
J.~F.~Gunion and H.~E.~Haber,
``The CP conserving two Higgs doublet model: The Approach to the decoupling limit,''
Phys. Rev. D \textbf{67}, 075019 (2003)
[arXiv:hep-ph/0207010 [hep-ph]].



\bibitem{Lu:2022bgw}
C.~T.~Lu, L.~Wu, Y.~Wu and B.~Zhu,
``Electroweak precision fit and new physics in light of the W boson mass,''
Phys. Rev. D \textbf{106}, no.3, 035034 (2022)
[arXiv:2204.03796 [hep-ph]].

\bibitem{Asadi:2022xiy}
P.~Asadi, C.~Cesarotti, K.~Fraser, S.~Homiller and A.~Parikh,
``Oblique Lessons from the $W$ Mass Measurement at CDF II,''
[arXiv:2204.05283 [hep-ph]].

\bibitem{Thomas:2022gib}
A.~W.~Thomas and X.~G.~Wang,
``Constraints on the dark photon from parity violation and the W mass,''
Phys. Rev. D \textbf{106}, no.5, 056017 (2022)
[arXiv:2205.01911 [hep-ph]].

\bibitem{Harigaya:2023uhg}
K.~Harigaya, E.~Petrosky and A.~Pierce,
``Precision Electroweak Tensions and a Dark Photon,''
[arXiv:2307.13045 [hep-ph]].

\bibitem{Lee:2022nqz}
H.~M.~Lee and K.~Yamashita,
``A model of vector-like leptons for the muon $g-2$ and the W boson mass,''
Eur. Phys. J. C \textbf{82}, no.8, 661 (2022)
[arXiv:2204.05024 [hep-ph]].

\bibitem{Lee:2021gnw}
H.~M.~Lee, J.~Song and K.~Yamashita,
``Seesaw lepton masses and muon $g-2$ from heavy vector-like leptons,''
J. Korean Phys. Soc. \textbf{79}, no.12, 1121-1134 (2021)
[arXiv:2110.09942 [hep-ph]].

\bibitem{Song:2022xts}
H.~Song, W.~Su and M.~Zhang,
``Electroweak phase transition in 2HDM under Higgs, Z-pole, and W precision measurements,''
JHEP \textbf{10}, 048 (2022)
[arXiv:2204.05085 [hep-ph]].

\bibitem{Heo:2022dey}
Y.~Heo, D.~W.~Jung and J.~S.~Lee,
``Impact of the CDF W-mass anomaly on two Higgs doublet model,''
Phys. Lett. B \textbf{833}, 137274 (2022)
[arXiv:2204.05728 [hep-ph]].

\bibitem{Ahn:2022xax}
Y.~H.~Ahn, S.~K.~Kang and R.~Ramos,
``Implications of New CDF-II $W$ Boson Mass on Two Higgs Doublet Model,''
Phys. Rev. D \textbf{106}, no.5, 055038 (2022)
[arXiv:2204.06485 [hep-ph]].

\bibitem{Lee:2022gyf}
S.~Lee, K.~Cheung, J.~Kim, C.~T.~Lu and J.~Song,
``Status of the two-Higgs-doublet model in light of the CDF mW measurement,''
Phys. Rev. D \textbf{106}, no.7, 075013 (2022)
[arXiv:2204.10338 [hep-ph]].

\bibitem{Bian:2017xzg}
L.~Bian, H.~M.~Lee and C.~B.~Park,
``$B$-meson anomalies and Higgs physics in flavored $U(1)'$ model,''
Eur. Phys. J. C \textbf{78}, no.4, 306 (2018)
[arXiv:1711.08930 [hep-ph]].

\bibitem{ATLAS:2019erb}
G.~Aad \textit{et al.} [ATLAS],
``Search for high-mass dilepton resonances using 139 fb$^{-1}$ of $pp$ collision data collected at $\sqrt{s}=$13 TeV with the ATLAS detector,''
Phys. Lett. B \textbf{796}, 68-87 (2019)
[arXiv:1903.06248 [hep-ex]].

\bibitem{CMS:2019gwf}
A.~M.~Sirunyan \textit{et al.} [CMS],
``Search for high mass dijet resonances with a new background prediction method in proton-proton collisions at $\sqrt{s} =$ 13 TeV,''
JHEP \textbf{05}, 033 (2020)
[arXiv:1911.03947 [hep-ex]].

\bibitem{Aboubrahim:2022qln}
A.~Aboubrahim, M.~M.~Altakach, M.~Klasen, P.~Nath and Z.~Y.~Wang,
``Combined constraints on dark photons and discovery prospects at the LHC and the Forward Physics Facility,''
JHEP \textbf{03}, 182 (2023)
[arXiv:2212.01268 [hep-ph]].

\bibitem{Alloul:2013bka}
A.~Alloul, N.~D.~Christensen, C.~Degrande, C.~Duhr and B.~Fuks,
``FeynRules  2.0 - A complete toolbox for tree-level phenomenology,''
Comput. Phys. Commun. \textbf{185}, 2250-2300 (2014)
[arXiv:1310.1921 [hep-ph]].

\bibitem{digfile}
\url{quark.phy.bnl.gov/Digital_Data_Archive/DarkZ_Wmass}.

\bibitem{Sirlin:1989qz}
A.~Sirlin,
``Considerations Concerning the Renormalization of the Electroweak Sector of the Standard Model,''
Nucl. Phys. B \textbf{332}, 20-38 (1990)

\bibitem{Jegerlehner1991}
F.~Jegerlehner, in ``Testing the Standard Model'', eds. M.~Cvetic and P.~Langacker, (World Scientific, Singapore 1991) p.~476. 

\bibitem{Marciano:1993jd}
W.~J.~Marciano,
``Spin and precision electroweak physics,''
BNL-60177 Lectures given at 21st Annual SLAC Summer Institute on Particle Physics: Spin Structure in High Energy Processes, Stanford, CA, 26 July\text{--}6 August 1993.

\bibitem{Cornwall:1981zr}
J.~M.~Cornwall,
``Dynamical Mass Generation in Continuum QCD,''
Phys. Rev. D \textbf{26}, 1453 (1982)

\bibitem{Cornwall:1989gv}
J.~M.~Cornwall and J.~Papavassiliou,
``Gauge Invariant Three Gluon Vertex in QCD,''
Phys. Rev. D \textbf{40}, 3474 (1989)

\bibitem{Papavassiliou:1989zd}
J.~Papavassiliou,
``Gauge Invariant Proper Selfenergies and Vertices in Gauge Theories with Broken Symmetry,''
Phys. Rev. D \textbf{41}, 3179 (1990)

\bibitem{Degrassi:1992ue}
G.~Degrassi and A.~Sirlin,
``Gauge invariant selfenergies and vertex parts of the Standard Model in the pinch technique framework,''
Phys. Rev. D \textbf{46}, 3104-3116 (1992)

\bibitem{Marciano:1980pb}
W.~J.~Marciano and A.~Sirlin,
``Radiative Corrections to Neutrino Induced Neutral Current Phenomena in the $SU(2)_L \times U(1)$ Theory,''
Phys. Rev. D \textbf{22}, 2695 (1980)
[erratum: Phys. Rev. D \textbf{31}, 213 (1985)]

\bibitem{Sirlin:1991fd}
A.~Sirlin,
``Theoretical considerations concerning the Z0 mass,''
Phys. Rev. Lett. \textbf{67}, 2127-2130 (1991)

\end{thebibliography}
\end{document}